\documentclass[aps,prd,superscriptaddress,preprint,tightenlines,nofootinbib,floatfix]{revtex4}

\usepackage{graphicx} % Include figure files
\usepackage{dcolumn}  % Align table columns on decimal point
\usepackage{bm}       % bold math
\usepackage{lscape}   % JL added for EPAPS addendums
\newcommand{\Kspipi}{\ensuremath{K^0_S \pi^+\pi^-}}
\newcommand{\kspp}{\ensuremath{K^0_S \pi^+\pi^-}}
\newcommand{\KsKK}{\ensuremath{K^0_S K^+K^-}}
\newcommand{\Kshh}{\ensuremath{K^0_S h^+h^-}}
\newcommand{\Klpipi}{\ensuremath{K^0_L \pi^+\pi^-}}
\newcommand{\klpp}{\ensuremath{K^0_L \pi^+\pi^-}}
\newcommand{\KlKK}{\ensuremath{K^0_L K^+K^-}}
\newcommand{\Klhh}{\ensuremath{K^0_L h^+h^-}}
\begin{document}

%\preprint line(s) will be ignored for PRL/PRD
%\preprint{CPDRAFT2010-004}
% For paper draft CBX YY-NN -> Draft YY-NNA
%\preprint{CLEO CONF YY-NN}   % For conference papers
%\preprint{ICHEP ABSnnn}      % For conference papers
\preprint{CLNS 10/2070}       % for CLNS notes
\preprint{CLEO 10-07}         % for CLNS notes

% Use this form if you DO NOT have mathematical symbols in the title
\title{\boldmath Model-independent determination of the strong-phase difference between
  $D^0$ and $\overline{D}^0\to K^{0}_{S,L}h^+h^-$ $(h=\pi,K)$ and its
  impact on the measurement of the CKM angle $\gamma/\phi_3$}

% Add \boldmath if you DO have mathematical symbols in the title
%\title{\boldmath Your Title Goes Here}

% for conference papers (ask CLEOAC for appropriate text)
%\thanks{Submitted to the 31$^{\rm st}$ International Conference on High Energy
%Physics, July 2002, Amsterdam}

%-------- INSERT HERE ------------
% Your author list goes here  REMOVE EVERYTHING to END INSERT and
% replace with your authorlist (ask cleoac).
\author{J.~Libby}
\affiliation{Indian Institute of Technology Madras, Chennai, Tamil Nadu 600036, India}
\author{M.~Kornicer}
\author{R.~E.~Mitchell}
\author{M.~R.~Shepherd}
\author{C.~M.~Tarbert}
\affiliation{Indiana University, Bloomington, Indiana 47405, USA }
\author{D.~Besson}
\affiliation{University of Kansas, Lawrence, Kansas 66045, USA}
\author{T.~K.~Pedlar}
\author{J.~Xavier}
\affiliation{Luther College, Decorah, Iowa 52101, USA}
\author{D.~Cronin-Hennessy}
\author{J.~Hietala}
\author{P.~Zweber}
\affiliation{University of Minnesota, Minneapolis, Minnesota 55455, USA}
\author{S.~Dobbs}
\author{Z.~Metreveli}
\author{K.~K.~Seth}
\author{A.~Tomaradze}
\author{T.~Xiao}
\affiliation{Northwestern University, Evanston, Illinois 60208, USA}
\author{S.~Brisbane}
\author{S.~Malde}
\author{L.~Martin}
\author{A.~Powell}
\author{P.~Spradlin}
\author{G.~Wilkinson}
\affiliation{University of Oxford, Oxford OX1 3RH, UK}
\author{H.~Mendez}
\affiliation{University of Puerto Rico, Mayaguez, Puerto Rico 00681}
\author{J.~Y.~Ge}
\author{D.~H.~Miller}
\author{I.~P.~J.~Shipsey}
\author{B.~Xin}
\affiliation{Purdue University, West Lafayette, Indiana 47907, USA}
\author{G.~S.~Adams}
\author{D.~Hu}
\author{B.~Moziak}
\author{J.~Napolitano}
\affiliation{Rensselaer Polytechnic Institute, Troy, New York 12180, USA}
\author{K.~M.~Ecklund}
\affiliation{Rice University, Houston, Texas 77005, USA}
\author{J.~Insler}
\author{H.~Muramatsu}
\author{C.~S.~Park}
\author{L.~J.~Pearson}
\author{E.~H.~Thorndike}
\author{F.~Yang}
\affiliation{University of Rochester, Rochester, New York 14627, USA}
\author{S.~Ricciardi}
\affiliation{STFC Rutherford Appleton Laboratory, Chilton, Didcot, Oxfordshire, OX11 0QX, UK}
\author{C.~Thomas}
\affiliation{University of Oxford, Oxford OX1 3RH, UK}
\affiliation{STFC Rutherford Appleton Laboratory, Chilton, Didcot, Oxfordshire, OX11 0QX, UK}
\author{M.~Artuso}
\author{S.~Blusk}
\author{N.~Horwitz}
\author{R.~Mountain}
\author{T.~Skwarnicki}
\author{S.~Stone}
\author{J.~C.~Wang}
\author{L.~M.~Zhang}
\affiliation{Syracuse University, Syracuse, New York 13244, USA}
\author{T.~Gershon}
\affiliation{University of Warwick, Coventry CV4 7AL, United Kingdom}
\author{G.~Bonvicini}
\author{D.~Cinabro}
\author{A.~Lincoln}
\author{M.~J.~Smith}
\author{P.~Zhou}
\author{J.~Zhu}
\affiliation{Wayne State University, Detroit, Michigan 48202, USA}
\author{P.~Naik}
\author{J.~Rademacker}
\affiliation{University of Bristol, Bristol BS8 1TL, UK}
\author{D.~M.~Asner}
\altaffiliation[Now at: ]{Pacific Northwest National Laboratory, Richland, WA 99352}
\author{K.~W.~Edwards}
\author{K.~Randrianarivony}
\author{G.~Tatishvili}
\altaffiliation[Now at: ]{Pacific Northwest National Laboratory, Richland, WA 99352}
\affiliation{Carleton University, Ottawa, Ontario, Canada K1S 5B6}
\author{R.~A.~Briere}
\author{H.~Vogel}
\affiliation{Carnegie Mellon University, Pittsburgh, Pennsylvania 15213, USA}
\author{P.~U.~E.~Onyisi}
\author{J.~L.~Rosner}
\affiliation{University of Chicago, Chicago, Illinois 60637, USA}
\author{J.~P.~Alexander}
\author{D.~G.~Cassel}
\author{S.~Das}
\author{R.~Ehrlich}
\author{L.~Fields}
\author{L.~Gibbons}
\author{S.~W.~Gray}
\author{D.~L.~Hartill}
\author{B.~K.~Heltsley}
\author{D.~L.~Kreinick}
\author{V.~E.~Kuznetsov}
\author{J.~R.~Patterson}
\author{D.~Peterson}
\author{D.~Riley}
\author{A.~Ryd}
\author{A.~J.~Sadoff}
\author{X.~Shi}
\author{W.~M.~Sun}
\affiliation{Cornell University, Ithaca, New York 14853, USA}
\author{J.~Yelton}
\affiliation{University of Florida, Gainesville, Florida 32611, USA}
\author{P.~Rubin}
\affiliation{George Mason University, Fairfax, Virginia 22030, USA}
\author{N.~Lowrey}
\author{S.~Mehrabyan}
\author{M.~Selen}
\author{J.~Wiss}
\affiliation{University of Illinois, Urbana-Champaign, Illinois 61801, USA}
\collaboration{CLEO Collaboration}
\noaffiliation
%-------- END INSERT ------------

%please hard code the date when you have a final draft and submit to CLEOAC
\date{13th October, 2010}

\begin{abstract} 
% Insert abstract here.
We report the first determination of the relative strong-phase difference between $D^{0}\to K^{0}_{S,L}K^{+}K^{-}$ %%@
and $\overline{D}^{0}\to K^{0}_{S,L}K^{+}K^{-}$. In addition, we present updated measurements of the relative %%@
strong-phase difference between $D^{0}\to K^{0}_{S,L}\pi^{+}\pi^{-}$ and $\overline{D}^{0}\to %%@
K^{0}_{S,L}\pi^{+}\pi^{-}$.
Both measurements exploit the quantum coherence between a pair of $D^{0}$ and $\overline{D}^{0}$ mesons produced %%@
from $\psi(3770)$ decays. The strong-phase differences measured are important for determining the %%@
Cabibbo-Kobayashi-Maskawa angle $\gamma/\phi_3$ in $B^{-}\to K^{-}\widetilde{D}^{0}$ decays, where %%@
$\widetilde{D}^{0}$ is a $D^{0}$ or $\overline{D}^{0}$ meson decaying to $K^{0}_{S}h^{+}h^{-}$ $(h=\pi,K)$, in a %%@
manner independent of the model assumed to describe the $D^{0}\to K^{0}_{S}h^{+}h^{-}$ decay. Using our results, %%@
the uncertainty in $\gamma/\phi_3$ due to the error
on the strong-phase difference is expected to be between $1.7^{\circ}$
and $3.9^{\circ}$ for an analysis using $B^- \to K^- \widetilde{D}^0$,
$\widetilde{D}^0 \to K^0_S \pi^+ \pi^-$ decays, and between $3.2^{\circ}$
and $3.9^{\circ}$ for an analysis based on
$B^- \to K^- \widetilde{D}^0$, $\widetilde{D}^0 \to K^0_S K^+ K^-$ decays. A measurement is also
 presented of the $CP$-odd fraction, $\mathcal{F}_{-}$,  of the decay $D^{0}\to K^{0}_{S}K^{+}K^{-}$ in
 the region of the $\phi\to K^{+}K^{-}$ resonance. We find that in a region within 0.01 $\mathrm{GeV}^2/c^4$
 of the nominal $\phi$ mass squared  $\mathcal{F}_{-} > 0.91$ at the 90\% confidence level.
 \end{abstract}

\pacs{13.25.Ft,12.15.Hh,14.40.Lb}
\maketitle
\section{Introduction}

A central goal of flavor physics is the determination of all elements of the CKM matrix~\cite{CKM},
both magnitudes and phases. Of the three angles of the $b-d$ CKM triangle, denoted $\alpha$, $\beta$,
and $\gamma$ by some, $\phi_2$, $\phi_1$, and $\phi_3$ by others, the least well determined is $\gamma/\phi_3$,
the phase of $V_{cb}$ relative to $V_{ub}$. It is of great interest to determine $\gamma/\phi_3$ using
the decay $B^\pm \to K^\pm \widetilde{D}^0$, since in this mode, the $\gamma/\phi_3$ value is expected to be
insensitive to new physics effects in $B$ decay; here, $\widetilde{D}^0$
is either a $D^0$ or $\overline{D}^0$ meson decaying to the same final state. This is in contrast with most %%@
measurements of $CP$ violation, which are dominated by processes that have significant contributions from loop %%@
diagrams that can be influenced by new physics \cite{CKMFITTER,UTFIT}. Therefore, precise measurements of %%@
$\gamma/\phi_3$ from the decay $B^\pm \to K^\pm \widetilde{D}^0$ compared to the predictions for $\gamma/\phi_3$ %%@
from loop-dominated processes provide a stringent test of the origin of $CP$ violation in the Standard Model.  %%@
Sensitivity to the angle $\gamma/\phi_3$ 
comes from the interference between two Cabibbo-suppressed diagrams: $b \to c \overline{u}s$, giving rise
to $B^- \to K^- D^0$,\footnote{Here and throughout this paper the charge-conjugate state
is implied unless otherwise stated.} and the color and CKM suppressed process $b \to u \overline{c}s$, giving rise
to $B^- \to K^- \overline{D}^0$ \cite{GLW}.  Promising $\widetilde{D}^0$ decays for measuring $\gamma/\phi_3$ using %%@
this
method are $\widetilde{D}^0\to\Kspipi$~\cite{BONDAR1,GIRI} and $\widetilde{D}^0 \to \KsKK$, 
here designated collectively as $\widetilde{D}^0 \to\Kshh$.  To make use of these decays, however,
it is necessary to understand the interference effects between $D^0 \to\Kshh$ and $\overline{D}^0 \to\Kshh$.
These effects can be measured using CLEO-c data. A study of the decay $\widetilde{D}^0 \to\Kspipi$ has 
already been published~\cite{BRIERE}. Here we present an update of that analysis, and first results from the
decay $\widetilde{D}^0 \to\KsKK$.

Let us write the amplitude for the $B^- \to K^- \widetilde{D}^0$, $\widetilde{D}^0 \to K^0_S h^+h^-$ decay as %%@
follows:
\begin{equation}
f_{B^{-}}(m^{2}_{+},m^{2}_{-})\propto f_D(m^{2}_{+},m^{2}_{-}) + r_B e^{i(\delta_B-\gamma)} %%@
f_{\overline{D}}(m^{2}_{+},m^{2}_{-}).
\label{eq:ampbdec1}
\end{equation}
Here, $m^{2}_{+}$ and $m^2_{-}$ are the invariant-mass squared of the $K^{0}_{S}h^{+}$ and $K^{0}_{S}h^{-}$ pairs %%@
from the  $\widetilde{D}^0\to\Kshh$ decay, which define the Dalitz plot, 
$f_D(m^{2}_{+},m^{2}_{-})(f_{\overline{D}}(m^{2}_{+},m^{2}_{-}))$ is the amplitude for the $D^0 (\overline{D}^0)$ %%@
decay to $\Kshh$ at $(m^{2}_{+},m^{2}_{-})$ in the Dalitz plot, $r_B$ is the ratio
of the suppressed to favored amplitudes, and $\delta_B$ is the
strong-phase difference between the color-favored and color-suppressed amplitudes.  Ignoring the second-order %%@
effects
of charm mixing and $CP$ violation in charm~\cite{GIRI,GROSS,BONDARCHARM}, we have %%@
$f_{\overline{D}}(m^{2}_{+},m^{2}_{-}) = f_D(m^{2}_{-},m^{2}_{+})$, and Eq.~(\ref{eq:ampbdec1}) can then
be written as:
\begin{equation}
f_{B^-}(m^{2}_{+},m^{2}_{-}) \propto f_D(m^{2}_{+},m^{2}_{-}) + r_B e^{i(\delta_B-\gamma)} %%@
f_{D}(m^{2}_{-},m^{2}_{+}).
\label{eq:ampbdec2}
\end{equation}
The square of the amplitude clearly depends on the phase difference $\Delta \delta_D \equiv %%@
\delta_D(m^{2}_{+},m^{2}_{-}) - \delta_D(m^{2}_{-},m^{2}_{+})$,
where $\delta_D(m^{2}_{+},m^{2}_{-})$ is the phase of $f_D(m^{2}_{+},m^{2}_{-})$.  Thus, for the determination of %%@
$\gamma/\phi_3$, one must know $\Delta \delta_D$.

Analyses of $B^{-}\to K^{-}\widetilde{D}^{0}$ decays to date extracted $\Delta \delta_D(m^{2}_{+},m^{2}_{-})$ for %%@
each final state by fitting flavor-tagged $D^0 \to$\Kspipi~\cite{BABAR1,BABAR2,BABAR3,BELLE,BELLE2} and $D^0 %%@
\to$\KsKK~\cite{BABAR2,BABAR3} Dalitz plots to $D^0$-decay models
involving various two-body intermediate states. The systematic uncertainty associated with this modeling is hard to 
 estimate; the assigned values vary between $3^\circ$ and $9^\circ$ for 
 the more recent analyses.  In order to exploit fully the high statistics
 expected at LHCb \cite{LHCB,TABS} and future $e^{+}e^{-}$ $B$-factory experiments \cite{BELLEII,SFF} it is highly 
 desirable to avoid this modeling uncertainty, and to do it in a manner which
 keeps all other error sources small compared with the foreseen statistical
 precision.

In the analysis presented here, we employ a model-independent approach to obtain $\Delta %%@
\delta_D(m^{2}_{+},m^{2}_{-})$ as suggested
in Refs.~\cite{GIRI,BONDAR2}, by exploiting the quantum coherence of $D^0 - \overline{D}^0$ pairs at the %%@
$\psi(3770)$.  Because
of this quantum correlation, $K^0_S h^+h^-$ and $K^0_L h^+ h^-$ decays recoiling against flavor tags, $CP$-tags, %%@
and
$D^0 \to K_S^0 h^+h^-$ tags, taken together, provide direct sensitivity to the quantities $\cos \Delta \delta_D$ 
and $\sin \Delta \delta_D$ for each final state.  The analysis is performed in discrete bins of $D^0 \to$\Kshh\
Dalitz space.  We have updated the  $D^0 \to$\Kspipi\ analysis reported in Ref.~\cite{BRIERE} by providing results %%@
in
alternative sets of Dalitz-plot bins, and by reducing some of the systematic uncertainties. 

 In addition measurements of the time-dependent evolution of the $D^{0}\to K^{0}_{S}h^{+}h^{-}$ Dalitz plot provide %%@
some of the most precise constraints on charm-mixing parameters \cite{K0SPIPICHARM}. These measurements also rely %%@
on $D^{0}$-decay models that introduce significant systematic uncertainties. A model-independent determination of %%@
the charm-mixing parameters from $D^{0}\to K^{0}_{S}h^{+}h^{-}$ that uses the same strong-phase difference %%@
parameters as the $\gamma/\phi_3$ analysis of $B^{-}$ decay has been proposed \cite{BONDARCHARM}. The advantage of %%@
the model-independent approach is again the elimination of model-dependent assumptions about the strong-phase %%@
differences.   

 We also present the first model-independent measurement of the $CP$ content of the decay $D^{0}\to K^{0}_{S} %%@
K^{+}K^{-}$ in the region of the $\phi\to K^{+}K^{-}$ resonance. The decay $D^{0}\to K^{0}_{S}\phi$ is a $CP$-odd %%@
eigenstate and has been used as such in several analyses; see for example Refs. \cite{LOWREY,BELLE_GLW}. The %%@
$\phi\to K^{+}K^{-}$ resonance is usually defined by a mass window about the nominal $\phi$ mass. Despite its %%@
narrow natural width of $4.26~\mathrm{MeV}/c^2$ \cite{PDG}, the potential contributions from $CP$-even final states %%@
beneath the $\phi$ resonance, such as $D^{0}\to K^{0}_{S}a_{0}(980)$ and non-resonant $D^{0}\to %%@
K^{0}_{S}K^{+}K^{-}$ decays, must be accounted for. Using $D^{0}\to K^{0}_{S,L}K^{+}K^{-}$ decays recoiling against %%@
$CP$ eigenstates we determine the $CP$-odd fraction of decays, $\mathcal{F}_{-}$, in the region close to the $\phi$ %%@
resonance. A measurement of $\mathcal{F}_{-}$ allows a systematic uncertainty related to the $CP$-even %%@
contamination to $D^{0}\to K^{0}_{S}\phi$ decays to be assigned without assuming an amplitude model for the decay  %%@
$D^{0}\to K^{0}_{S}K^{+}K^{-}$.     

This paper is organized as follows.   The formalism for the measurement of the strong-phase difference and %%@
$\mathcal{F}_{-}$ is outlined
in Sec.~\ref{sec:formalism}. The choice of Dalitz-plot bins
is given in Sec.~\ref{sec:bindef}. The event selection is described in Sec.~\ref{sec:eventsel}. %%@
Sections~\ref{sec:cisiextract} and \ref{sec:systematic} present the extraction
of the variables associated with the strong-phase differences and the assignment of systematic uncertainties, %%@
respectively.
The impact of these results on the measurement of $\gamma/\phi_3$ is discussed in Sec.~\ref{sec:gamma}, along with %%@
the measurement of $\mathcal{F_{-}}$. A summary
is given in Sec.~\ref{sec:summary}.  Throughout this article the $D^0 \to$\Kspipi\ and  $D^0 \to$\KsKK\ analyses %%@
are described in
parallel, but more weight is given to the latter as it has not been presented previously.

\section{Formalism}
\label{sec:formalism}

Giri {\it et al.} proposed~\cite{GIRI} a model-independent procedure for obtaining $\Delta %%@
\delta_D(m^{2}_{+},m^{2}_{-})$
as follows.  The Dalitz plot is divided into $2{\cal N}$ bins, symmetrically about the line $m_{+}^{2}=m_{-}^2$.
The bins are indexed with $i$, running from $-{\mathcal N}$ to ${\mathcal N}$ excluding zero. Thus the coordinate %%@
exchange $m^{2}_{+} \leftrightarrow m^{2}_{-}$ 
corresponds to the exchange of the bins $i \leftrightarrow -i$. The number of events $(K_i)$ in the %%@
$i^{\mathrm{th}}$ bin of 
a flavor-tagged \Kshh\ Dalitz plot from a $D^0$ decay is then expressed as
\begin{equation}
K_i = A_D \int_i |f_D(m^{2}_{+},m^{2}_{-})|^2 dm^{2}_{+} dm^{2}_{-} = A_D F_i,
\label{eq:kidef}
\end{equation}
where the integral is performed over the $i^{\mathrm{th}}$ bin. Here $A_D$ is a normalization factor and $F_i$ is %%@
the fraction of $D^{0}\to K^{0}_{S}h^{+}h^{-}$ events in the $i^{\mathrm{th}}$ bin. The interference between the %%@
$D^0$ and $\overline{D}^0$ amplitudes is parameterized by two quantities
\begin{equation}
c_i \equiv \frac{1}{\sqrt{F_i F_{-i}}} \int_i |f_D (m^{2}_{+},m^{2}_{-})||f_D(m^{2}_{-},m^{2}_{+})|\cos[\Delta %%@
\delta_D (m^{2}_{+},m^{2}_{-})]dm^{2}_{+} dm^{2}_{-},
\end{equation}
and
\begin{equation}
s_i \equiv \frac{1}{\sqrt{F_i F_{-i}}} \int_i |f_D (m^{2}_{+},m^{2}_{-})||f_D(m^{2}_{-},m^{2}_{+})|\sin[\Delta %%@
\delta_D (m^{2}_{+},m^{2}_{-})]dm^{2}_{+} dm^{2}_{-}.
\end{equation}
The parameters $c_i$ and $s_i$ are the amplitude-weighted
averages of $\cos\Delta \delta_D$ and $\sin\Delta \delta_D$ over each Dalitz-plot bin.

\begin{figure}
\includegraphics*[width=5.75in]{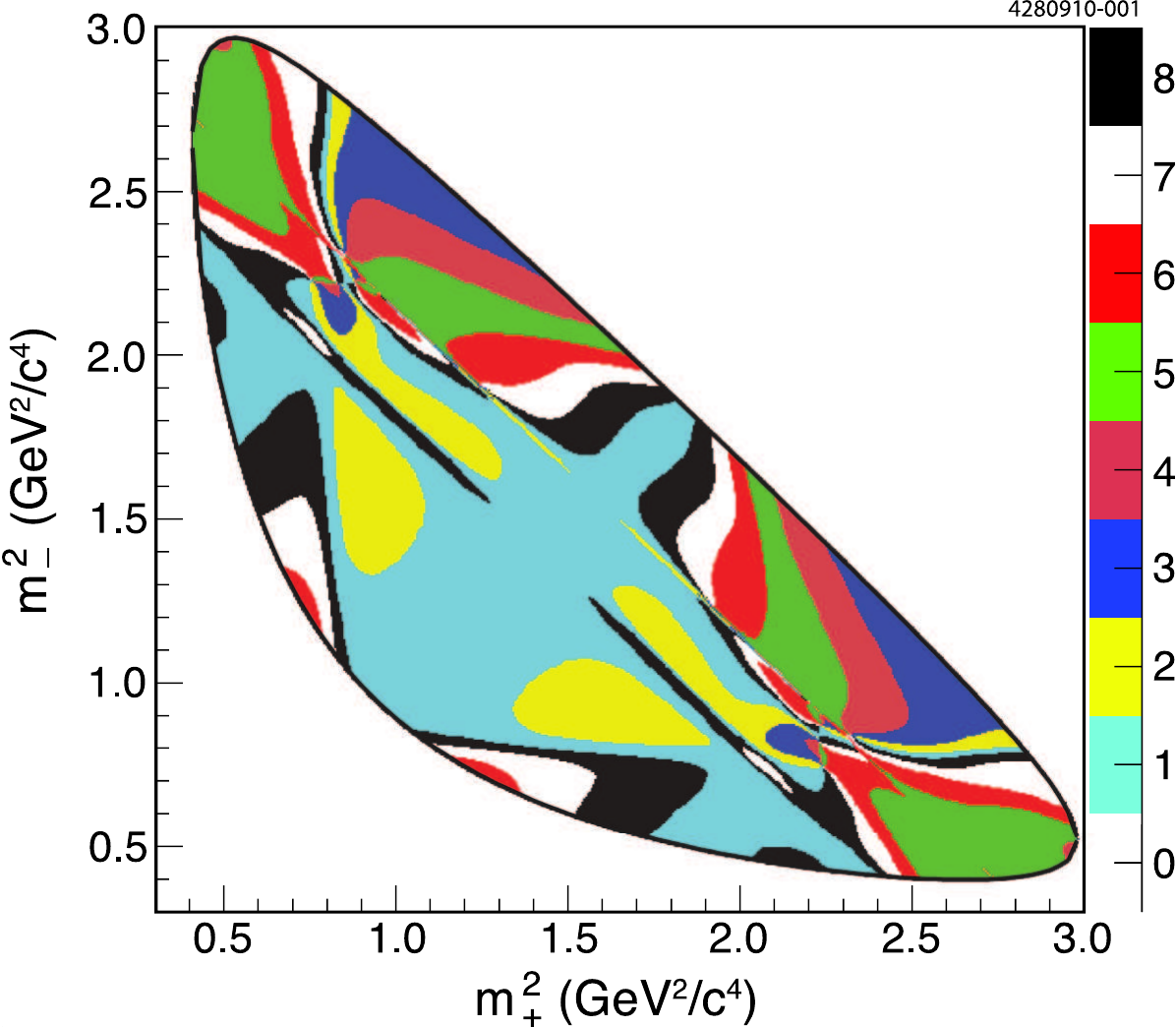}
\caption{Equal $\Delta\delta_D$ binning of the $D^0 \to$\Kspipi\ Dalitz plot with ${\cal N}=8$ based on
the model from Ref.~\cite{BABAR2}. The color scale represents the absolute value of the bin number, $|i|$.}
\label{fig:babar1bins}
\end{figure}

Though the original idea of Giri {\it et al.} was to divide the Dalitz plot into square bins,
Bondar and Poluektov noted~\cite{BONDAR2} that alternative bin definitions will lead
to significantly increased sensitivity.  In particular, one can choose to minimize
the variation in $\Delta \delta_D$ over each bin according to the predictions of one of
the models developed on flavor-tagged data~\cite{BELLE,BABAR1,BABAR2,BELLE2,BABAR3}. Note that this
approach does {\it not} introduce a model-dependence in the final result for $\gamma/\phi_3$.
This result will remain unbiased by the choice of an incorrect model, but will have 
less  statistical sensitivity than expected.  If  we divide the Dalitz plot into
${\cal N}$ bins of equal size with respect to $\Delta \delta_D$ as predicted by one of
these models, then in the half of the Dalitz plot $m^2_+<m^2_-$, the
$i^{\mathrm{th}}$ bin is defined by the condition
\begin{equation}
2\pi(i - 3/2)/{\cal N} < \Delta \delta_D (m^{2}_{+},m^{2}_{-}) < 2 \pi (i - 1/2)/{\cal N},
\label{eq:binning}
\end{equation}
and the $-i^{\mathrm{th}}$ bin is defined symmetrically in the lower portion of the Dalitz plot.
The choice of $D^0 \to$\Kspipi\ binning with ${\cal N}=8$ as obtained from the model presented in %%@
Ref.~\cite{BABAR2}
is shown in Fig.~\ref{fig:babar1bins}.  A discussion on alternative choices of binning for $D^0 \to$\Kspipi\, and
those for $D^0 \to$\KsKK\ can be found in Sec.~\ref{sec:bindef}.

We now describe how CLEO-c data can be used to determine $c_i$ and $s_i$.  The event yields in
the $i^{\mathrm{th}}$ bin of both flavor-tagged and $CP$-tagged ${\widetilde D}^0 \to \Kshh$ Dalitz plot are 
required.  Because the $\psi(3770)$ has $C=-1$, the $CP$ eigenvalue of one $D$ meson can be determined by
reconstructing the companion $D$ meson in a $CP$ eigenstate. With a $CP$-tagged
${\widetilde D}^0 \to\Kshh$ decay, the amplitude is given by
\begin{equation}
f_{CP\pm}(m^{2}_{+},m^{2}_{-}) = \frac{1}{\sqrt{2}}[f_D(m^{2}_{+},m^{2}_{-}) \pm f_D(m^{2}_{-},m^{2}_{+})], 
\end{equation}
for $CP$-even $(+)$ and $CP$-odd $(-)$ states of a
${\widetilde D}^0 \to$\Kshh\ decay. Since the event rate is proportional to the square of this
amplitude, the number of events in the $i^{\mathrm{th}}$ bin of a $CP$-even  or $CP$-odd  tagged Dalitz plot is %%@
then
\begin{equation}
M_i^{\pm} = h_{CP\pm}(K_i \pm 2 c_i \sqrt{K_i K_{-i}} + K_{-i}),
\label{eq:kscptag}
\end{equation}
where $h_{CP\pm} = S^\pm/2S_f$ is a normalization factor that depends on the number, $S_f$, of
flavor-tagged signal decays, and the number, $S^\pm$, of $D^{0}$-mesons decaying to a $CP$ eigenstate in the sample %%@
irrespective of the decay of the other $D$ meson; this is referred to as a single-tagged (ST) sample. Alternatively %%@
the normalization factor can be defined in terms of branching fractions, $\mathcal{B}$, as 
$h_{CP^{\pm}}=\mathcal{B}_{CP^{\pm}}/2\mathcal{B}_{f}$, where $\mathcal{B}_{CP^{\pm}}$ $(\mathcal{B}_{f})$ is the %%@
branching fraction of $D^{0}$ to $CP$ eigenstates (flavor tags).
Thus, access to $c_i$ is enabled by measuring the number of events, $M_i^\pm$, in a $CP$-tagged
\Kshh\ Dalitz plot, and the number of events, $K_i$, in a flavor-tagged \Kshh\ Dalitz plot.

Important additional information can be gained through analysis of $D^0 \to$\Kshh\ {\it vs.}
${\overline D}^0 \to$\Kshh\ data.  The amplitude for $\psi(3770)$ decaying to a double \Kshh\ final state
is as follows:
\begin{equation}
f(m^{2}_{+},m^{2}_{-},m^{\prime 2}_{+},m^{\prime 2}_{-}) = \frac{f_D(m^{2}_{+},m^{2}_{-})f_D(m^{\prime %%@
2}_{-},m^{\prime 2}_{+}) - f_D(m^{\prime 2}_{+},m^{\prime 2}_{-})f_D(m^{2}_{-},m^{2}_{+})}{\sqrt{2}}.
\end{equation}
The primed and unprimed Dalitz-plot coordinates correspond to the Dalitz-plot
variables of the two $\widetilde{D}^0 \to$\Kshh\ decays. Defining $M_{ij}$ as the
event rate in the $i^{\mathrm{th}}$ bin of the first and the $j^{\mathrm{th}}$ bin of the
second ${\widetilde D} \to$\Kshh\ Dalitz plots, respectively, we have:
\begin{equation}
M_{ij} = h_{\mathrm{corr}}(K_iK_{-j} + K_{-i}K_{j} - 2 \sqrt{K_i K_{-j}K_{-i}K_j} (c_i c_j + s_is_j)).
\label{eq:kskstag}
\end{equation}
Here, $h_{\mathrm{corr}} = N_{D \overline{D}}/2S_f^2 = N_{D\overline{D}}/8\mathcal{B}_f^2$, where %%@
$N_{D\overline{D}}$ is the number of $D\overline{D}$ pairs, and as before $S_f$ is the number of flavor-tagged %%@
signal decays.
Thus analysis of both
$D^0 \to$\Kshh\ {\it vs.} ${\overline D}^0 \to$\Kshh\ data and $CP$-tagged $D^0 \to$\Kshh\ decays
allows $c_i$ and $s_i$ to be determined.  The ambiguity in the sign of $s_i$ can
be resolved using weak model assumptions.

The decay $D^0 \to$\Klhh, due to its close relations with $D^0 \to$\Kshh, can be used
to improve further the $c_i$ and $s_i$ determination. We assume the convention that $A(D^0 \to K^0_S h^+h^-) = %%@
A(\overline{D}^0 \to K^0_S h^-h^+)$. Then, since the $K^0_S$ and $K^0_L$ mesons
are of opposite $CP$,
it follows that $A(D^0 \to K^0_L h^+h^-) = -A(\overline{D}^0 \to K^0_L h^-h^+)$. Hence for \Klhh\ the
Dalitz-plot rates of Eqs.~(\ref{eq:kscptag}) and (\ref{eq:kskstag}) become
\begin{equation}
M_i^{\prime\pm} = h_{CP\pm}(K_i' \mp 2 c_i' \sqrt{K_i' K'_{-i}} + K_{-i}'),
\label{eq:klcptag}
\end{equation}
and
\begin{equation}
M^{\prime}_{ij} = h_{corr}(K_iK_{-j}' + K_{-i}K_{j}' + 2 \sqrt{K_i K_{-j}'K_{-i}K_j'} (c_i c_j' + s_i s_j')),
\label{eq:kskltag}
\end{equation}
for $CP$ $vs.$ $D^0 \to$\Klhh\ and $D^0 \to$\Kshh\ $vs.$ $\overline{D}^0 \to$\Klhh\ respectively, where %%@
$K^{\prime}_{i}$
$c_i^{\prime}$, and $s_i^{\prime}$ are associated with the decay $D^0 \to$\Klhh.   In the analysis all four %%@
parameters
$c_i$, $s_i$, $c_i'$, and $s_i'$ are determined for each channel, but in order to improve
the precision on $c_i$ and $s_i$ constraints are imposed on the differences $\Delta c_i \equiv c_i'-c_i$
and $\Delta s_i \equiv s_i' - s_i$.  These constraints are discussed in Sec.~\ref{sec:cisiextract}.

Because the branching fraction of $D^0 \to K^0 \pi^+\pi^-$ is around five times larger than $D^0 \to K^0 K^+ K^-$ 
it is advantageous to first determine the coefficients $c_i$, $s_i$, $c_i'$, and $s_i'$ for the former decay,
and then use these to help improve our knowledge of the coefficients for $D^0 \to K^0 K^+ K^-$.  This is achieved
through measuring the bin-by-bin rates for $\KsKK$ $vs.$ $\Kspipi$, $\KlKK$ $vs.$ $\Kspipi$, and $\KsKK$ $vs.$ %%@
$\Klpipi$ Dalitz plots, and using suitably modified forms of Eqs.~(\ref{eq:kskstag}) and (\ref{eq:kskltag}).

The expression for the $CP$-odd fraction in the region of the $\phi\to K^{+}K^{-}$ resonance in $D^{0}\to %%@
K^{0}_{S}K^{+}K^{-}$ decays follows from Eq.~(\ref{eq:kscptag}). We note that %%@
$M^{+}_{i}+M^{-}_{i}=M^{+}_{-i}+M^{-}_{-i}$; in addition, this sum is proportional to $K_{i}+K_{-i}$, the total %%@
rate of $D^{0}\to K^{0}_{S} K^{+}K^{-}$ decays in the combined $i$ and $-i$ bins. Therefore, if bin $i$ defines an %%@
interval of $K^{+}K^{-}$ invariant-mass squared, $m_{K^{+}K^{-}}^2$, about the nominal $\phi$ mass squared, then %%@
the $CP$-odd fraction of $D^{0}\to K^{0}_{S}K^{+}K^{-}$ decays, $\mathcal{F}_{-}$, in that region is
\begin{equation}
\mathcal{F}_{-} = \frac{{M}^{-}_{i}+{M}^{-}_{-i}}{{M}^{-}_{i}+{M}^{-}_{-i}+{M}^{+}_{i}+{M}^{+}_{-i}} \; .
\label{eq:fminus}
\end{equation}  
We determine $\mathcal{F}_{-}$ for four different invariant-mass squared intervals: 0.006, 0.010, 0.014, and 0.018 %%@
$\mathrm{GeV^2}/c^4$. 

\section{Dalitz Plot Bin Definitions}
\label{sec:bindef}
Measurements of $c_i^{(\prime)}$ and $s_i^{(\prime)}$ are presented for three and four alternative binnings for %%@
$D^{0}\to K^{0}_{S} K^{+}K^{-}$ and $D^{0}\to K^{0}_{S}\pi^{+}\pi^{-}$, respectively. The motivation for these %%@
choices and the resulting binning definitions are presented in this section.

\subsection{\boldmath Binnings of the $D^{0}\to\KsKK$ Dalitz plot}
\label{subsec:k0skkbin}

We use an amplitude model determined by {\it BABAR}~\cite{BABAR3} for which a lookup table of the results in Dalitz %%@
space has been provided by the authors \cite{BITMAPS}. The amplitude model uses the isobar formalism and consists %%@
of eight intermediate resonances, of which five are parameterized with Breit-Wigner lineshapes and three, %%@
$a_0(980)^0K^0_S$ and $a_0(980)^{\pm}K^{\mp}$, are parameterized by a coupled-channel Breit-Wigner function %%@
\cite{FLATTE}.

 We consider binnings in which the Dalitz plane is divided into $\mathcal{N}=2$, $\mathcal{N}=3$, and %%@
$\mathcal{N}=4$ equal $\Delta\delta_D$ bins, according to Eq.~(\ref{eq:binning}). A smaller number of bins provides %%@
superior statistical precision on the parameters associated with $\Delta \delta_D$ but a reduced sensitivity to %%@
$\gamma$. Using a larger number of bins is not feasible due to the limited statistics available in the CLEO-c data; %%@
the fit to the parameters (Sec.~\ref{sec:cisiextract}) fails to converge if $\mathcal{N}>4$. However, these %%@
alternatives will allow flexibility in matching an appropriate number of bins for the size of the available %%@
$B$-decay sample when the values $c_{i}$ and $s_{i}$ are used to extract $\gamma$. 

\begin{figure}[htb]
 \begin{center}
\includegraphics*[width=1.0\columnwidth]{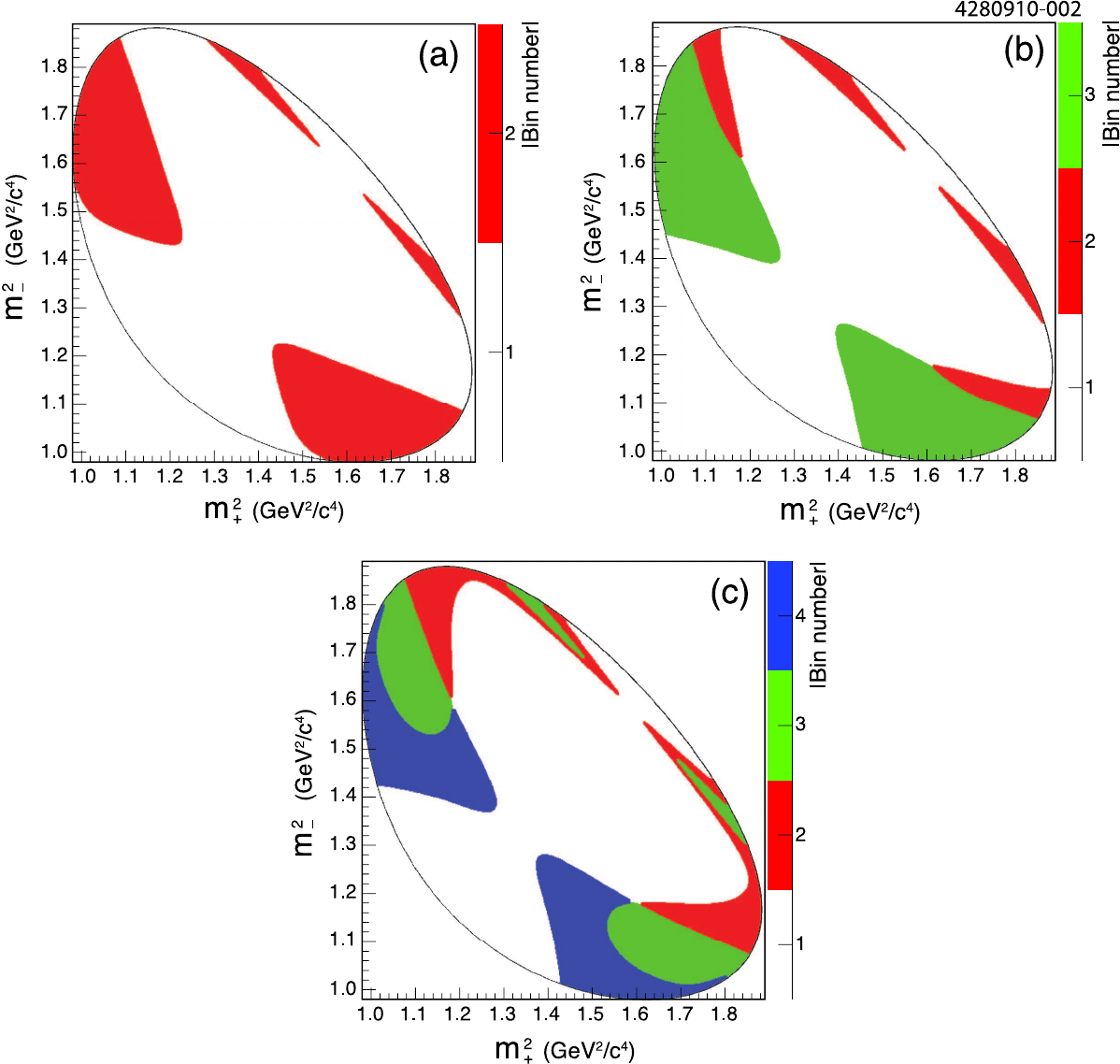}
 \caption{Equal $\Delta\delta_D$ division of the $D^{0}\to\KsKK$ Dalitz plot into (a) $\mathcal{N}=2$, (b) %%@
$\mathcal{N}=3$, and (c) $\mathcal{N}=4$ bins.}\label{fig:KsKKbinsequal}
 \end{center}
\end{figure}

The three cases considered are shown in Fig.~\ref{fig:KsKKbinsequal}. In each case, there is a narrow bin located %%@
at low values of $m^2_{K^+K^-}$, which is close to the diagonal boundary of the Dalitz plot. This bin encompasses %%@
the $\phi$ intermediate resonance and typically contains the largest number of events. For three and four bins %%@
there are `lobes' at high values of $m^2_{K^+K^-}$ that contain relatively few events.

\subsection{\boldmath Binnings of the $D^{0}\to\Kspipi$ Dalitz plot}
The four binnings used in the updated analysis of $D^{0}\to K^{0}_{S}\pi^{+}\pi^{-}$ are described in this section. %%@
The {\it BABAR} model \cite{BABAR2} that is used to define the bin choices is described in %%@
Sec.~\ref{subsec:babarmodel}. Then the binning in equal intervals of the strong-phase difference is presented in %%@
Sec.~\ref{subsec:Kmatrixbinning}. In Sec.~\ref{subsec:optimal_binning} the procedure to optimize the binning for %%@
maximal sensitivity to $\gamma$ is described and the resulting binning is presented. In %%@
Sec.~\ref{subsec:alt_binning} we describe a modified procedure of optimization which takes into account expected %%@
background levels at LHCb. In Sec. \ref{subsec:belle_model} the binning in equal intervals of the strong-phase %%@
difference as defined by the latest Belle model \cite{BELLE2} is given.   Finally, in %%@
Sec.~\ref{subsec:ultimate_babar} we summarize the differences between the various bin definitions and also assess %%@
what consequences a very recent {\it BABAR}
$D^0 \to K^0_S \pi^+ \pi^-$ model \cite{BABAR3}, not available originally for our analysis, would have for the bin %%@
definitions.

\subsubsection{{\it BABAR} $K$-matrix model}
\label{subsec:babarmodel}
The amplitude models used by Belle \cite{BELLE,BELLE2} and the first {\it BABAR} analysis \cite{BABAR1} are %%@
parameterized in terms of a Breit-Wigner isobar model. However, the broad $\pi\pi$ and $K\pi$ $S$-wave components %%@
are not well described by such Breit-Wigner lineshapes. In particular an additional intermediate state,  %%@
$\sigma^\prime$, is required to fit the $\pi\pi$ $S$-wave even though it is known not to be a physical resonance. %%@
Furthermore, parameterizing these broad overlapping states in terms of Breit-Wigner functions is unphysical in that %%@
unitarity can be violated. Therefore, a more recent {\it BABAR} model \cite{BABAR2} uses the $K$-matrix %%@
\cite{KMATRIX} ansatz to parameterize the $\pi\pi$ $S$-wave contributions, which does not violate unitarity. In %%@
addition, the empirical LASS lineshape \cite{LASS} of the $K^{*}_{0}(1430)$ is used to improve the fit to the %%@
$K\pi$ $S$-wave component. We refer to this approach as the {\it BABAR 2008 model}. The reduced $\chi^2$ for the  %%@
{\it BABAR} 2008 model fit to the $D^{*+}\to D^{0}\pi^{+}$ data is significantly improved over the first {\it %%@
BABAR} model \cite{BABAR1} and is much better than that for the Belle model. Therefore, this model was considered %%@
the best available and is used to define the nominal binnings implemented in this analysis. The results of the %%@
model have been made available by the {\it BABAR} collaboration in the form of a lookup table of the amplitude and %%@
strong-phase difference, $\delta_D$, in a fine grid of $0.0054~\mathrm{GeV^2}/c^4 \times 0.0054~\mathrm{GeV^2}/c^4$ %%@
sub-bins of the Dalitz plot variables $m^{2}_{\pm}$ \cite{BITMAPS}. The $m_{\pm}^{2}$ resolution, estimated from %%@
simulation, is of the 
same order as the sub-bin dimensions; the resolution is 0.006~$\mathrm{GeV}^2/c^4$ (0.015~$\mathrm{GeV}^2/c^4$) for %%@
$D^{0}\to K^{0}_{S}\pi^{+}\pi^{-}$ ($D^{0}\to K^{0}_{L}\pi^{+}\pi^{-}$). (The lookup table for $D^{0}\to %%@
K^{0}_{S}K^{+}K^{-}$ has a grid of $0.0018~\mathrm{GeV^2}/c^4 \times 0.0018~\mathrm{GeV^2}/c^4$ sub-bins. The %%@
resolution is 0.005~$\mathrm{GeV}^2/c^4$ (0.010~$\mathrm{GeV}^2/c^4$) for $D^{0}\to K^{0}_{S}K^{+}K^{-}$ ($D^{0}\to %%@
K^{0}_{L}K^{+}K^{-}$).)

Since performing the analysis of the CLEO-c data using the bin choices described here,
which are based on the {\it BABAR} 2008 model, a new {\it BABAR} measurement \cite{BABAR3} has been published which
presents an updated version of the $D^0\to K^0_S \pi^+\pi^-$ decay model that we 
term the {\it BABAR 2010 model}. This model is derived from a larger data sample that has been reprocessed and the %%@
analysis has been improved with respect to experimental systematic uncertainties. In %%@
Sec.~\ref{subsec:ultimate_babar} we assess the consequences on the bin choices of the differences between the two %%@
models and conclude that they are minor.

\subsubsection{Equal $\Delta\delta_D$ binning of the {\it BABAR} 2008  model}
\label{subsec:Kmatrixbinning}
The first binning of the Dalitz space for the {\it BABAR} 2008 model we consider follows the proposal in %%@
Ref.~\cite{BONDAR2}, which was used in the previous CLEO-c analysis \cite{BRIERE} and in the analysis of $D^{0}\to %%@
K^{0}_{S}K^{+}K^{-}$. The binning is such that there are $\mathcal{N}=8$ bins of $\Delta\delta_D$ in each half the %%@
Dalitz plot as defined in Eq.~(\ref{eq:binning}). This equal $\Delta\delta_D$ binning for the {\it BABAR} 2008 %%@
model is shown in Fig.~\ref{fig:babar1bins}.

\subsubsection{Optimal binning of the {\it BABAR} 2008 model} 
\label{subsec:optimal_binning}
 Following Ref.~\cite{BONDAR2} the ratio of sensitivity to $\gamma/\phi_3$ of a binned compared to an unbinned %%@
method is given by
 \begin{equation}
 \label{eq:originalQ}
Q^{2} = \frac{\displaystyle\sum_i \left[\left(\frac{1}{\sqrt{\Gamma_i}}\frac{d\Gamma_i}{dx}\right)^2 +  %%@
\left(\frac{1}{\sqrt{\Gamma_i}}\frac{d\Gamma_i}{dy}\right)^2\right]}{\displaystyle \int \left[ %%@
\left(\frac{1}{\sqrt{|f_{B^-}|^2}} \frac{d|f_{B^-}|^2}{dx}\right)^2 +  \left(\frac{1}{\sqrt{|f_{B^-}|^2}} %%@
\frac{d|f_{B^-}|^2}{dy}\right)^2\right] dm^{2}_{+}dm^{2}_{-}} \;,  
\end{equation}
where
\begin{equation}
 \Gamma_i = \int_i{|f_{B^{-}}|^2 dm^{2}_{+}dm^{2}_{-}} \; .
\end{equation}
Here, $f_{B^-}$ is expressed as 
\begin{equation}
 f_{B^{-}}  =  f_D(m^2_+,m^2_-) + (x+iy)f_D(m^2_-,m^2_+) \;, 
\end{equation}
where $x=r_B\cos{(\delta_B-\gamma)}$ and $y=r_B\sin{(\delta_B-\gamma)}$. The parameter, $Q$, is the ratio of the %%@
number of standard deviations difference in the yields of $B^{-}\to K^{-}\widetilde{D}^{0}$ events as $x$ and $y$ %%@
change in a finite number of bins with respect to an infinite number of bins. The sensitivity to $x$ and $y$ is %%@
largely independent of their values. Therefore, again following Ref.~\cite{BONDAR2}, Eq.~(\ref{eq:originalQ}) can %%@
be simplified assuming $x=y=0$ to become
\begin{equation}
\label{eq:Q}
 Q^2|_{x=y=0} = \frac{\sum_{i} N_i (c_{i}^2+s_{i}^2)}{\sum_{i} N_i} \; ,
\end{equation} 
where $N_i$ is the number of $B^{-}\to K^{-}\widetilde{D}^{0}(K^{0}_{S}\pi^{+}\pi^{-})$ events in the %%@
$i^{\mathrm{th}}$ bin when $r_B$ is zero. Recalling that $c_i$ and $s_i$ are the amplitude-weighted averages of %%@
$\cos{\Delta\delta_{D}}$ and $\sin{\Delta\delta_{D}}$ over each bin, respectively, 
it is clear that regions of similar $\Delta\delta_D$ will yield reasonable, though not necessarily optimal, values %%@
of $Q$. The $Q$ value of the equal $\Delta\delta_D$ binning presented in Sec.~\ref{subsec:Kmatrixbinning} is 0.786 %%@
indicating that this binning choice is over 20\% less sensitive statistically than an unbinned approach.
(The values of $Q$ are also computed for the different $D^{0}\to K^{0}_{S}K^{+}K^{-}$ binnings reported in %%@
Sec.~\ref{subsec:k0skkbin}; the values are 0.771, 0.803, and 0.822 for $\mathcal{N}=2$, $\mathcal{N}=3$, and %%@
$\mathcal{N}=4$ equal $\Delta\delta_D$ binnings, respectively.)
 
In Ref.~\cite{BONDAR2} it was shown that the binning can be optimized to increase the value of $Q$. The %%@
optimization algorithm was provided by the authors of Ref. \cite{BONDAR2} and was adapted to use the lookup table %%@
of the {\it BABAR} 2008 model. The optimization is iterative and starts from the equal $\Delta\delta_D$ binning %%@
presented in Sec.~\ref{subsec:Kmatrixbinning}. Each iteration starts with the random selection of a sub-bin from %%@
the lookup table. In 90\% of iterations the sub-bin is first tested to see if it lies on the boundary of a bin. If %%@
the sub-bin is not at a boundary the next iteration begins. Otherwise the sub-bin is moved from its current %%@
assignment to that of the neighboring bin and the value of $Q$ is computed with the new assignment. If the value of %%@
$Q$ is increased by this migration, the new assignment for this bin is kept and the next iteration begins. If the %%@
value of $Q$ does not increase the assignment reverts to that originally given and the next iteration begins. In %%@
$10\%$ of iterations, the selected sub-bin is given an assignment at random, irrespective of whether it is on a bin %%@
boundary. Again the reassignment of the sub-bin is kept if there is an improvement in $Q$; this allows part of one %%@
bin to `grow' inside another bin if there is an improvement in the sensitivity. The procedure terminates when no %%@
further significant increase in the value of $Q$ can be found. 

\begin{figure}[hbt]
 \begin{center}
\includegraphics*[width=1.0\columnwidth]{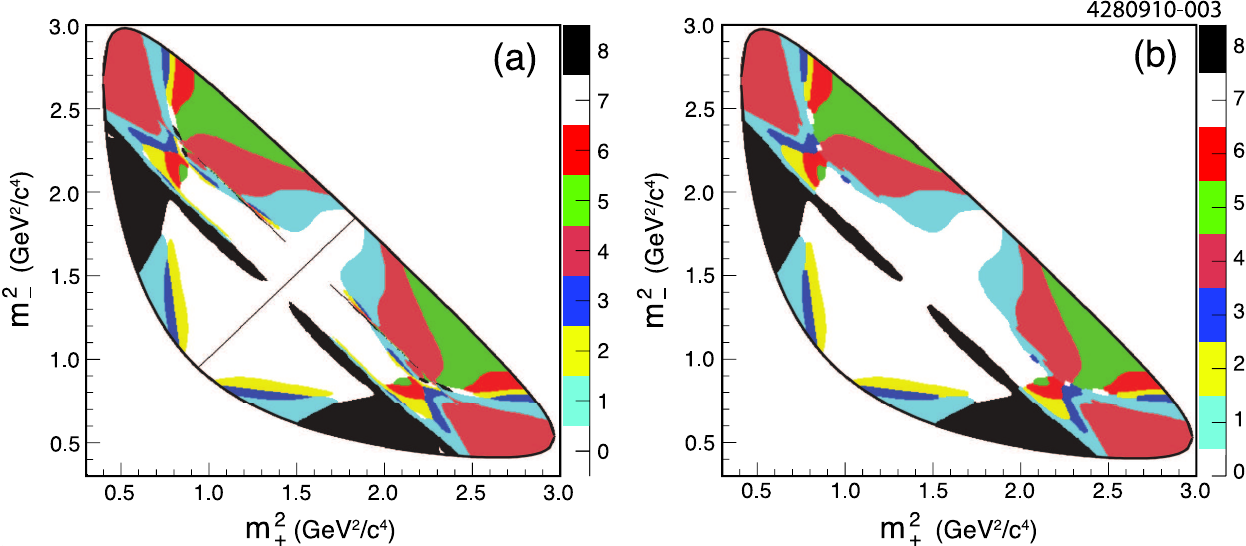}
\caption{(a) Optimal binning of the $D^{0}\to K^{0}_{S}\pi^{+}\pi^{-}$ Dalitz plot that was found to exploit best %%@
the $B$-statistics according to the {\it BABAR} 2008 model. (b) The same binning after the smoothing procedure %%@
described in the text has been applied. The color scale represents the value $|i|$.}\label{fig:optimalbinning}
\end{center}
\end{figure}

 The binning that results from this optimization procedure is shown in Fig.~\ref{fig:optimalbinning}(a) and is %%@
significantly different from that of the equal $\Delta\delta_D$ binning (Fig.~\ref{fig:babar1bins}). The optimized %%@
$Q$ value is 0.892, which is a 13\% relative increase in sensitivity. However, there are many structures that are %%@
only a few sub-bins in size. Such regions are smaller than the experimental resolution and may result in systematic %%@
effects related to asymmetric migration of events from one bin to another. Furthermore, the position and shape of %%@
this fine structure depends critically on the components in the model, which may be realized differently in nature. %%@
Therefore, a smoothing procedure is implemented  to remove these structures. The smoothing procedure starts by %%@
defining an $11\times 11$ square of sub-bins centred about the sub-bin that is being tested. The number of sub-bins %%@
with the same bin assignment as the central sub-bin within the square is found. If the fraction of sub-bins of the %%@
same assignment is less than 30\% the sub-bin assignment is changed to the modal bin assignment within the square. %%@
This procedure is performed for each sub-bin with the bin assignments from the original optimization to prevent %%@
bias. Figure~\ref{fig:optimalbinning}(b) shows the binning after this smoothing procedure; the value of $Q$ only %%@
decreases by 0.005. The smoothed optimal binning is used to calculate $c^{(\prime)}_i$ and $s^{(\prime)}_i$.

The same optimization procedure is applied to the three $D^{0}\to K^{0}_{S}K^{+}K^{-}$ binnings in terms of equal %%@
intervals of $\Delta \delta_D$. However, the improvements in $Q$ are found to be negligible compared to the equal  %%@
binnings. Therefore, the $c^{(\prime)}_i$ and $s^{(\prime)}_i$ parameters for $D^{0}\to K^{0}_{S}K^{+}K^{-}$ decay %%@
are not reported here for optimized binnings.  

\subsubsection{Modified-optimal binning of the {\it BABAR} 2008 model for the presence of background}
\label{subsec:alt_binning}

The $Q$ values for the binnings provided are computed assuming that there is no background present. There is a %%@
clear advantage to using the optimal binning in such a case, and simulation studies of a measurement of $\gamma$ %%@
using the observed values of $c_i$, $s_i$, and the number of flavor-tagged $D^{0}\to K^{0}_{S}\pi^{+}\pi^{-}$ %%@
events in each bin, $K_i$, have confirmed the improved sensitivity in comparison to the equal $\Delta\delta_D$ %%@
binning. However, when background is added to the simulation studies the sensitivity to $\gamma$ using the optimal %%@
binning can be worse than that for the equal $\Delta\delta_D$ binning (see Sec.~\ref{sec:gamma}). The addition of %%@
background events naturally reduces the sensitivity to $\gamma$. The measurement of $\gamma$ is most sensitive when %%@
there are significant differences between yields in the bins for positive and negative $B$ decays. In simulations %%@
there are two observed effects that can dilute the sensitivity. For the optimal binning there are bins where the %%@
asymmetry is large while the expected yields are low. If there is a large background yield in such a bin then the %%@
size of the asymmetry can be diluted to the point where the sensitivity gained by the optimal binning choice is %%@
significantly reduced by the presence of background. Also, very large bins contain a large fraction of the %%@
combinatoric background, which follows a reasonably uniform distribution over the Dalitz plot, which dilutes the %%@
asymmetry in that bin. With an assumed background model, it is possible to find a binning choice that maximizes the %%@
sensitivity to $\gamma$ in the presence of background. 

The background model assumed is determined from simulation studies of LHCb described in Ref.~\cite{TABS}. In this %%@
work three distinct types of background are considered. The first type of background is pure combinatoric, where %%@
the $D^{0}$ is reconstructed from a random combination of pions; the background-to-signal ratio, B/S, is expected %%@
to be less than 1.1 at the 90$\%$ confidence level. The second type of background is where a $D$ meson is %%@
reconstructed correctly, and is then subsequently combined with a random kaon candidate to form a $B$ candidate. %%@
The reconstruction of the $D^0$ and $\overline{D}^0$ is approximately equally likely and hence the distribution of %%@
this type of background in the $i^{\mathrm{th}}$ bin will be proportional to $(K_i + K_{-i})$. For this type of %%@
background, B/S is expected to be 0.35 $\pm$ 0.03. The third type of background, which has the smallest %%@
contribution to the total background, involves real $B$ decays, predominantly $B^{-}\to \pi^{-}\widetilde{D}^{0} $ %%@
where the pion is misidentified as a kaon. In total the real $B$ background has B/S less than 0.24 at the 90$\%$ %%@
confidence level. Sensitivity studies have shown that this type of background causes only a minor degradation in %%@
the sensitivity to $\gamma$. Therefore, this background type is ignored. In summary, the data sample is assumed to %%@
be composed of 41$\%$ signal events, 45$\%$ combinatoric background and 14$\%$ fully-reconstructed $D$ background.

In the presence of background the calculation of $Q$ changes and will be written as $Q^{\prime}$ to distinguish it %%@
from the no background case. The value of $Q^{\prime}$ is still related to the number of standard deviations by %%@
which the number of events in each bin is changed by varying parameters $x$ and $y$, to the number of standard %%@
deviations if the Dalitz plot is divided into infinitely small regions as defined in Eq.~(\ref{eq:originalQ}), %%@
however the definition of $|f_{B^-}|^2$ is now 
\begin{equation}
|f_{B^-}|^2 = f_s  |f_D(m^2_+,m^2_-) +(x + iy)f_D(m^2_-,m^2_+)|^2 + f_1 \mathcal{B}_1(m^2_+,m^2_-) + f_2 %%@
\mathcal{B}_2(m^2_+,m^2_-) ,
\end{equation}
where $\mathcal{B}_1(m^2_+,m^2_-)$ and $\mathcal{B}_2(m^2_+,m^2_-)$ are the probability density functions for the %%@
combinatoric and fully-reconstructed $D$ backgrounds, respectively, and $f_s$, $f_1$, and $f_2$ are the fractions %%@
of signal, combinatoric background, and fully-reconstructed $D$ backgrounds, respectively. The assumed values of %%@
$f_s$, $f_1$ and $f_2$ are 0.41, 0.45, and 0.14.

As before the precision of $x$ and $y$ weakly depends on their values, therefore, the simplification that $x=y=0$ %%@
is once more made. In this case the expression analogous to Eq.~(\ref{eq:Q}) is given by
\begin{equation}
\label{eq:Qprime}
Q^{'2}|_{x=y=0} = \frac{\displaystyle \sum_i \frac{f_s^2 F_i F_{-i}}{f_s F_i + f_1 B_{1i} + f_2 B_{2i}} (c_i^2 + %%@
s_i^2)}{\displaystyle \int \frac{f_s^2 |f_D(m^2_+,m^2_-)|^2 |f_D(m^2_-,m^2_+)|^2}{f_s |f_D(m^2_+,m^2_-)|^2 + f_1 %%@
\mathcal{B}_1 + f_2\mathcal{B}_2} \,dm^{2}_{+}dm^{2}_{-}} \; ,
\end{equation}
where $B_{1i}$~$(B_{2i})$ is the integrated probability density functions for the combinatoric (fully-reconstructed %%@
$D$) background over the $i^{\mathrm{th}}$ bin. 

\begin{figure}
\includegraphics*[width=0.5\columnwidth]{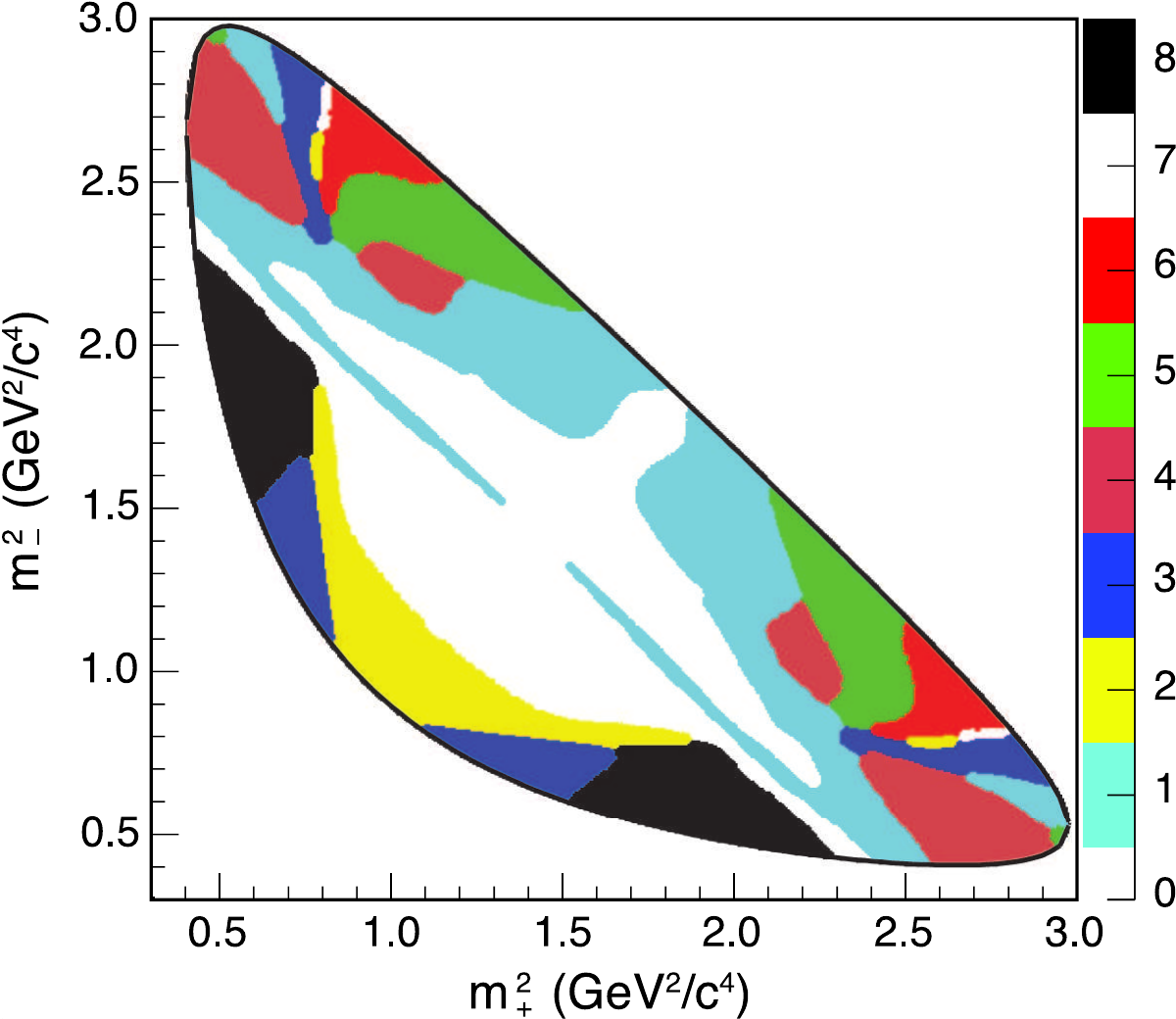}
\caption{Modified-optimal binning of the $D^0 \to$\Kspipi\ Dalitz plot based on
the {\it BABAR} 2008 model. The color scale represents the absolute value of the bin number, $|i|$.}
\label{fig:binning_modopt}
\end{figure}

The optimization algorithm to find the modified-optimal binning with the highest $Q^{\prime}$ is the same as %%@
described in Sec.~\ref{subsec:optimal_binning}. The modified-optimal binning $Q^{\prime}$ value is 0.910. In %%@
comparison, the equal $\Delta\delta_D$ binning has $Q^{\prime}=0.882$ and the optimal binning has $Q^{\prime} = %%@
0.867$. 
The fine structure of the binning is smoothed out using the same technique as described for the optimal binning; %%@
the $Q^{'}$ value drops by 0.006. The binning after the smoothing procedure is given in %%@
Fig.~\ref{fig:binning_modopt}.  
In addition, we performed studies that show this binning choice retains the highest values of $Q^{\prime}$ even %%@
when the assumptions of the background model are modified. The alternative background models tested contain %%@
combinatoric background with a B/S between 0.8 and 1.3, and fully-reconstructed $D^{0}$ background with a B/S %%@
between 0.26 and 0.44. 

\begin{figure}
\includegraphics*[width=0.5\columnwidth]{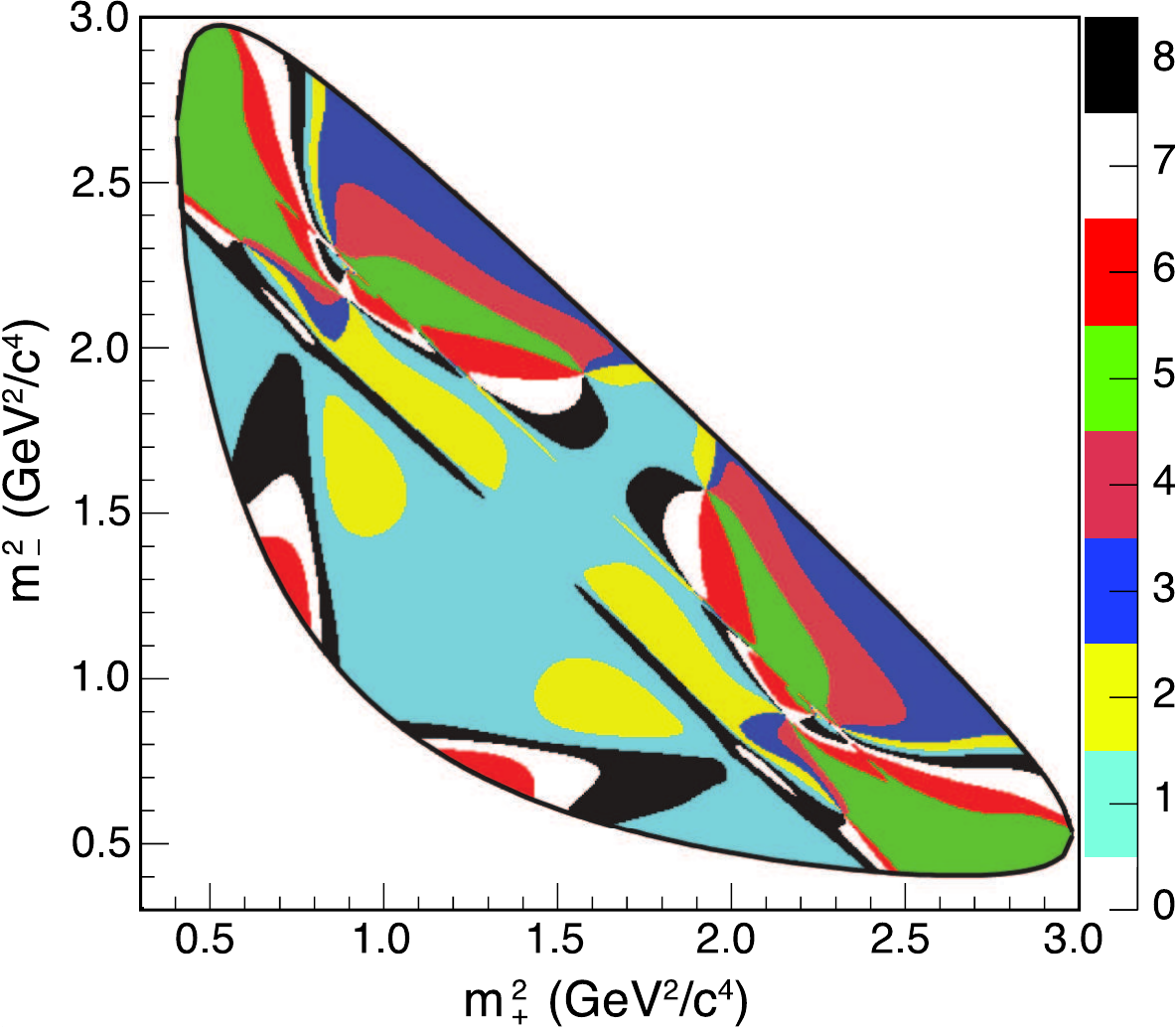}
\caption{Equal $\Delta\delta_D$ binning of the $D^0 \to$\Kspipi\ Dalitz plot with ${\cal N}=8$ based on
the Belle model \cite{BELLE2}. The color scale represents the absolute value of the bin number, $|i|$.}
\label{fig:binning_belle}
\end{figure}

\subsubsection{Belle model binning}
\label{subsec:belle_model}
The shape of the bins are dependent on the details of the amplitude model for the decay. There is another model %%@
available from the Belle experiment with which to define the bins \cite{BELLE2}. This model does not use the same %%@
descriptions of the $\pi\pi$ and $K\pi$ S-wave as the {\it BABAR} 2008 and 2010 models \cite{BABAR2,BABAR3}. A %%@
lookup table of this model has been provided by the Belle collaboration \cite{ANTON}. Therefore, for a %%@
completeness, and to cross check our baseline results derived from the {\it BABAR} 2008 model, an equal %%@
$\Delta\delta_D$ interval binning is derived from the latest Belle model. The binning over the Dalitz space is %%@
given in Fig.~\ref{fig:binning_belle}.  

\subsubsection{Bin choice comparisons and the {\it BABAR} 2010 model}
 \label{subsec:ultimate_babar}
 \begin{table}[t]
 \begin{center}
 \caption{Comparison of the figure of merit values $(Q,Q^{\prime})$ calculated using the different %%@
models.}\label{fig:bincomparison}
 \begin{tabular}{lcccc} \hline\hline
 	 Binning & Figure of merit & {\it BABAR} 2008 \cite{BABAR2} & {\it BABAR} 2010 \cite{BABAR3} & Belle %%@
\cite{BELLE2} \\ \hline
 	 Equal $\Delta\delta_D$& $Q$ &         0.786    & 0.780 & 0.762\\ 
	 Optimal & $Q$ &      0.887    & 0.880 & 0.857\\ 
     Modified optimal & $Q^{\prime}$& 0.904 & 0.903 & 0.886\\
     Belle & $Q$ &              0.754    & 0.755 & 0.773  \\ 
 	 \hline \hline
 \end{tabular}
 \end{center}
 \end{table}
 
The 2010 {\it BABAR} model only became available \cite{BITMAPS} after the completion of the $D^{0}\to %%@
K^{0}_{S}\pi^{+}\pi^{-}$ analysis. Therefore, to determine the possible impact on $\gamma/\phi_3$ precision related %%@
to differences between the 2008 and 2010 models we compute the values of the respective figure of merit, $Q$ or %%@
$Q^{\prime}$, for each {\it BABAR} model. This is done with the binnings fixed to those described above, which are %%@
derived from the {\it BABAR} 2008 model or the Belle model. The resulting figures of merit are given in %%@
Table~\ref{fig:bincomparison} for the 2008 and 2010 {\it BABAR} models; the values of  $Q^{(\prime)}$ are also %%@
given when computed with the Belle model. The values of $Q^{(\prime)}$ computed using the {\it BABAR} 2010 model %%@
are slightly smaller than those for the 2008 model, but the difference is never greater than 0.007. In comparison, %%@
when calculating $Q^{(\prime)}$ with the Belle model, for binnings derived from the {\it BABAR} 2008 model, the %%@
decrease is between 0.018 and 0.030. Therefore, we conclude that using the {\it BABAR} 2008 model, rather than the %%@
{\it BABAR} 2010 model, to derive the binnings in this analysis will not result in a significant degradation in %%@
sensitivity to $\gamma/\phi_3$.

\section{Event Selection}
\label{sec:eventsel}
This section summarizes the event selection for the two analyses. Section~\ref{subsec:K0SKK_selection} describes %%@
the selection of $D^0\to K^{0}_{S,L}K^{+}K^{-}$ events. Section~\ref{subsec:K0Spipi_selection} briefly summarizes %%@
the changes to the selection of $D^0\to K^{0}_{S,L}\pi^{+}\pi^{-}$ events with respect to the previous analysis %%@
\cite{BRIERE}.

\subsection{\boldmath Selection of $D^{0}\to K^{0}_{S,L}K^{+}K^{-}$}
\label{subsec:K0SKK_selection} 
We perform the analysis on $e^{+}e^{-}$ collision data produced by the Cornell Electron Storage Ring at a %%@
center-of-mass energy, $E_{\mathrm{cm}}$, of 3.77~GeV. The data were collected by the CLEO-c detector and %%@
correspond to an integrated luminosity of $818~\mathrm{pb}^{-1}$. The CLEO-c detector is a solenoidal detector %%@
which includes a gaseous tracking system for the measurement of charged particle momenta and ionization energy %%@
loss, a ring-imaging Cherenkov detector to aid in particle identification, and a CsI crystal calorimeter to measure %%@
the energy of electromagnetic showers. The CLEO-c detector is described in detail elsewhere \cite{CLEOC}.   

Samples of Monte Carlo (MC) simulated data are used to develop selection criteria, determine selection %%@
efficiencies, and to estimate certain types of background. EVTGEN \cite{EVTGEN} is used to generate the decays and %%@
GEANT \cite{GEANT} is used to simulate the CLEO-c detector response. Efficiency estimations are made on samples of %%@
signal events generated according to the $D^{0}\to K^{0}_{S} K^{+}K^{-}$ resonance model reported in %%@
Ref.~\cite{BABAR2}. Separate signal samples are generated for each exclusive final state considered in the analysis %%@
and comprise 40,000 events per final state. Quantum correlations in the $D^{0}\overline{D}^{0}$ system are also %%@
simulated for each tag mode; this is particularly important for the $CP$-tagged $D^{0}$ decays. In addition a %%@
sample of generic $D\overline{D}$ decays corresponding to an integrated luminosity approximately 25 times greater %%@
than the data is used to estimate backgrounds. Quantum correlations are accounted for in the generic simulation. 

We adopt standard CLEO-c selection criteria for $\pi^{+}$, $\pi^{0}$, and $K^{0}_{S}$ mesons, which are described %%@
in Ref.~\cite{DHAD}. The standard CLEO-c $K^{+}$ selection \cite{DHAD} is used for all final states apart from %%@
$D^{0}\to K^{0}_{S,L}K^{+}K^{-}$. For this final state, which has much smaller yields than %%@
$K^{0}_{S,L}\pi^{+}\pi^{-}$, the significance of the signal is found to increase if the impact parameter criteria %%@
are loosened by a factor of four and the requirement on the fraction of associated tracking chamber hits compared %%@
to the expectation is removed.  We require candidate $K^{0}_{S}\to \pi^{+}\pi^{-}$ decays to have a mass within %%@
$7.5~\mathrm{MeV}/c^2$ of the nominal mass and the $K^0_{S}$ decay vertex is required to be separated from the %%@
interaction region by at least half a standard deviation. We reconstruct $\eta\to\gamma\gamma$ candidates in a %%@
similar fashion to $\pi^{0}\to\gamma\gamma$ candidates, with the requirement that the invariant mass is within %%@
$42~\mathrm{MeV}/c^2$ of the nominal mass; the same requirement is applied to $\eta\to \pi^{+}\pi^{-}\pi^{0}$ %%@
candidates. Candidates for $\omega\to\pi^{+}\pi^{-}\pi^{0}$ decays are required to be within $20~\mathrm{MeV}/c^2$ %%@
of the nominal $\omega$ mass. We require $\eta^{\prime}\to \eta\pi^{+}\pi^{-}$ candidates to have an invariant mass %%@
in the range 950 to $964~\mathrm{MeV}/c^2$.  All nominal masses are taken from Ref.~\cite{PDG}.

We consider $K^{0}_{S,L}K^{+}K^{-}$ candidates reconstructed against the different final states listed in %%@
Table~\ref{tab:yields}. These are referred to as double-tagged (DT) events. More $CP$-tag final states are used in %%@
the analysis of $K^{0}_{S,L} K^{+}K^{-}$ than the $K^{0}_{S,L}\pi^{+}\pi^{-}$ analysis \cite{BRIERE} to increase %%@
the statistics available to determine $c_i$ for this decay. (These modes are not included in the analysis of %%@
$K^{0}_{S}\pi^{+}\pi^{-}$ because in this measurement the principal statistical limitation is the number of %%@
$K^{0}_{S}\pi^{+}\pi^{-}~vs.~K^{0}_{L,S}\pi^{+}\pi^{-}$ events used to determine $s_i$.) We do not reconstruct %%@
final states containing two missing particles, such as $K^{0}_{L}K^{+}K^{-}~vs.~K^{0}_{L}\pi^{0}$.

\begin{table*}[t]
 \begin{center}
 \caption{Single-tag (ST) and $D^{0}\to K^{0}_{S,L}h^{+}h^{-}$ double-tag (DT) yields. The single tag yields and %%@
uncertainties are computed following the method reported in Ref.~\cite{BRIERE} and are not corrected for %%@
efficiency. The DT yields are the observed number of events in the signal region prior to background subtraction %%@
and before efficiency correction.}\label{tab:yields}
  \begin{tabular}{lccccc} \hline\hline
  Mode & ST yield & \multicolumn{4}{c}{DT yields} \\
       &          &$K^{0}_{S}\pi^{+}\pi^{-}$ & $K^{0}_{L}\pi^{+}\pi^{-}$ & $K^{0}_{S}K^{+}K^{-}$ & %%@
$K^{0}_{L}K^{+}K^{-}$ \\ \hline
Flavor tags & & & & & \\
$K^{-}\pi^{+}$ & $144 563\pm 403$ & 1444 & 2857 & 168 & 302 \\
$K^{-}\pi^{+}\pi^{0}$ & $258 938 \pm 581 $& 2759 & 5133 & 330 & 585 \\
$K^{-}\pi^{+}\pi^{+}\pi^{-}$ & $220 831 \pm 541$ & 2240 & 4100 & 248 & 287 \\
$K^{-}e^{+}\nu$ & & 1191 & & 100 & \\ \hline
$CP$-even tags & & & & & \\
$K^{+}K^{-}$ & $13349\pm 128$ & 124 & 357 & 12 & 32 \\
$\pi^{+}\pi^{-}$ & $6177\pm 114$& 61 & 184 & 4 & 13 \\
$K^{0}_{S}\pi^{0}\pi^{0}$ & $6838\pm 134$& 56 & & 7 & 14 \\
$K^{0}_{L}\pi^{0}$ & & 237 & & 17 &  \\
$K^{0}_{L}\eta(\gamma\gamma)$ & & & & 4 & \\
$K^{0}_{L}\eta(\pi^{+}\pi^{-}\pi^{0})$ & & & & 1 & \\
$K^{0}_{L}\omega$ & & & & 4 & \\
$K^{0}_{L}\eta^{\prime}$ & & & & 1 & \\ \hline
$CP$-odd tags & & & & & \\
$K^{0}_{S}\pi^{0}$ & $19753\pm 153$ & 189 & 288 & 18 & 43 \\
$K^{0}_{S}\eta(\gamma\gamma)$ & $2886\pm 71$& 39 & 43 & 4 & 6 \\
$K^{0}_{S}\eta(\pi^{+}\pi^{-}\pi^{0})$ & & & & 2 & 1 \\
$K^{0}_{S}\omega$ & $8830\pm 110$& 83 & & 14 & 10 \\
$K^{0}_{S}\eta^{\prime}$ & & & & 3 & 4\\
$K^{0}_{L}\pi^{0}\pi^{0}$ & & & & 5 & \\\hline

$K^{0}_{S}\pi^{+}\pi^{-}$ & & 473 & 1201 & 56 & 126 \\
$K^{0}_{L}\pi^{+}\pi^{-}$ & & & & 140 & \\
$K^{0}_{S}K^{+}K^{-}$ & & & & 4 & 9 \\ \hline\hline
  \end{tabular}
 \end{center}
\end{table*}

 \begin{figure}[htb]
 \begin{center}
 \includegraphics*[width=1.0\columnwidth]{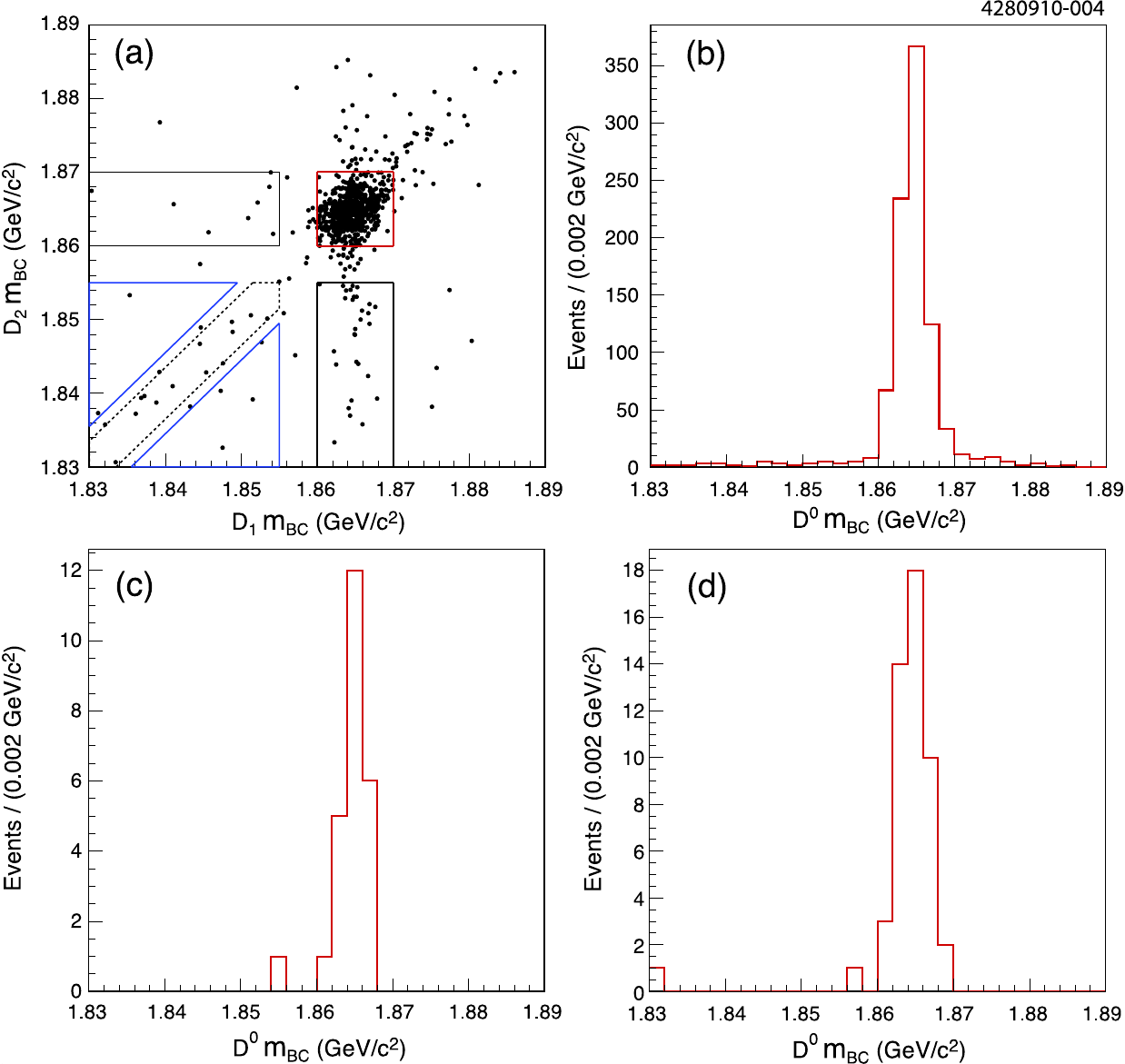}
	\caption{(a) Distribution of $m_{bc}$ for $D^0\to K^{0}_{S}K^{+}K^{-}$ candidates $(D_1)$ against the $m_{bc}$ %%@
for flavor-tag candidates ($D_2$): $\overline{D}^{0}\to K^{+}\pi^{-}$, $K^{+}\pi^{-}\pi^{0}$ and %%@
$K^{+}\pi^{-}\pi^{-}\pi^{+}$. The square signal region (red online) and four sideband regions are shown. %%@
Distributions of $m_{bc}$ for $D^0\to K^{0}_{S}K^{+}K^{-}$ candidates tagged by (b) flavor, (c) $CP$-even, and (d) %%@
$CP$-odd decays.}  \label{fig:mbc_KsKK}
\end{center}
\end{figure}

Final states that do not contain a $K^{0}_{L}$ meson or neutrino are fully reconstructed via two kinematic %%@
variables: the beam-constrained candidate mass, %%@
$m_{bc}\equiv\sqrt{E_{\mathrm{cm}}^{2}/(4c^{4})-\mathbf{p}_{D}^{2}/c^{2}}$, where $\mathbf{p}_{D}$ is the $D$ %%@
candidate momentum, and $\Delta E\equiv E_{D}-E_{\mathrm{cm}}/2$, where $E_{D}$ is the sum of the $D$ daughter %%@
candidate energies. Signal decays will peak at the nominal $D^{0}$ mass and zero in $m_{bc}$ and $\Delta E$, %%@
respectively. Mode-dependent requirements are placed on $K^{0}_{S}K^{+}K^{-}$  and tag candidates such that $\Delta %%@
E$ is less than three standard deviations from zero. The DT yield is determined from counting events in signal and %%@
sideband regions of the $(m_{bc}(D^{0}),m_{bc}(\overline{D}^{0}))$ plane in a manner similar to that presented in %%@
Refs. \cite{ASNER,LOWREY}. The signal region is defined as %%@
$1.86~\mathrm{GeV}/c^2<m_{bc}(D^{0})<1.87~\mathrm{GeV}/c^2$ and %%@
$1.86~\mathrm{GeV}/c^2<m_{bc}(\overline{D}^{0})<1.87~\mathrm{GeV}/c^2$. An example of the two-dimensional %%@
distribution of $(m_{bc}(D^{0}),m_{bc}(\overline{D}^{0}))$ is shown in Fig.~\ref{fig:mbc_KsKK}(a) for 
$K^{0}_{S}K^{+}K^{-}$ candidates reconstructed against $K^{+}\pi^{-}$, $K^{+}\pi^{-}\pi^{0}$, and %%@
$K^{+}\pi^{-}\pi^{-}\pi^{+}$ decays. The four different sidebands contain contributions from distinct types of %%@
combinatorial background. The yields in these sidebands are scaled and subtracted from the yield in the signal %%@
region. The $m_{bc}$ distributions for $\widetilde{D}^{0}\to K^0_{S}K^{+}K^{-}$ candidates tagged by flavor, %%@
$CP$-even, and $CP$-odd final states are shown in Figs. \ref{fig:mbc_KsKK}(b), \ref{fig:mbc_KsKK}(c), and %%@
\ref{fig:mbc_KsKK}(d), respectively. These figures show clearly that the combinatorial backgrounds are small. The %%@
background-to-signal ratio for combinatoric background is less than 7.3\% for all modes.
 
\begin{figure}[htb]
 \begin{center}
\includegraphics*[width=0.5\columnwidth]{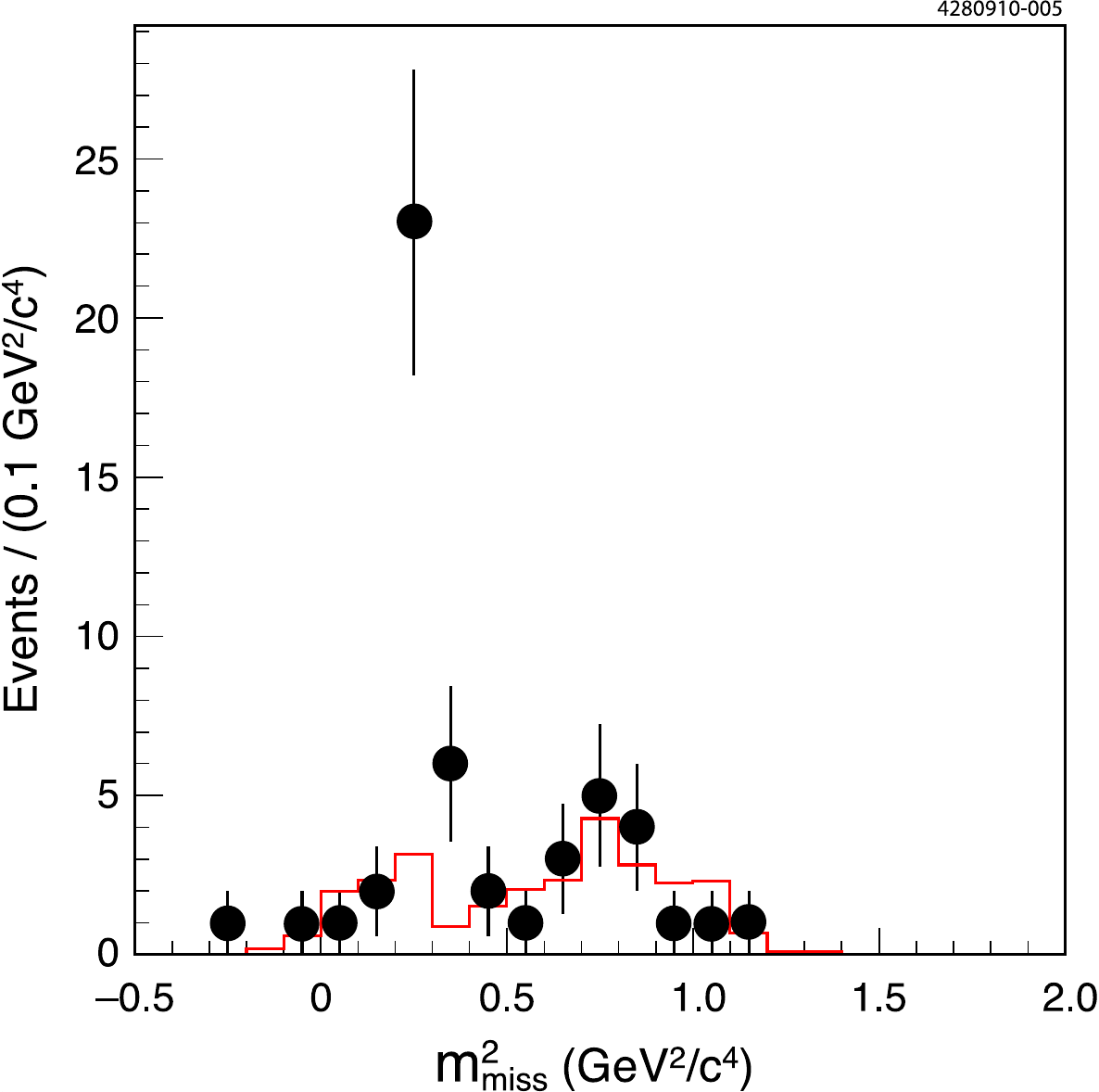}	
	\caption{Missing-mass squared distribution for $CP$-tagged $D^{0}\to K^{0}_{S}K^{+}K^{-}$ candidates where the %%@
$CP$-eigenstate contains a $K^{0}_{L}$ meson. The points are data and the solid histogram is the background %%@
estimated from simulation.}\label{fig:mm2}
 \end{center} 
\end{figure}

To identify final states containing a single $K^0_L$ meson, we compute the missing-mass squared recoiling against %%@
the fully-reconstructed $D$ candidate and the particles from the other $D$ decay containing the $K^{0}_{L}$ meson. %%@
We select events consistent with the mass of the $K^0_L$ meson squared. This technique was introduced in %%@
Ref.~\cite{HE}. We reject events with additional charged tracks, $\pi^{0}$, and $\eta$ candidates that are %%@
unassigned to the final state of interest. In addition, requirements are placed on any calorimeter energy deposits %%@
not associated with the charged or neutral particles that make up the final state of interest. The angle, $\alpha$, %%@
between each unassigned shower and the missing-momentum direction is computed. Criteria are chosen to maximize %%@
signal sensitivity based on simulated samples of signal and background events. We retain events where %%@
$\cos{\alpha}\geq 0.98$, which indicates that the deposit is likely to be due to the interaction of the $K^{0}_{L}$ %%@
meson with the calorimeter.   When $\cos{\alpha}< 0.98$ mode-by-mode requirements are placed on the unassociated %%@
shower energy. The unassociated shower energy is required to be below a
  certain value which varies from 200~MeV for $D^{0}\to K^{0}_{L}\pi^{0}\pi^{0}$ candidates to 370~MeV for $D^0\to %%@
K^{0}_{L}\omega$ and $D^{0}\to K^{0}_{L}\eta^{\prime}$ candidates. Finally, criteria are placed on the momenta of %%@
$\pi^{0}$ and $\eta$ candidates in tags containing a $K^{0}_{L}$ meson to reduce background further.
The combinatoric background yield in the signal region is estimated from the population in the lower and upper %%@
missing-mass squared sidebands. Information from the generic background simulation is used to determine the %%@
relative composition of the sidebands and the signal region to estimate better the combinatorial background. %%@
Figure~\ref{fig:mm2} is the distribution of missing-mass squared for $CP$-tagged $D^{0}\to K^{0}_{S}K^{+}K^{-}$ %%@
candidates for data and simulated background, where the $CP$-eigenstate used to tag the event contains a %%@
$K^{0}_{L}$ meson. 

\begin{figure}[htb]
 \begin{center}
\includegraphics*[width=0.5\columnwidth]{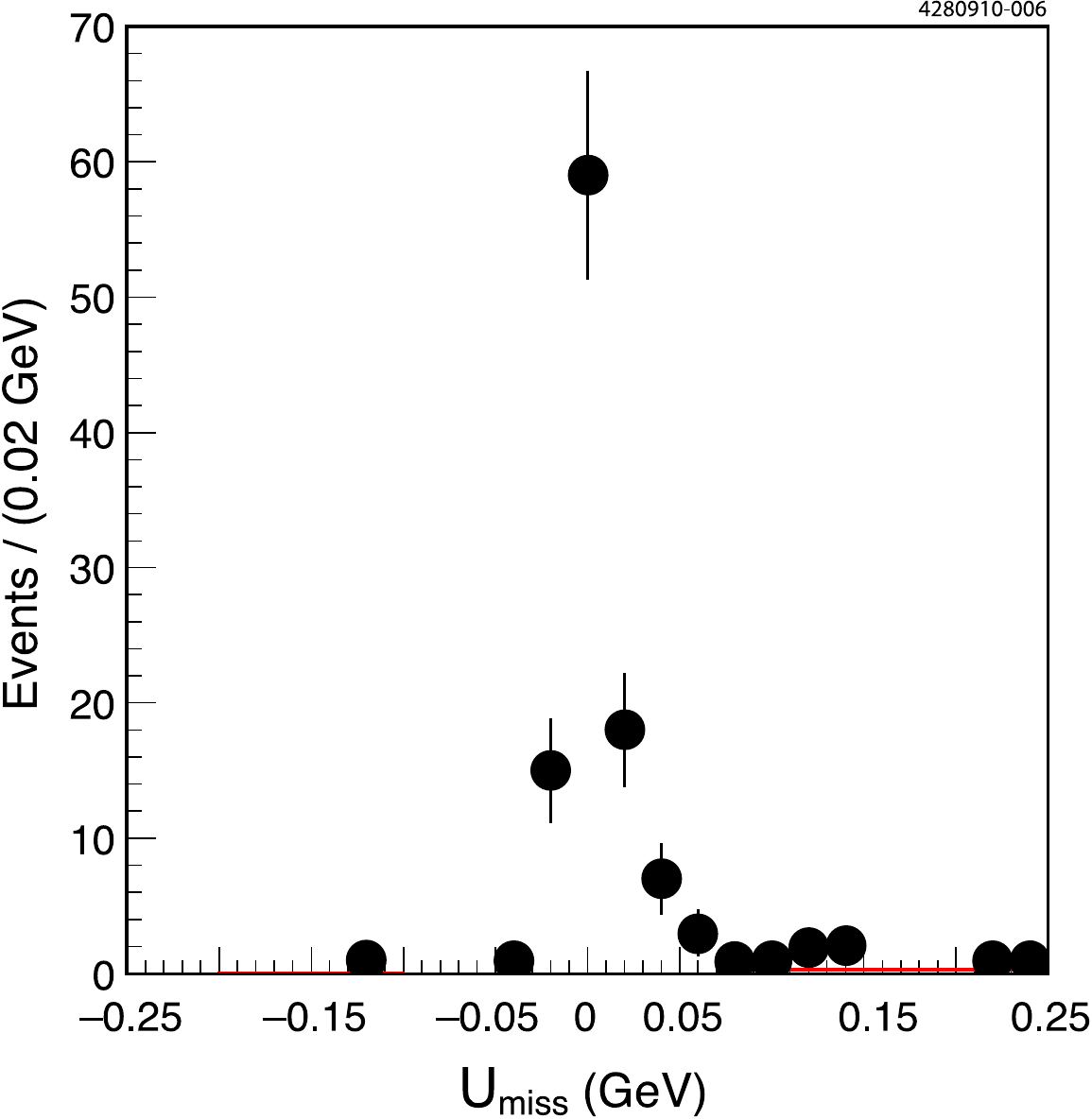}
	\caption{$U_{miss}$ distribution for $K^+e^-\nu$-tagged $K^0_SK^+K^-$ candidates.
          The points are data and the solid histogram is the background
          estimated from simulation. The background histogram (red online) is barely visible indicating that the %%@
background is negligible.
		  }\label{fig:Umiss}
 \end{center} 
\end{figure}

We reconstruct the final state $\overline{D}^{0}\to K^+e^-\nu$ by fully reconstructing a $\widetilde{D}^{0}\to %%@
K^0_SK^+K^-$ candidate and
requiring that the rest of the event contains both a kaon and an electron candidate of opposite charge.
The quantity $U_{\mathrm{miss}} \equiv E_{\mathrm{miss}} - c|\mathbf{p}_{\mathrm{miss}}|$ is used as a %%@
discriminating variable, where 
$E_{\mathrm{miss}}$ and $\mathbf{p}_{\mathrm{miss}}$ are the missing energy and momentum in the event, determined
using the momenta of the fully reconstructed particles. The neutrino is the only particle not detected,
so for a correctly reconstructed event $U_{\mathrm{miss}}$ will equal zero.
Figure~\ref{fig:Umiss} is the distribution of $U_{\mathrm{miss}}$ in the data and simulated background.  
The DT event yields for all final states are given in Table~\ref{tab:yields}.

There are backgrounds to the signal that peak in $m_{BC}$ and missing-mass squared at the same values as the %%@
signal, which can not be evaluated by examining the sidebands.  These peaking backgrounds are estimated from the %%@
generic $D\overline{D}$ MC data samples. The largest peaking background to $D^{0}\to K^{0}_{S}K^{+}K^{-}$ decays is %%@
from $D^{0}\to K^{+}K^{-}\pi^+\pi^{-}$ decays where the $\pi^{+}\pi^{-}$ pair form a $K^{0}_{S}$ candidate. The %%@
peaking background from this source is estimated to be approximately 3.2\% of the signal. For $D^{0}\to %%@
K^{0}_{L}K^{+}K^{-}$ decays there are two significant peaking background contributions. The first source is %%@
$D^{0}\to K^{0}_{S}K^{+}K^{-}$ decays where the $K^{0}_{S}$ is not reconstructed, usually in the case where it %%@
decays to $\pi^{0}\pi^{0}$. The second significant source is from $D^{0}\to K^{+}K^{-}\pi^{0}\pi^{0}$ decays where %%@
both the $\pi^{0}$ daughters are not reconstructed and the missing mass corresponds to that of the $K^{0}_{L}$ %%@
meson. This background is not strictly peaking in that the $\pi^{0}\pi^{0}$ invariant mass does not always %%@
correspond to that of the $K^{0}_{L}$ meson. However, the missing-mass squared distribution for $D^{0}\to %%@
K^{+}K^{-}\pi^{0}\pi^{0}$ background events determined from simulated data is not distributed linearly in the %%@
signal region and low missing-mass squared sideband, which would lead to a biased estimate of the background level %%@
if the sideband is used to determine the background level in the signal region. Therefore, the absolute level of %%@
this background is determined from the simulation and subtracted from both the signal and low missing-mass squared %%@
sideband. The $D^{0}\to K^{0}_{S}K^{+}K^{-}$ and $D^{0}\to K^{+}K^{-}\pi^{0}\pi^{0}$ peaking backgrounds are %%@
estimated to be 6.7\% and 4.4\% of the $D^{0}\to K^{0}_{L}K^{+}K^{-}$ signal, respectively. Tag modes that contain %%@
a $K^{0}_{L}$, $\overline{D}^{0}\to K^{0}_{L}X$, also have a significant peaking background from %%@
$\overline{D}^{0}\to K^{0}_{S}X$ decays where the $K^{0}_{S}$ is not reconstructed. The peaking background to %%@
$\overline{D}^{0}\to K^{0}_{L}X$ events is between 4.0\% and 6.9\%. The estimated peaking backgrounds are %%@
subtracted from the measured yields. The differing $CP$ eigenvalues of the $\overline{D}^{0}\to K^{0}_{S}X$ and %%@
$\overline{D}^{0}\to K^{0}_{L}X$ tags means that the distributions over Dalitz space of the signal and background %%@
can be significantly different. The effect of differing distributions of background is treated as a systematic %%@
uncertainty; the procedure to evaluate the uncertainty is described in Sec.~\ref{sec:syst_K0SKK}. 
The background to $K^+e^-\nu$ events is estimated as 1.8\% in the signal region, defined as $U_{\mathrm{miss}} < %%@
50\textrm{ MeV}$.

We apply a kinematic fit to determine more reliably the position of a candidate in the Dalitz space. For final %%@
states containing a $D^{0}\to\KsKK$ decay, the fit constrains the invariant mass of both the signal and tag $D^{0}$ %%@
meson candidates to be the nominal $D^{0}$ mass and the $K^{0}_{S}$ daughters to the nominal $K^{0}_S$ mass. For %%@
$D^{0}\to K^{0}_{L}K^{+}K^{-}$ decays there are several stages to the fit. The first stage constrains both the %%@
$D^{0}$ daughter kaons in the $D^{0}\to\KlKK$ decay to originate from a common vertex and the tag decay candidate %%@
to the $D^{0}$ mass. The second stage uses the resulting four-momenta to estimate the mass of the missing %%@
$K^{0}_{L}$ meson. The energy of the $K^{0}_{L}$ candidate is rescaled such that the invariant mass is the nominal %%@
$K^{0}_{L}$ meson mass. In the final stage the $K^{0}_{L}$ candidate and the daughter kaon pair are constrained to %%@
the nominal $D^{0}$ meson mass. The introduction of the kinematic fit improves the resolution on $m^2_{+}$ and %%@
$m^{2}_{-}$ by up to a factor of three for fully reconstructed $K^{0}_{S}K^{+}K^{-}$ DT candidates. Although the %%@
fit only gives  small improvements in core resolution for $K^{0}_{L}$ final states there is a significant reduction %%@
in the number of events that lie in the non-Gaussian tails of the resolution distribution.

The kinematic fit fails to converge in $1\%$ to $3\%$ of events, depending on the decay mode.  In these cases the %%@
measured values of $m^{2}_{+}$ and $m^{2}_{-}$ are rescaled such that they give the nominal $D^{0}$ mass. Even %%@
after the fit a small fraction of events are reconstructed outside the kinematic boundaries of the Dalitz space. %%@
The amplitude model used to assign events to a bin is undefined outside the Dalitz space. Therefore, the invariant %%@
masses of the candidate are changed to correspond to the point in the Dalitz space that is closest to the measured %%@
value in terms of the sum of $(\Delta m_{+}^2)^2$ and $(\Delta m_{-}^2)^2$, where $\Delta m_{\pm}^2$ is the %%@
residual between the measured value and a point within the Dalitz plot.

\begin{figure}[htb]
 \begin{center}
\includegraphics*[width=1.0\columnwidth]{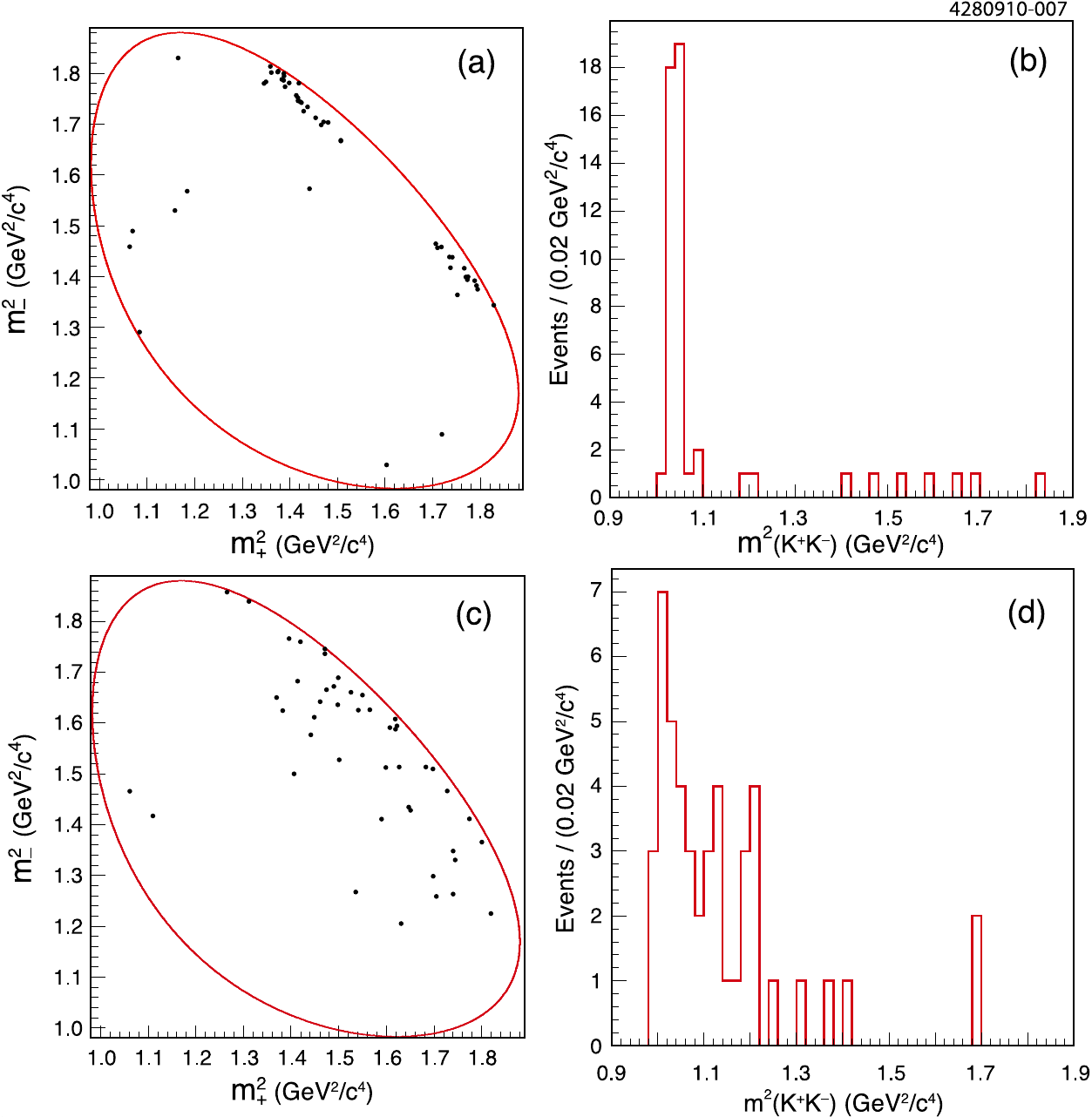}
	\caption{(a) Dalitz plot and (b) $m^2_{K^+K^-}$ distributions of $D^{0}\to K^{0}_{S}K^{+}K^{-}$ candidates %%@
tagged by a $CP$-even eigenstate. (c) Dalitz plot and (d) $m^2_{K^+K^-}$ distributions of $D^{0}\to %%@
K^{0}_{S}K^{+}K^{-}$ candidates tagged by a $CP$-odd eigenstate.}\label{fig:K0SKK_Dalitz_CP}
 \end{center}
\end{figure}

\begin{figure}[htb]
 \begin{center}
\includegraphics*[width=1.0\columnwidth]{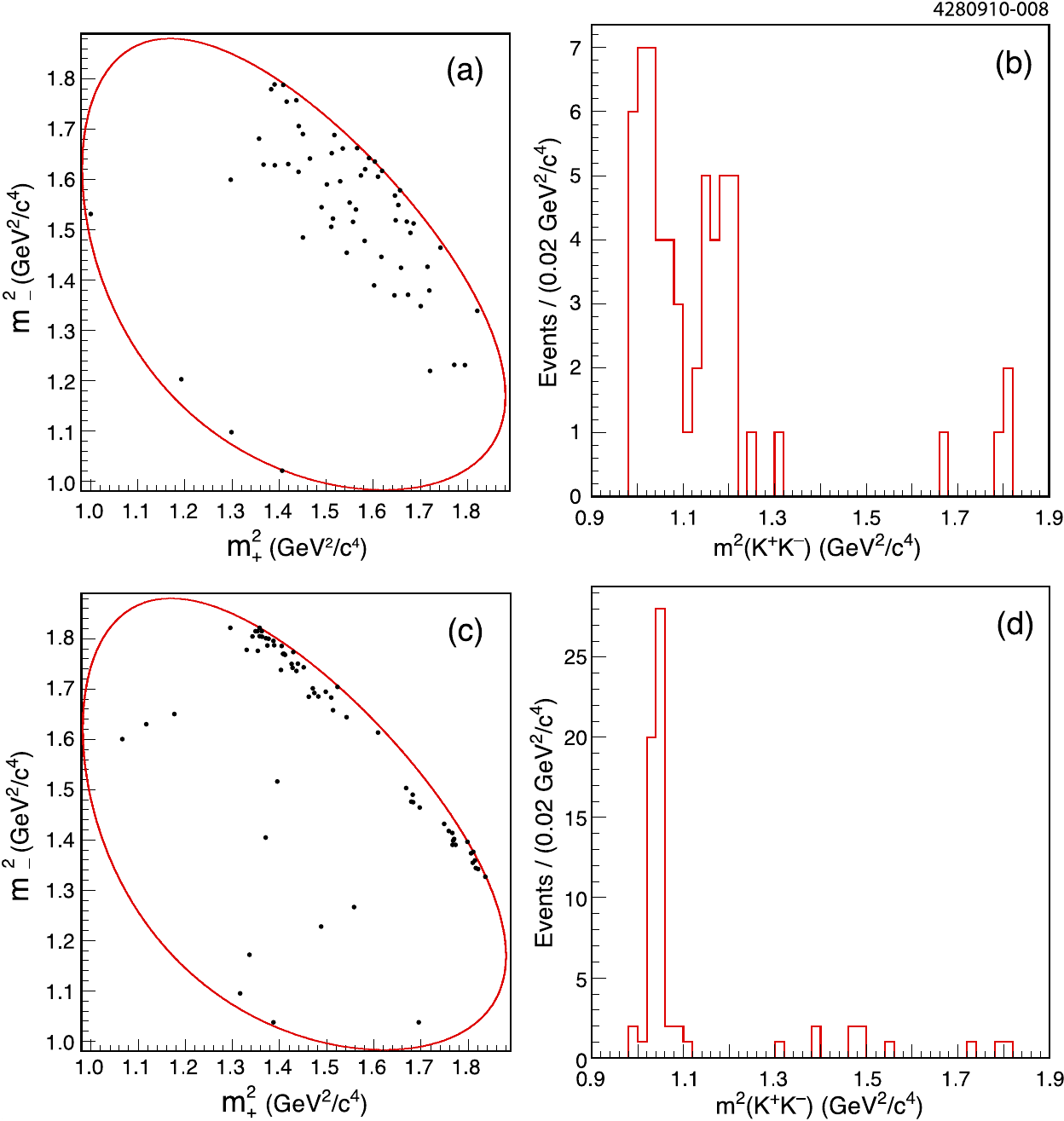}
	\caption{(a) Dalitz plot and (b) $m^2_{K^+K^-}$ distributions of $D^{0}\to K^{0}_{L}K^{+}K^{-}$ candidates %%@
tagged by a $CP$-even eigenstate. (c) Dalitz plot and (d) $m^2_{K^+K^-}$ distributions of $D^{0}\to %%@
K^{0}_{L}K^{+}K^{-}$ candidates tagged by a $CP$-odd eigenstate.}\label{fig:K0LKK_Dalitz_CP}
 \end{center}
\end{figure}

\begin{figure}[htb]
 \begin{center}
\includegraphics*[width=1.0\columnwidth]{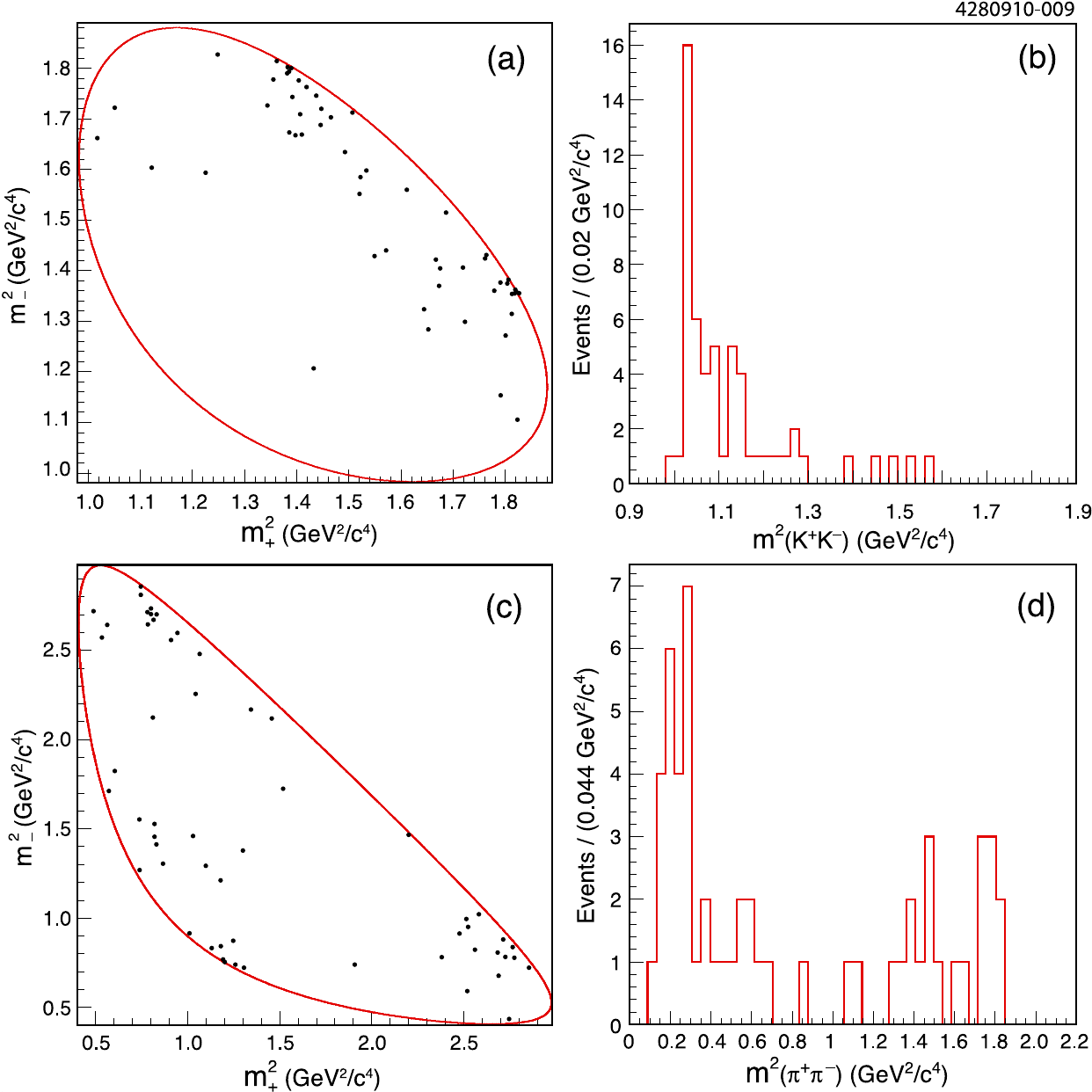}
	\caption{(a) Dalitz plot and (b) $m^2_{K^+K^-}$ distributions of $D^{0}\to K^{0}_{S}K^{+}K^{-}$ candidates %%@
tagged by  $\overline{D}^{0}\to K^{0}_{S}\pi^{+}\pi^{-}$ decays. (c) Dalitz plot and (d) $m^2_{\pi^+\pi^-}$ %%@
distributions of the $D^{0}\to K^{0}_{S}\pi^{+}\pi^{-}$ candidates in the same events.}\label{fig:pseudotag}
 \end{center}
\end{figure}

\begin{figure}[htb]
 \begin{center}
\includegraphics*[width=1.0\columnwidth]{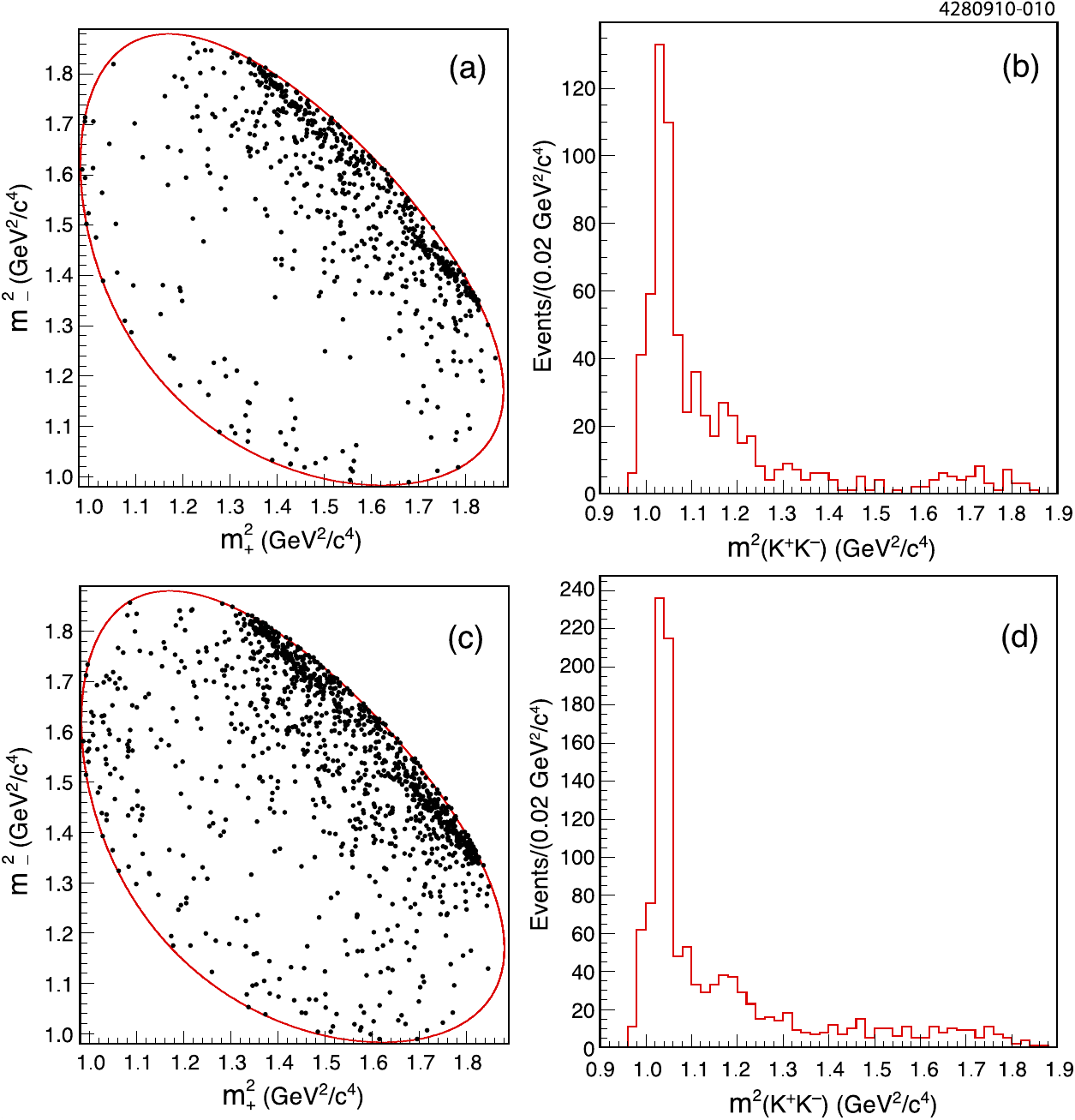}
	\caption{(a) Dalitz plot and (b) $m^2_{K^+K^-}$ distributions of flavor-tagged $D^{0}\to K^{0}_{S}K^{+}K^{-}$ %%@
candidates. (c) Dalitz plot and (d) $m^2_{K^+K^-}$ distributions of flavor-tagged $D^{0}\to K^{0}_{L}K^{+}K^{-}$ %%@
candidates.}\label{fig:flavourtag}
 \end{center}
\end{figure}

The distributions of events across the Dalitz plane and as a function of $m^{2}_{K^{+}K^{-}}$ for  $K^0_S K^+K^-$ %%@
reconstructed against $CP$-even tags are shown in Fig.~\ref{fig:K0SKK_Dalitz_CP}. In addition, %%@
Fig.~\ref{fig:K0SKK_Dalitz_CP} shows the distributions across the Dalitz plane and as a function of %%@
$m^{2}_{K^{+}K^{-}}$ for  $K^0_S K^+K^-$ candidates reconstructed against $CP$-odd tags. The equivalent %%@
distributions for $K^0_LK^+K^-$ candidates reconstructed against a $CP$ eigenstate are shown in %%@
Fig.~\ref{fig:K0LKK_Dalitz_CP}. The $m_{K^{+}K^{-}}^{2}$ distribution of $K^0_SK^+K^-$ ($K^{0}_LK^{+}K^{-}$) %%@
candidates tagged with $CP$-even ($CP$-odd) eigenstates exhibits a peak due to the $\phi$ resonance; as expected %%@
from $CP$ conservation, this peak is not present for $K^0_SK^+K^-$ $(K^{0}_{L}K^{+}K^{-})$ candidates tagged with %%@
$CP$-odd ($CP$-even) eigenstates. Figure~\ref{fig:pseudotag} shows the distribution of events across the Dalitz %%@
plane and as a function of $m^{2}_{K^{+}K^{-}}$ for $K^{0}_{S}K^{+}K^{-}$ candidates reconstructed against %%@
$K^{0}_{S}\pi^{+}\pi^{-}$ decays. Furthermore, Fig. \ref{fig:pseudotag} shows the distribution of the same events %%@
across the  $K^{0}_{S}\pi^{+}\pi^{-}$ Dalitz plane and as a function of the $\pi^{+}\pi^{-}$ invariant-mass %%@
squared, $m_{\pi^{+}\pi^{-}}^2$. The increased statistics available from using events in which both $D$ mesons %%@
decay to $K^{0}_{S,L}h^{+}h^{-}$ is clear. The distributions of flavor-tagged $D^{0}\to K^{0}_{S}K^{+}K^{-}$ and %%@
$D^{0}\to K^{0}_{L}K^{+}K^{-}$ candidates across the Dalitz plane and as a function of the $m_{K^{+}K^{-}}$ are %%@
shown in Fig.~\ref{fig:flavourtag}. The flavor-tagged samples are used to determine $K^{(\prime)}_{i}$ for each %%@
Dalitz-plot bin.

The efficiency in each bin is evaluated from the signal MC for each individual tag mode. Table~\ref{tab:eff} gives %%@
the total efficiency for each tag mode; these vary between $(0.51 \pm 0.04)\%$ for %%@
$K^{0}_{S}K^{+}K^{-}~vs.~K^0_L\pi^0\pi^0$ to $(29.4 \pm 0.3)\%$ for $K^{0}_{L}K^{+}K^{-}~vs.~\pi^+\pi^-$. The %%@
uncertainty on the efficiency is that due to MC statistics.

\begin{table}[htb]
\begin{center}
\caption{Selection efficiency for the different DT $K^{0}_{S,L}K^{+}K^{-}$ modes. The uncertainty is that due to MC %%@
simulation statistics.}\label{tab:eff}
\begin{tabular}{lcc}\hline\hline
Tag & \multicolumn{2}{c}{Efficiency (\%)} \\ 
    & \hspace{1cm} $K^0_SK^+K^-$ \hspace{1cm} & \hspace{1cm} $K^0_LK^+K^-$ \hspace{1cm} \\ \hline  
Flavor tags & & \\
$K^{-}\pi^{+}$ & $14.6\pm 0.2$ & $25.2\pm 0.3$\\
$K^{-}\pi^{+}\pi^0$ & $8.5\pm 0.2$ & $14.3\pm 0.2$\\
$K^{-}\pi^{+}\pi^{+}\pi^{-}$ & $10.8 \pm 0.2$ & $15.9\pm 0.2$\\
$K^- e^+ \nu$ & $11.9 \pm 0.2$ & \\ \hline
$CP$-even tags & & \\
$K^+K^-$ & $12.2 \pm 0.2$ & $23.7\pm 0.3$\\
$\pi^+\pi^-$ & $15.1 \pm 0.2$ & $29.4 \pm 0.3$ \\
$K^0_S\pi^0\pi^0$ & $2.8 \pm 0.1$ & $5.9 \pm 0.1$\\
$K^0_L\pi^0$ & $8.0\pm 0.1$ & \\
$K^0_L\eta(\gamma\gamma)$ & $7.9\pm 0.1$ & \\
$K^0_L\eta(\pi^+\pi^-\pi^0)$ & $1.6 \pm 0.1$ & \\
$K^0_L\omega$ & $3.1\pm 0.1$ & \\
$K^0_L\eta'$ & $1.7 \pm 0.1$ & \\ \hline\hline
$CP$-odd tags & & \\
$K^0_S\pi^0$ & $7.1\pm 0.1$ & $10.6\pm 0.2$\\ 
$K^0_S\eta(\gamma\gamma)$ & $6.5\pm 0.1$ & $9.7 \pm 0.2$\\
$K^0_S\eta(\pi^+\pi^-\pi^0)$ & $4.4\pm 0.1$ & $6.8\pm 0.1$ \\
$K^0_S\omega$ & $3.4\pm 0.1$ & $5.0 \pm 0.1$\\
$K^0_S\eta'$ & $1.4 \pm 0.1$ & $2.0\pm 0.1$ \\
$K^0_L\pi^0\pi^0$ & $0.51\pm 0.04$ & \\ \hline
\Kspipi\ & $7.9\pm 0.1$ & $13.0 \pm 0.2$\\
\Klpipi\ & $14.0 \pm 0.2$ & \\ 
\KsKK\ & $4.9\pm 0.1$ & $7.0\pm 0.1$ \\
\hline\hline
\end{tabular}
\end{center}
\end{table}

The finite detector resolution causes events to migrate between Dalitz-plane bins after reconstruction. %%@
Occasionally there is a significant asymmetric migration from one bin to another. The effect is more pronounced in %%@
$K^0_{S,L}K^+K^-$ decays than in $K^0_{S,L}\pi^+\pi^-$ decays because of the presence of a relatively narrow and %%@
densely populated bin that encloses the $\phi\to K^{+}K^{-}$ resonance. We correct for this 
migration using MC data to determine the size and nature of the effect. For each binning we define a %%@
$2\mathcal{N}\times 2\mathcal{N}$ matrix $\mathbf{U}$ for each DT mode as follows 
\begin{equation}
 U_{i,j} \equiv \frac{m_{j,i}}{\displaystyle\sum_{k=-\mathcal{N},~k\neq 0}^{\mathcal{N}} m_{j,k}} \; ,
\end{equation} 
where $m_{j,i}$ is the number of signal MC events that are generated in bin $j$ and reconstructed in bin $i$. The %%@
vector of migration-corrected yields in each bin, $\mathbf{D}^{\mathrm{corr}}$, is determined from the vector of %%@
reconstructed yields in each bin, $\mathbf{D}^{\mathrm{rec}}$, using the relation: 
\begin{equation}
 \mathbf{D}^{\mathrm{corr}} = \mathbf{U}^{-1}\mathbf{D}^{\mathrm{rec}}. 
\end{equation}
As an example the migration matrix for $K^{0}_{L}K^{+}K^{-}~vs.~K^{0}_{S}\pi^{0}$ events when the $D^{0}\to %%@
K^{0}_{S} K^{+}K^{-}$ Dalitz space is divided into $\mathcal{N}=3$ bins is given in Table~\ref{tab:migmatrix}. %%@
Typically the migration out of the bin containing the $\phi$ resonance is 5\% for $D^{0}\to K^{0}_{S}K^{+}K^{-}$ %%@
modes and between $10\%$ to $25\%$ for $D^{0}\to K^{0}_{L}K^{+}K^{-}$ modes. The errors on the elements of %%@
$\mathbf{U}$ due to the limited MC statistics are treated as a systematic uncertainty.

\begin{table}[htb]
\begin{center}
\caption{Migration matrix {\bf U} elements $(\%)$ for $K^{0}_{L}K^{+}K^{-}~vs.~K^{0}_{S}\pi^{0}$ events when the %%@
$D^{0}\to K^{0}_{L} K^{+}K^{-}$ Dalitz space is divided into $\mathcal{N}=3$ bins.}\label{tab:migmatrix}
\begin{tabular}{lcccccc} \hline\hline 
$i$ \hspace{0.5cm} & $U_{i,1}$ & $U_{i,2}$ & $U_{i,3}$ & $U_{i,-1}$ & $U_{i,-2}$ & $U_{i,-3}$ \\ \hline 
1 & \hspace{0.5cm} 86.2 \hspace{0.5cm} & \hspace{0.5cm} 11.8 \hspace{0.5cm} & \hspace{0.5cm} \phantom{0}0.5 %%@
\hspace{0.5cm} & \hspace{0.5cm} \phantom{0}0.4 \hspace{0.5cm} & \hspace{0.5cm} \phantom{0}0.0 \hspace{0.5cm} & %%@
\hspace{0.5cm} \phantom{0}0.0 \hspace{0.5cm} \\
2 & 11.3 & 88.1 & \phantom{0}0.5 & \phantom{0}0.0 & \phantom{0}0.0 & \phantom{0}0.0 \\
3 & \phantom{0}1.5 & \phantom{0}0.1 & 98.9 & \phantom{0}0.0 & \phantom{0}0.0 & \phantom{0}0.0 \\
-1 & \phantom{0}1.0 & \phantom{0}0.1 & \phantom{0}0.0 & 85.2 & 10.6 & \phantom{0}0.0 \\
-2 & \phantom{0}0.0 & \phantom{0}0.0 & \phantom{0}0.0 & 13.9 & 89.3 & \phantom{0}1.0 \\
-3 & \phantom{0}0.0 & \phantom{0}0.0 & \phantom{0}0.0 & \phantom{0}0.4 & \phantom{0}0.1 & 99.0 \\ \hline\hline
\end{tabular} 
\end{center}
\end{table}  

\subsection{\boldmath Updates to the $K^{0}_{S,L}\pi^{+}\pi^{-}$ selection}
\label{subsec:K0Spipi_selection}

Single tags are used in the analysis of $D^{0}\to \Kspipi$  to determine the normalization factors in %%@
Eqs.~(\ref{eq:kscptag}) and (\ref{eq:kskltag}) with limited systematic uncertainties. (In the $D^{0}\to \KsKK$ %%@
analysis these systematic uncertainties are not as important given the available statistics and therefore the %%@
normalizations are determined from the number of $D\overline{D}$ pairs and the measured branching fractions.) 
The ST selection is identical to that described in Ref.~\cite{BRIERE}. However, an error in the previous analysis %%@
led to a small fraction (3.6\%) of the data being excluded from the ST analysis of $CP$ eigenstates. The updated %%@
yields are given in Table~\ref{tab:yields}.
 
The selection requirements of DTs are identical to that in the previous analysis \cite{BRIERE}. 
However, there are several changes to the yields reported. The same data excluded from the original ST analysis %%@
were missing from the evaluation of yields of final states containing a $K^{0}_{L}$ candidate.\footnote{In %%@
addition, the yields of $\Kspipi~vs.~K^{-}e^{+}\nu$ and $\Kspipi~vs.~\Kspipi$ were incorrectly documented in %%@
Table~II of Ref.~\cite{BRIERE}.} Also, the kinematic fit procedure described in Sec.~\ref{subsec:K0SKK_selection} %%@
has also been applied to the $K^{0}\pi^{+}\pi^{-}$ events. However, events that lie outside the Dalitz plot after %%@
the fit procedure are rejected rather than migrated into the physical region, which changes some of the yields %%@
compared to those reported in Ref. \cite{BRIERE}; in the previous analysis all events were retained. The updated DT %%@
yields are shown in Table~\ref{tab:yields}.

\section{\boldmath Extraction of $c_i$ and $s_i$}
\label{sec:cisiextract}
The efficiency-corrected, background-subtracted, and migration-corrected bin yields for each DT mode need to be %%@
normalized to determine the measured values of $M^{(\prime)\pm}_{i}$ and $M^{(\prime)}_{ij}$ so that they can be %%@
used to evaluate $c_i^{(\prime)}$ and $s_i^{(\prime)}$ via Eqs.~(\ref{eq:kscptag}), (\ref{eq:kskstag}), %%@
(\ref{eq:klcptag}), and (\ref{eq:kskltag}). In addition, 
the values of flavor-tagged yields in each bin must also be normalized appropriately to obtain $K^{(\prime)}_{i}$. %%@
For the $K^{0}_{S}K^{+}K^{-}$ analysis the single-tag yields in the normalization factors $h_{CP}$ and %%@
$h_{\mathrm{corr}}$ in Eqs. (\ref{eq:kscptag}), (\ref{eq:kskstag}), (\ref{eq:klcptag}), and (\ref{eq:kskltag}) are %%@
determined from the number of $D^{0}\overline{D}^{0}$ pairs in the sample, $N_{D^{0}\overline{D}^{0}}$, multiplied %%@
by the branching fractions of the modes taken from Ref.~\cite{PDG}. The value of $N_{D^{0}\overline{D}^{0}}$ is %%@
calculated from the integrated luminosity and the value of the cross section for $e^{+}e^{-}\to\psi(3770)\to %%@
D^{0}\overline{D}^{0}$ reported in Ref.~\cite{DHAD}. For the $K^{0}_{S,L}\pi^{+}\pi^{-}$ analysis the measured %%@
single-tag yields are used for all modes apart from those tagged by $D^{0}\to K^{0}_{L}\pi^{0}$ and $D^{0}\to K^- %%@
e^{+}\nu$, which cannot be reconstructed exclusively; in these cases the normalization is performed as in the %%@
$D^{0}\to K^{0}_{S,L}K^{+}K^{-}$ analysis.
\begin{table}[htb]
\begin{center}    
\caption{Values of the parameters used to make the corrections to the pseudo-flavor tag yields and the references %%@
from which they are taken.}
\begin{tabular}{lccc} \hline \hline
    $F$ & $r_{D}^{F}~(\%)$ & $\delta_{D}^F~(^\circ)$  &     $R_F$  \\
\hline
$K\pi$ & \hspace{3mm} $5.80 \pm 0.08$ \cite{HFAG}  \hspace{3mm} & \hspace{3mm} $202 \pm 10$ \cite{HFAG} %%@
\hspace{3mm} & 1 \\
$K\pi\pi^0$ &  $4.8 \pm 0.2$ \cite{TIAN} & $227 \pm 17$ \cite{LOWREY} & \hspace{3mm} $0.84 \pm 0.07$ \cite{LOWREY} %%@
\hspace{3mm}\\
$K\pi\pi\pi$ & $5.7 \pm 0.2$ \cite{TIAN} & $114 \pm 26$ \cite{LOWREY} & $0.33 \pm 0.26$ \cite{LOWREY}\\ \hline %%@
\hline
\label{tab:pseudoftinputs}
\end{tabular}
\end{center}
\end{table}

\begin{table}[htb]
\begin{center}
\caption{Values of $F_{(-)i}$ (\%) measured from the flavor-tagged $D^{0}\to K^{0}_{S}K^{+}K^{-}$ data for the %%@
$\mathcal{N}=2$, $\mathcal{N}=3$, and $\mathcal{N}=4$ equal $\Delta\delta_D$ binnings. Predicted values from the %%@
{\it BABAR} 2010 model of $D^{0}\to K^{0}_{S} K^{+}K^{-}$ are also given.}\label{tab:ki_K0SKK}
\begin{tabular}{lcccc} \hline\hline
$i$ \hspace{1cm} & \multicolumn{2}{c}{$F_i$ (\%)} & \multicolumn{2}{c}{$F_{-i}$ (\%)} \\
  & Measured      & Predicted     & Measured      & Predicted      \\ \hline
\multicolumn{5}{c}{$\mathcal{N}=2$ equal $\Delta\delta_D$ binning} \\ \hline
1 & \hspace{2mm} $23.9\pm 1.6$ \hspace{2mm} & \hspace{2mm} $22.5\pm 4.2$ \hspace{2mm} & \hspace{2mm} $35.5\pm 1.9$ %%@
\hspace{2mm} & \hspace{2mm} $28.6\pm 1.1$ \hspace{2mm} \\
2 & $17.3\pm 1.5$ & $21.1\pm 1.2$ & $23.3\pm 1.7$ & $27.8\pm 4.1$  \\ \hline
\multicolumn{5}{c}{$\mathcal{N}=3$ equal $\Delta\delta_D$ binning} \\ \hline
1 & $22.0\pm 1.5$ & $19.8\pm 3.8$ & $33.0\pm 1.7$ & $25.6\pm 1.0$  \\
2 & $18.1\pm 1.4$ & $22.7\pm 1.4$ & $22.8\pm 1.6$ & $26.1\pm 5.3$  \\
3 & $ 1.2\pm 0.4$ & $ 1.4\pm 0.7$ & $ 3.0\pm 0.6$ & $ 3.8\pm 1.6$  \\ \hline
\multicolumn{5}{c}{$\mathcal{N}=4$ equal $\Delta\delta_D$ binning} \\ \hline
1 & $20.0\pm 1.5$ & $18.3\pm 3.3$ & $30.5\pm 1.7$ & $23.0\pm 1.1$  \\
2 & $ 7.2\pm 1.1$ & $ 8.5\pm 1.0$ & $ 7.6\pm 1.3$ & $ 8.6\pm 1.3$  \\ 
3 & $13.3\pm 1.4$ & $16.3\pm 1.3$ & $17.7\pm 1.4$ & $21.3\pm 4.0$  \\
4 & $ 0.8\pm 0.4$ & $ 0.5\pm 0.4$ & $ 2.8\pm 0.6$ & $ 3.5\pm 1.3$  \\ \hline\hline
\end{tabular}
\end{center}
\end{table}

\begin{table}[htb]
\begin{center}
\caption{Values of $F_{(-)i}$ (\%) measured from the flavor-tagged $D^{0}\to K^{0}_{S}\pi^{+}\pi^{-}$ data for the %%@
optimal binning. Predicted values from the {\it BABAR} 2008 model of $D^{0}\to K^{0}_{S} \pi^{+}\pi^{-}$ are also %%@
given.}\label{tab:ki_optimal}
\begin{tabular}{lcccc} \hline\hline
$i$ \hspace{0.5cm} & \multicolumn{2}{c}{$F_i$ (\%)}        & \multicolumn{2}{c}{$F_{-i}$ (\%)} \\
  &  Measured     & Predicted & Measured & Predicted \\ \hline
1 & $9.0\pm 0.4$  & 9.1       & $2.6\pm 0.2$  & 2.2 \\ 
2 & $14.4\pm 0.5$ & 14.2      & $0.5\pm 0.1$  & 0.5\\
3 & $14.7\pm 0.5$ & 14.1      & $0.3\pm 0.1$  & 0.4\\
4 & $9.9\pm 0.4$  & 10.0      & $5.9\pm 0.3$  & 6.6\\
5 & $5.7\pm 0.3$  &  5.4      & $3.3\pm 0.2$  & 3.2\\
6 & $7.5\pm 0.4$  & 7.4       & $0.5\pm 0.1$  & 0.4\\
7 & $10.9\pm 0.4$ & 11.5      & $5.5\pm 0.3$  & 5.4 \\
8 & $2.2\pm 0.2$  & 2.2       & $6.9\pm 0.3$  & 6.3\\ \hline\hline
\end{tabular}
\end{center}
\end{table}

\begin{table}[htb]
\begin{center}
\caption{Values of $F_{(-)i}$ (\%) measured from the flavor-tagged $D^{0}\to K^{0}_{S}\pi^{+}\pi^{-}$ data for the %%@
equal $\Delta\delta_D$ binning derived from the {\it BABAR} 2008 model. Predicted values from the {\it BABAR} 2008 %%@
model of $D^{0}\to K^{0}_{S} \pi^{+}\pi^{-}$ are also given.}\label{tab:ki_equal}
\begin{tabular}{lcccc} \hline\hline
$i$ \hspace{0.5cm} & \multicolumn{2}{c}{$F_i$ (\%)}        & \multicolumn{2}{c}{$F_{-i}$ (\%)} \\
  &  Measured     & Predicted & Measured & Predicted \\ \hline
1 & $17.0\pm 0.5$  & 17.2       & $8.3\pm 0.4$  & 8.3 \\ 
2 & $8.4\pm 0.4$ & 8.4      & $2.4\pm 0.2$  & 1.9\\
3 & $7.2\pm 0.3$ & 6.9      & $2.3\pm 0.2$  & 2.0\\
4 & $2.4\pm 0.2$  & 2.5      & $1.6\pm 0.2$  & 1.7\\
5 & $7.6\pm 0.4$  &  8.6      & $4.8\pm 0.3$  & 5.3\\
6 & $5.9\pm 0.3$  & 5.9      & $1.3\pm 0.2$  & 1.5\\
7 & $12.8\pm 0.5$ & 12.4      & $1.6\pm 0.2$  & 1.4 \\
8 & $13.0\pm 0.5$  & 13.0      & $3.1\pm 0.2$  & 2.8\\ \hline\hline
\end{tabular}
\end{center}
\end{table}

\begin{table}[htb]
\begin{center}
\caption{Values of $F_{(-)i}$ (\%) measured from the flavor-tagged $D^{0}\to K^{0}_{S}\pi^{+}\pi^{-}$ data for the %%@
modified-optimal binning. Predicted values from the {\it BABAR} 2008 model of $D^{0}\to K^{0}_{S} \pi^{+}\pi^{-}$ %%@
are also given.}\label{tab:ki_modopt}
\begin{tabular}{lcccc} \hline\hline
$i$ \hspace{0.5cm} & \multicolumn{2}{c}{$F_i$ (\%)}        & \multicolumn{2}{c}{$F_{-i}$ (\%)} \\
  &  Measured     & Predicted & Measured & Predicted \\ \hline
1 & $5.4\pm 0.3$  & 5.1       & $1.5\pm 0.2$  & 1.6 \\ 
2 & $16.0\pm 0.5$ & 16.2      & $2.1\pm 0.3$  & 2.0\\
3 & $22.0\pm 0.6$ & 21.6      & $2.3\pm 0.3$  & 2.1\\
4 & $7.8\pm 0.4$  & 8.8      & $4.9\pm 0.5$  & 5.3\\
5 & $3.8\pm 0.3$  &  3.9      & $3.1\pm 0.4$  & 2.8\\
6 & $8.3\pm 0.4$  & 8.1      & $1.1\pm 0.2$  & 1.2\\
7 & $8.7\pm 0.4$ & 9.0      & $4.4\pm 0.5$  & 4.6 \\
8 & $2.2\pm 0.2$  & 2.3      & $6.0\pm 0.6$  & 5.5\\ \hline\hline
\end{tabular}
\end{center}
\end{table}

\begin{table}[htb]
\begin{center}
\caption{Values of $F_{(-)i}$ (\%) measured from the flavor-tagged $D^{0}\to K^{0}_{S}\pi^{+}\pi^{-}$ data for the %%@
equal $\Delta\delta_D$ binning derived from the Belle model. Predicted values from the BABAR 2008 model of %%@
$D^{0}\to K^{0}_{S} \pi^{+}\pi^{-}$ are also given.}\label{tab:ki_belle}
\begin{tabular}{lcccc} \hline\hline
$i$ \hspace{0.5cm} & \multicolumn{2}{c}{$F_i$ (\%)}        & \multicolumn{2}{c}{$F_{-i}$ (\%)} \\
  &  Measured     & Predicted & Measured & Predicted \\ \hline
1 & $16.5\pm 0.5$  & 16.5       & $8.8\pm 0.4$  & 8.0 \\ 
2 & $7.7\pm 0.4$ & 7.6          & $2.0\pm 0.2$  & 1.6\\
3 & $9.8\pm 0.4$ & 10.2         & $3.2\pm 0.2$  & 2.8\\
4 & $3.0\pm 0.2$  & 3.0         & $1.3\pm 0.1$  & 1.2\\
5 & $8.0\pm 0.4$  &  9.2        & $4.0\pm 0.3$  & 4.6\\
6 & $7.1\pm 0.3$  & 7.3         & $1.8\pm 0.2$  & 1.7\\
7 & $9.9\pm 0.4$ & 10.0         & $1.6\pm 0.2$  & 1.3 \\
8 & $12.4\pm 0.4$  & 12.2       & $2.9\pm 0.2$  & 2.6\\ \hline\hline
\end{tabular}
\end{center}
\end{table}

Yields of $K^0_{S,L}h^+h^-$ events selected against Cabibbo-favored (CF) hadronic flavor tags are contaminated with %%@
doubly-Cabibbo suppressed (DCS) decays. We refer to these hadronic flavor modes as {\it pseudo-flavor tags}. This %%@
introduces a bias in the extracted values of $K^{(\prime)}_i$ \cite{BRIERE}. To account for this effect, the %%@
flavor-tagged yields in each bin are scaled by a correction factor, which is estimated using the $D^0\rightarrow %%@
K^0_{S,L}K^+K^-$ and $D^{0}\to K^{0}_{S,L}\pi^{+}\pi^{-}$ decay models reported in Refs.~\cite{BABAR3} and   %%@
\cite{BABAR1}, respectively. The correction factor for pseudo-flavor tag, %%@
$F\in(K^{-}\pi^{+},~K^{-}\pi^{+}\pi^{0},~K^{-}\pi^{+}\pi^{+}\pi-)$, is
\begin{equation}
\frac{\int_i|f(m_{+}^{2},m_{-}^{2})|^2 dm^{2}_{+}dm^{2}_{-}}{\int_i\big(|f(m_{+}^{2},m_{-}^{2})|^2 + %%@
(r^F_D)^2|f(m_{-}^{2},m_{+}^{2})|^2 + %%@
2r^F_DR_F\Re[e^{-i\delta^F_D}f(m_{+}^{2},m_{-}^{2})f^{*}(m_{-}^{2},m_{+}^{2})]\big)dm_{+}^{2}dm_{-}^{2}} \; ,
\end{equation}
where $r^F_D$ is the ratio of the DCS to CF decay amplitudes and $\delta^F_D$ is the associated average %%@
strong-phase difference.  $R_F$ is the coherence factor for decays to three or more particles \cite{AS} and equals %%@
unity for two-body decays. The values of these parameters and the references from which they are taken are given in %%@
Table~\ref{tab:pseudoftinputs}. 

Equation~(\ref{eq:kidef}) defines $F_{(-)i}$, the normalized values of the flavor-tagged yields in each bin, such %%@
that $F_{(-)i}=K_{(-)i}/A_D$, where $A_{D}=\sum_{i=1}^{\mathcal{N}}(K_{i}+K_{-i})$. The fully-corrected values of %%@
$F_{(-)i}$ measured for the $\mathcal{N}=2$, $\mathcal{N}=3$, and $\mathcal{N}=4$ equal $\Delta\delta_D$ binnings %%@
are given in Table~\ref{tab:ki_K0SKK}. The results are the average of the pseudo-flavor tag modes with the  $D^0\to %%@
K^{-}e^{+}\nu$ tagged data. Also given are the predicted values for $F_{(-)i}$ from the {\it BABAR} 2010 amplitude %%@
model of $D^{0}\to K^{0}_{S} K^{+}K^{-}$ decays. The error on the predicted value of $F_{(-)i}$ is determined from %%@
the uncertainties on the amplitude-model parameters. The agreement between the measured and predicted values is %%@
reasonable in all bins except $F_{-1}$, which is different by more than three standard deviations for all binnings. %%@
This discrepancy may be a statistical fluctuation or indicate a deficiency in the model in this region. In order to %%@
ascertain whether this effect can lead to a significant bias in our analysis, we perform the fit to determine %%@
$c^{(\prime)}_{i}$ and $s^{(\prime)}_{i}$ with the predicted rather than measured values of $K_{(-)i}$; the %%@
difference in fit results is negligible. Therefore, we conclude it is reasonable to use our measured values of %%@
$K_{(-)i}$ to determine the parameters. Tables~\ref{tab:ki_optimal}, \ref{tab:ki_equal}, \ref{tab:ki_modopt}, and %%@
\ref{tab:ki_belle} show the measured values of $F_{(-)i}$ for the $D^{0}\to K^{0}_{S}\pi^{+}\pi^{-}$ data divided %%@
according to the optimal, {\it BABAR} 2008 model with equal $\Delta\delta_D$, modified-optimal, and Belle equal %%@
$\Delta\delta_D$ binnings, respectively. Again these results are the average of pseudo-flavor and semileptonic %%@
tagged data. Predicted values are also given. In this case the uncertainty due to the amplitude-model parameters is %%@
negligible compared to the uncertainties on the measurements. The agreement between measured and predicted values %%@
is reasonable in all cases.

We use the corrected and normalized values of the bin yields to determine $c_i^{(\prime)}$ and $s_i^{(\prime)}$. %%@
The fits to the $K^{0}_{S,L}\pi^{+}\pi^{-}$ and $K^{0}_{S,L}K^{+}K^{-}$ data are made separately. We perform the %%@
fit to $K^{0}_{S,L}\pi^{+}\pi^{-}$ data first because the $K^{0}_{S,L}K^{+}K^{-}$ fit depends upon the values of %%@
the $c^{(\prime)}_i$ and $s^{(\prime)}_i$ for $D^{0}\to K^{0}_{S,L}\pi^{+}\pi^{-}$ decays when including %%@
$K^{0}_{S}K^{+}K^{-}~vs.~K^{0}_{S,L}\pi^{+}\pi^{-}$ and $K^{0}_{L}K^{+}K^{-}~vs.~K^{0}_{S}\pi^{+}\pi^{-}$ %%@
candidates. The fit results from the equal $\Delta\delta_D$ binning derived from the {\it BABAR} 2008 model are %%@
used in the fit to $K^{0}_{S,L}K^{+}K^{-}$ data; the $K^{0}_{S,L}\pi^{+}\pi^{-}$ strong-phase difference parameters %%@
are fixed to the measured values in the nominal fit, these are then varied within their errors to determine the %%@
related systematic uncertainty (Sec.~\ref{sec:syst_K0SKK}). In the $K^0_S\pi^+\pi^-$ analysis we obtain values of %%@
$c_i^{(\prime)}$ and $s_i^{(\prime)}$ by minimizing the log-likelihood expression
\begin{eqnarray}
-2\log\mathcal{L}& = & -2\sum_i \log P(M^\pm_i,\langle M^\pm_i\rangle)_{K^0_S\pi^+\pi^-,CP} \nonumber \\
& & -2\sum_i \log P(M^{\prime\pm}_i,\langle M^{\prime\pm}_i\rangle)_{K^0_L\pi^+\pi^-,CP} \nonumber \\
& & -2\sum_{i,j} \log P(M_{ij},\langle M_{ij}\rangle)_{K^0_S\pi^+\pi^-,K^0_S\pi^+\pi^-} \nonumber \\
& & -2\sum_{i,j} \log P(M^{\prime}_{ij},\langle M^{\prime}_{ij}\rangle)_{K^0_S\pi^+\pi^-,K^0_L\pi^+\pi^-} \nonumber %%@
\\
& & + \chi^2 \; .
\label{eq:kspipiextract}
\end{eqnarray}
In the $K^0_SK^+K^-$ analysis we minimize the expression
\begin{eqnarray}
-2\log\mathcal{L} & = &  -2\sum_i \log P(M^\pm_i,\langle M^\pm_i\rangle)_{K^0_SK^+K^-,CP} \nonumber \\
&& -2\sum_i \log P(M^{\prime\pm}_i,\langle M^{\prime\pm}_i\rangle)_{K^0_LK^+K^-,CP} \nonumber \\
&& -2\sum_{i,j} \log P(M_{ij},\langle M_{ij}\rangle)_{K^0_SK^+K^-,K^0_SK^+K^-} \nonumber \\
&& -2\sum_{i,j} \log P(M^{\prime}_{ij},\langle M^{\prime}_{ij}\rangle)_{K^0_SK^+K^-,K^0_LK^+K^-} \nonumber \\
&& -2\sum_{i,j} \log P(M_{ij},\langle M_{ij}\rangle)_{K^0_SK^+K^-,K^0_S\pi^+\pi^-} \nonumber \\
&& -2\sum_{i,j} \log P(M^{\prime}_{ij},\langle M^{\prime}_{ij}\rangle)_{K^0_SK^+K^-,K^0_L\pi^+\pi^-} \nonumber \\
&& -2\sum_{i,j} \log P(M^{\prime}_{ij},\langle M^{\prime}_{ij}\rangle)_{K^0_LK^+K^-,K^0_S\pi^+\pi^-} \nonumber \\
&& + \chi^2 \; .
\label{eq:kskkextract}
\end{eqnarray}
Here the expected number of $CP$-tagged $K^0_Sh^+h^-$ $(K^0_Lh^+h^-)$ events in the $i^{\mathrm{th}}$ bin, $\langle %%@
M^\pm_i\rangle$ ($\langle M^{\prime\pm}_i\rangle$), is determined from Eq.~(\ref{eq:kscptag}) %%@
(Eq.~(\ref{eq:klcptag})). Similarly the expected number of events where both $D$ mesons decay to %%@
$K^{0}_{S,L}h^{+}h^{-}$, $\langle M_{ij}\rangle$ $(\langle M^{\prime}_{ij}\rangle)$ is determined using %%@
Eq.~(\ref{eq:kskstag}) (Eq.~(\ref{eq:kskltag})). The function $P(M, \langle M\rangle)$ is the Poisson probability %%@
of obtaining $M$ events given a mean of $\langle M\rangle$. There is an additional $\chi^2$ term 
\begin{equation}
\chi^2 = \sum_i \left(\frac{c_i'-c_i-\Delta c_i}{\delta\Delta c_i}\right)^2 + \sum_i \left(\frac{s_i'-s_i-\Delta %%@
s_i}{\delta\Delta s_i}\right)^2
\end{equation}
which constrains the extracted $c^{\prime}_i$ $(s^{\prime}_i)$ to differ within errors from $c_i$ $(s_i)$ by the %%@
predicted quantities $\Delta c_i$ $(\Delta s_i)$. 

We briefly discuss the estimation of $\Delta c_i$ $(\Delta s_i)$ and their uncertainties. An isobar resonance model %%@
must be used to determine these constraints; we use the $D^{0}\to K^{0}_{S}K^{+}K^{-}$ and $D^{0}\to %%@
K^{0}_{S}\pi^{+}\pi^{-}$ models reported in Refs.~\cite{BABAR3} and \cite{BABAR1}, respectively. The intermediate %%@
resonances contributing to $D^{0}\to K^{0}_{L}h^{+}h^{-}$ model differ in two ways from those contributing to %%@
$D^{0}\to K^{0}_{S}h^{+}h^{-}$. Firstly, DCS decays contribute with opposite sign. This can be seen by considering %%@
the $D^{0}\to K^{0}_{S,L}h^{+}h^{-}$ amplitudes, $\mathcal{A}$, in terms of those to the flavor eigenstates %%@
$D^{0}\to K^{0} h^{+}h^{-}$ and $D^{0}\to\overline{K}^{0}h^{+}h^{-}$
\begin{eqnarray}
 \mathcal{A}(D^{0}\to K^{0}_{S}h^{+}h^{-}) & = & \frac{1}{\sqrt{2}}\left[ \mathcal{A}(D^{0}\to K^{0}h^{+}h^{-}) + %%@
\mathcal{A}(D^{0}\to \overline{K}^{0}h^{+}h^{-})  \right] \\
\mathcal{A}(D^{0}\to K^{0}_{L}h^{+}h^{-}) & = & \frac{1}{\sqrt{2}}\left[ \mathcal{A}(D^{0}\to K^{0}h^{+}h^{-}) - %%@
\mathcal{A}(D^{0}\to \overline{K}^{0}h^{+}h^{-})  \right] \; .  
\end{eqnarray}
The relative minus sign between the terms in $\mathcal{A}(D^{0}\to K^{0}_{L}h^{+}h^{-})$ can be accommodated by %%@
introducing a $180^{\circ}$ phase difference for all DCS contributions to the $D^{0}\to K^{0}_{L}h^{+}h^{-}$ model %%@
compared to the same contribution to the $D^{0}\to K^{0}_{S}h^{+}h^{-}$ model. Secondly, for $CP$-eigenstate %%@
amplitudes, such as $D^{0}\to K^{0}_{S,L}\phi$, the $D^{0}\to K^{0}_{L}h^{+}h^{-}$ amplitude can be related to the %%@
$D^{0}\to K^{0}_{S}h^{+}h^{-}$ amplitude by multiplying the latter by a factor $(1-2re^{i\delta})$, where $r$ is of %%@
the order $\tan^2{\theta_{C}}$ and $\delta$ can take any value. Here, $\theta_{C}$ is the Cabibbo angle. The origin %%@
of this factor is again related to the sign difference between DCS contributions to the $D^{0}\to %%@
K^{0}_{S}h^{+}h^{-}$ and $D^{0}\to K^{0}_{L}h^{+}h^{-}$ amplitudes, and is analogous to the mechanism which induces %%@
the difference in rates for $D\to K^{0}_{S}\pi$ and $D\to K^{0}_{L}\pi$ decay \cite{BIGIYAMA,HE}. We determine %%@
central values of $\Delta c_i$ and $\Delta s_i$ by assuming $r=\tan^2{\theta_{C}}$ \cite{PDG} and %%@
$\delta=0^{\circ}$. Part of the uncertainty on $\Delta c_i$ and $\Delta s_i$ is evaluated by randomly choosing the %%@
assumed values of $r$ and $\delta$ 100 times, and recomputing the constraints for each set of parameters. The value %%@
of $\delta$ is assumed to have a equal probability to lie between $0^{\circ}$ and $360^{\circ}$ and that of $r$ to %%@
have a Gaussian distribution of mean $\tan^2{\theta_{C}}$ and width $0.5\times \tan^2{\theta_{C}}$.  The RMS of the %%@
resulting distributions of $\Delta c_i$ and $\Delta s_i$ are taken as the uncertainties from this source. A second %%@
source of uncertainty is related to the model choice. For the $D^{0}\to K^{0}_{S}K^{+}K^{-}$ model this is %%@
estimated by varying the isobar model parameters by their uncertainties \cite{BABAR3}, accounting for any %%@
correlations among the parameters \cite{BITMAPS}, then recomputing $\Delta c_i$ and $\Delta s_i$. The differences %%@
with respect to the values of the constraints computed with the nominal values of the parameters are taken as the %%@
uncertainty on $\Delta c_i$ and $\Delta s_i$ from this source. For $D^{0}\to K^{0}_{S}\pi^{+}\pi^{-}$ we follow %%@
Ref.~\cite{BRIERE} and consider two alternative models \cite{BELLE, CLEOAMP} to determine $\Delta c_i$ and $\Delta %%@
s_i$; the largest deviation of the central value from that computed with the default model \cite{BABAR1} is taken %%@
as a systematic uncertainty.  As examples of the constraints found by this procedure, the values of $\Delta c_i$ %%@
and $\Delta s_i$ for the $\mathcal{N}=3$ division of the $D^{0}\to K^{0}_{S}K^{+}K^{-}$ Dalitz plot and the optimal %%@
binning of the $D^{0}\to K^{0}_{S}\pi^{+}\pi^{-}$ Dalitz plot are given in Tables~\ref{tab:kskk_deltac} and %%@
\ref{tab:kspipi_deltac}, respectively. The size of $\Delta c_i$ and $\Delta s_i$ can be significant for bins %%@
dominated by either a DCS decay, such as the $K^{*+}(892)$ in bin three of the optimal $D^{0}\to %%@
K^{0}_{S}\pi^{+}\pi^{-}$ binning (Fig.~\ref{fig:optimalbinning}(b)), or a neutral resonance, such as the %%@
$a^{0}(1450)$ in bin three of the $\mathcal{N}=3$ division of the $D^{0}\to K^{0}_{S}K^{+}K^{-}$ Dalitz plot %%@
(Fig.~\ref{fig:KsKKbinsequal}).   

\begin{table}[htb]
\begin{center}
 \caption{Values of $\Delta c_i$ and $\Delta s_i$ constraints for the $\mathcal{N}=3$ equal $\Delta\delta_D$ %%@
binning of the $D^{0}\to K^{0}_{S}K^{+}K^{-}$ Dalitz plot.}\label{tab:kskk_deltac}
\begin{tabular}{l c c} \hline\hline
$i$ & $\Delta c_i$ & $\Delta s_i$ \\ \hline
$1$ & $\phantom{-}0.026 \pm 0.014$ & $-0.007 \pm 0.023$ \\
$2$ & $\phantom{-}0.041 \pm 0.019$ & $\phantom{-}0.012 \pm 0.010$ \\
$3$ & $-0.563 \pm 0.311$ & $\phantom{-}0.713 \pm 0.161$ \\ \hline\hline 
\end{tabular}
\end{center}
\end{table}

\begin{table}[htb]
\begin{center}
 \caption{Values of $\Delta c_i$ and $\Delta s_i$ constraints for the optimal $D^{0}\to K^{0}_{S}\pi^{+}\pi^{-}$ %%@
binning.}\label{tab:kspipi_deltac}
\begin{tabular}{lcc} \hline\hline
$i$ & $\Delta c_i$ & $\Delta s_i$    \\ \hline
1 & $\phantom{-}0.39\pm 0.17$ & $\phantom{-}0.07\pm 0.06$  \\ 
2 & $\phantom{-}0.18\pm 0.05$ & $\phantom{-}0.01\pm 0.10$  \\
3 & $\phantom{-}0.61\pm 0.15$ & $\phantom{-}0.30\pm 0.12$  \\
4 & $\phantom{-}0.09\pm 0.08$ & $\phantom{-}0.00\pm 0.08$  \\
5 & $\phantom{-}0.16\pm 0.17$ & $\phantom{-}0.06\pm 0.06$  \\
6 & $\phantom{-}0.57\pm 0.21$ & $-0.15\pm 0.24$  \\
7 & $\phantom{-}0.03\pm 0.01$ & $-0.04\pm 0.06$  \\
8 & $-0.10\pm 0.15$& $-0.15\pm 0.21$ \\ \hline\hline
\end{tabular}
\end{center}
\end{table}

The contribution of the $\chi^2$ to the likelihood is investigated to ensure that these constraints are not leading %%@
to any significant bias. For the fits to $D^{0}\to K^{0}_{S,L}K^{+}K^{-}$ and $D^{0}\to K^{0}_{S,L}\pi^{+}\pi^{-}$ %%@
the ranges of $\chi^2$ are 0.30 to 0.75 and 1.0 to 2.3, respectively. In addition, no individual bin contributes %%@
more than 0.9 to the total $\chi^2$. Therefore, we conclude that the constraint is not biasing our result %%@
significantly from the values favored by the data, and is principally improving the precision of the parameters.

The $D^{0}\to K^{0}_{S,L}K^{+}K^{-}$ fitting procedure has been tested using samples of signal MC
events. In the validation procedure,
the number of events for each tag is assumed to follow a Poisson distribution about the mean expectation, while the %%@
ratio between double-tagged events and single-tagged events is computed as
${\cal B}_{K^{0}_S K^+ K^-}/2$, ignoring quantum-correlations.
The means of the fitted $c_i$ and $s_i$ distributions exhibit small, but statistically significant, biases due to %%@
the assumptions made in the fit. The systematic uncertainty we associate to the bias is described in %%@
Sec.~\ref{sec:syst_K0SKK}.

The parameters that result from fits to the $D^{0}\to K^{0}_{S,L}K^{+}K^{-}$ data divided into $\mathcal{N}=2$, %%@
$\mathcal{N}=3$, and $\mathcal{N}=4$ equal $\Delta\delta_D$ bins are given in Tables~\ref{tab:K0SKK_results} and %%@
\ref{tab:K0SKK_results_prime}. The statistical uncertainty on the parameters includes that related to the $\Delta %%@
c_i$ and $\Delta s_i$ constraints. The statistical correlations among the parameters for the $\mathcal{N}=3$ equal %%@
$\Delta\delta_D$ binning are shown in Table~\ref{tab:K0SKK_correlations}. The other statistical correlations are %%@
given in Ref.~\cite{EPAPS}.

\begin{table}
\begin{center}
\caption{Measured values of $c_i$ and $s_i$ for the different $D^{0}\to K^{0}_{S}K^{+}K^{-}$ binnings. The first %%@
uncertainty is statistical, including that related to the $\Delta c_i$ and $\Delta s_i$ constraints, and the second %%@
uncertainty is systematic.}\label{tab:K0SKK_results}
\begin{tabular}{l c c c}
\hline\hline
 $i$ \hspace{1cm} & $c_i$ & \hspace{1cm} & $s_i$ \\ \hline
\multicolumn{4}{c}{$\mathcal{N}=2$ equal $\Delta\delta_D$ bins} \\\hline 
 $1$ & $ \phantom{-}0.818 \pm 0.107 \pm 0.037 $ & & $ \phantom{-}0.445 \pm 0.215 \pm 0.143 $ \\
 $2$ & $           -0.746 \pm 0.083 \pm 0.035 $ & & $ \phantom{-}0.229 \pm 0.220 \pm 0.079 $ \\ \hline
\multicolumn{4}{c}{$\mathcal{N}=3$ equal $\Delta\delta_D$ bins} \\\hline 
 $1$ & $ \phantom{-}0.793 \pm 0.063 \pm 0.038 $ & & $ \phantom{-}0.431 \pm 0.222 \pm 0.142 $ \\
 $2$ & $           -0.566 \pm 0.092 \pm 0.034 $ & & $ \phantom{-}0.413 \pm 0.234 \pm 0.094 $ \\
 $3$ & $           -0.096 \pm 0.329 \pm 0.131 $ & & $           -0.461 \pm 0.432 \pm 0.175 $ \\ \hline
\multicolumn{4}{c}{$\mathcal{N}=4$ equal $\Delta\delta_D$ bins} \\\hline
 $1$ & $ \phantom{-}0.858 \pm 0.059 \pm 0.034 $ & & $ \phantom{-}0.309 \pm 0.248 \pm 0.180 $ \\
 $2$ & $ \phantom{-}0.176 \pm 0.223 \pm 0.091 $ & & $ \phantom{-}0.992 \pm 0.473 \pm 0.403 $ \\
 $3$ & $           -0.819 \pm 0.095 \pm 0.045 $ & & $ \phantom{-}0.307 \pm 0.267 \pm 0.201 $ \\
 $4$ & $ \phantom{-}0.376 \pm 0.329 \pm 0.157 $ & & $           -0.133 \pm 0.659 \pm 0.323 $ \\ \hline\hline
\end{tabular} 
\end{center}
\end{table} 

\begin{table}
\begin{center}
\caption{Measured values of $c^{\prime}_i$ and $s^{\prime}_i$ for the different $D^{0}\to K^{0}_{S}K^{+}K^{-}$ %%@
binnings. The first uncertainty is statistical, including that related to the $\Delta c_i$ and $\Delta s_i$ %%@
constraints, and the second uncertainty is systematic.}\label{tab:K0SKK_results_prime}
\begin{tabular}{l c c c}
\hline\hline
 $i$ \hspace{1cm} & $c^{\prime}_i$ & \hspace{1cm} & $s^{\prime}_i$ \\ \hline
\multicolumn{4}{c}{$\mathcal{N}=2$ equal $\Delta\delta_D$ bins} \\\hline 
 $1$ & $ \phantom{-}0.839 \pm 0.108 \pm 0.073 $ & & $ \phantom{-}0.445 \pm 0.216 \pm 0.150 $ \\
 $2$ & $           -0.775 \pm 0.085 \pm 0.068 $ & & $ \phantom{-}0.298 \pm 0.220 \pm 0.093 $ \\ \hline
\multicolumn{4}{c}{$\mathcal{N}=3$ equal $\Delta\delta_D$ bins} \\\hline 
 $1$ & $ \phantom{-}0.814 \pm 0.063 \pm 0.064 $ & & $ \phantom{-}0.422 \pm 0.222 \pm 0.143 $ \\
 $2$ & $           -0.529 \pm 0.092 \pm 0.071 $ & & $ \phantom{-}0.426 \pm 0.234 \pm 0.098 $ \\
 $3$ & $           -0.583 \pm 0.349 \pm 0.197 $ & & $ \phantom{-}0.241 \pm 0.456 \pm 0.181 $ \\  \hline
\multicolumn{4}{c}{$\mathcal{N}=4$ equal $\Delta\delta_D$ bins} \\\hline
 $1$ & $ \phantom{-}0.874 \pm 0.059 \pm 0.055 $ & & $ \phantom{-}0.303 \pm 0.248 \pm 0.180 $ \\
 $2$ & $ \phantom{-}0.270 \pm 0.225 \pm 0.160 $ & & $ \phantom{-}0.965 \pm 0.473 \pm 0.406 $ \\
 $3$ & $           -0.810 \pm 0.096 \pm 0.060 $ & & $ \phantom{-}0.346 \pm 0.268 \pm 0.204 $ \\
 $4$ & $           -0.317 \pm 0.408 \pm 0.201 $ & & $ \phantom{-}0.770 \pm 0.696 \pm 0.350 $ \\ \hline\hline
\end{tabular} 
\end{center}
\end{table} 

\begin{table}
\begin{center}
\caption{Statistical correlations (\%) among the parameters for $\mathcal{N}=3$ equal $\Delta\delta_D$ binning of %%@
the $D^{0}\to K^{0}_{S}K^{+}K^{-}$ Dalitz plot.}\label{tab:K0SKK_correlations}
\begin{tabular}{lrrrrrrrrrrr} \hline\hline
                & $c_2 $ & $c_3$ & $s_1$ & $s_2$ & $s_3$ & $c^{\prime}_1$ & $c^{\prime}_2$ & $c^{\prime}_3$ & %%@
$s^{\prime}_1$ & $s^{\prime}_2$ & $s^{\prime}_3$ \\ \hline 
$c_1$           & $  0.6 $  & $ -2.9 $  & $ -1.3 $  & $ -0.6 $  & $ -0.2 $  & $  97.6 $ & $  0.6 $  & $ -2.1 $  & $ %%@
-1.3 $  & $ -0.6 $  & $ -0.1 $ \\ 
$c_2$           &           & $  0.8 $  & $  1.1 $  & $  3.2 $  & $  -0.0 $ & $  0.6 $  & $  98.0 $ & $  0.4 $  & $  %%@
1.1 $  & $  3.2 $  & $  -0.0 $ \\
$c_3$           &           &           & $ -0.3 $  & $  0.2 $  & $  2.5 $  & $ -2.8 $  & $  0.8 $  & $  66.4 $ & $ %%@
-0.3 $  & $  0.2 $  & $  2.4 $ \\ 
$s_1$           &           &           &           & $ -2.0 $  & $  6.1 $  & $ -1.3 $  & $  1.1 $  & $ -0.1 $  & $  %%@
99.4 $ & $ -2.0 $  & $  5.8 $ \\ 
$s_2$           &           &           &           &           & $ -3.3 $  & $ -0.5 $  & $  3.2 $  & $  0.1 $  & $ %%@
-1.9 $  & $  99.9 $ & $ -3.0 $ \\ 
$s_3$           &           &           &           &           &           & $ -0.2 $  & $  -0.0 $ & $  2.2 $  & $  %%@
6.1 $  & $ -3.3 $  & $  93.7 $ \\
$c^{\prime}_1$  &           &           &           &           &           &           & $  0.6 $  & $ -2.0 $  & $ %%@
-1.3 $  & $ -0.5 $  & $ -0.1 $ \\ 
$c^{\prime}_2$  &           &           &           &           &           &           &           & $  0.4 $  & $  %%@
1.1 $  & $  3.2 $  & $  -0.0 $ \\
$c^{\prime}_3$  &           &           &           &           &           &           &           &           & $ %%@
-0.1 $  & $  0.1 $  & $  2.2 $ \\ 
$s^{\prime}_1$  &           &           &           &           &           &           &           &           &           %%@
& $ -1.9 $  & $  5.8 $ \\ 
$s^{\prime}_2$  &           &           &           &           &           &           &           &           &           %%@
&           & $ -3.0 $ \\ 
\hline\hline
\end{tabular}
\end{center}
\end{table}

Tables~\ref{tab:K0Spipi_results} and \ref{tab:K0Spipi_resultsprime} give the fit results for the $D^{0}\to %%@
K^{0}_{S,L}\pi^{+}\pi^{-}$ data divided according to the four different binnings. The statistical correlation %%@
matrices for each binning are given in Ref.~\cite{EPAPS}. 

\begin{table}
\begin{center}
\caption{Measured values of $c_i$ and $s_i$ for the different $D^{0}\to K^{0}_{S}\pi^{+}\pi^{-}$ binnings. The %%@
first uncertainty is statistical, including that related to the $\Delta c_i$ and $\Delta s_i$ constraints, and  the %%@
second uncertainty is systematic.}\label{tab:K0Spipi_results}
\begin{tabular}{l|cc|cc}\hline\hline
    & \multicolumn{2}{c|}{Optimal}   & \multicolumn{2}{c}{Equal $\Delta\delta_{D}$ BABAR 2008} \\
$i$ & $c_i$ & $s_i$   & $c_i$   & $s_i$ \\ \hline
 1 &  $-0.009 \pm 0.088 \pm 0.094$& $-0.438 \pm 0.184 \pm 0.045$ & $\phantom{-}0.655 \pm 0.036 \pm 0.042$ & $-0.025 %%@
\pm 0.098 \pm 0.043$ \\     
 2 &  $\phantom{-}0.900 \pm 0.106 \pm 0.082$ & $-0.490 \pm 0.295 \pm 0.261$&  $\phantom{-}0.511 \pm 0.068 \pm %%@
0.063$ & $\phantom{-}0.141 \pm 0.183 \pm 0.066$  \\     
 3 &  $\phantom{-}0.292 \pm 0.168 \pm 0.139$ & $-1.243 \pm 0.341 \pm 0.123$ & $\phantom{-}0.024 \pm 0.140 \pm %%@
0.080$ & $\phantom{-}1.111 \pm 0.131 \pm 0.044$  \\    
 4 &  $-0.890 \pm 0.041 \pm 0.044$& $-0.119 \pm 0.141 \pm 0.038$ & $-0.569 \pm 0.118 \pm 0.098$ & $\phantom{-}0.328 %%@
\pm 0.202 \pm 0.072$ \\    
 5 &  $-0.208 \pm 0.085 \pm 0.080$& $\phantom{-}0.853 \pm 0.123 \pm 0.035$ & $-0.903 \pm 0.045 \pm 0.042$ & $-0.181 %%@
\pm 0.131 \pm 0.026$ \\    
 6 &  $\phantom{-}0.258 \pm 0.155 \pm 0.108$ & $\phantom{-}0.984 \pm 0.357 \pm 0.165$ & $-0.616 \pm 0.103 \pm %%@
0.072$ & $-0.520 \pm 0.196 \pm 0.059$ \\    
 7 &  $\phantom{-}0.869 \pm 0.034 \pm 0.033$ & $-0.041 \pm 0.132 \pm 0.034$ & $\phantom{-}0.100 \pm 0.106 \pm %%@
0.124$  & $-1.129 \pm 0.120 \pm 0.096$\\     
 8 &  $\phantom{-}0.798 \pm 0.070 \pm 0.047$ & $-0.107 \pm 0.240 \pm 0.080$ & $\phantom{-}0.422 \pm 0.069 \pm %%@
0.075$  & $-0.350 \pm 0.151 \pm 0.045$ \\ \hline
    & \multicolumn{2}{c|}{Modified optimal}   & \multicolumn{2}{c}{Equal $\Delta\delta_{D}$ Belle} \\
$i$ & $c_i$ & $s_i$   & $c_i$   & $s_i$ \\ \hline
 1 &  $-0.216 \pm 0.104 \pm 0.088$& $-0.399 \pm 0.204 \pm 0.049$ & $\phantom{-}0.710 \pm 0.034 \pm 0.038$ & $-0.013 %%@
\pm 0.097  \pm 0.031$ \\     
 2 &  $\phantom{-}0.827 \pm 0.060 \pm 0.053$ & $-0.031 \pm 0.172 \pm 0.084$&  $\phantom{-}0.481 \pm 0.080 \pm %%@
0.070$ & $-0.147 \pm 0.177  \pm 0.107$  \\     
 3 &  $\phantom{-}0.101 \pm 0.085 \pm 0.118$ & $-0.558 \pm 0.161 \pm 0.070$ & $\phantom{-}0.008 \pm 0.080 \pm %%@
0.087$ & $\phantom{-}0.938 \pm 0.120 \pm 0.047$  \\    
 4 &  $-0.955 \pm 0.038 \pm 0.034$& $-0.204 \pm 0.137 \pm 0.055$ & $-0.757 \pm 0.099 \pm 0.065$ & $\phantom{-}0.386 %%@
\pm 0.208 \pm 0.067$ \\    
 5 &  $-0.522 \pm 0.095 \pm 0.079$& $\phantom{-}0.911 \pm 0.130 \pm 0.067$ & $-0.884 \pm 0.056 \pm 0.054$ & $-0.162 %%@
\pm 0.130 \pm 0.041$ \\    
 6 &  $\phantom{-}0.291 \pm 0.102 \pm 0.075$ & $\phantom{-}1.030 \pm 0.196 \pm 0.065$ & $-0.462 \pm 0.100 \pm %%@
0.082$ & $-0.616 \pm 0.188 \pm 0.052$ \\    
 7 &  $\phantom{-}0.682 \pm 0.051 \pm 0.047$ & $-0.037 \pm 0.146 \pm 0.029$ & $\phantom{-}0.106 \pm 0.105 \pm %%@
0.100$  & $-1.063 \pm 0.174 \pm 0.066$\\     
 8 &  $\phantom{-}0.724 \pm 0.071 \pm 0.044$ & $-0.180 \pm 0.194 \pm 0.050$ & $\phantom{-}0.365 \pm 0.071 \pm %%@
0.078$  & $-0.179 \pm 0.166 \pm 0.048$ \\ \hline\hline
\end{tabular}
\end{center}
\end{table}

\begin{table}
\begin{center}
\caption{Measured values of $c^{\prime}_i$ and $s^{\prime}_i$ for the different $D^{0}\to K^{0}_{S}\pi^{+}\pi^{-}$ %%@
binnings. The first uncertainty is statistical, including that related to the $\Delta c_i$ and $\Delta s_i$ %%@
constraints, and the second uncertainty is systematic.}\label{tab:K0Spipi_resultsprime}
\begin{tabular}{l|cc|cc}\hline\hline
    & \multicolumn{2}{c|}{Optimal}   & \multicolumn{2}{c}{Equal $\Delta\delta_{D}$ BABAR 2008} \\
$i$ & $c^{\prime}_i$ & $s^{\prime}_i$   & $c^{\prime}_i$   & $s^{\prime}_i$ \\ \hline
1& $\phantom{-}0.470\pm 0.096 \pm 0.102 $& $-0.363\pm 0.185\pm 0.191 $& $\phantom{-}0.768 \pm 0.038\pm 0.051$ & %%@
$-0.079\pm 0.095\pm 0.108$\\
2& $\phantom{-}1.073\pm 0.102\pm 0.126$ & $-0.501\pm 0.297 \pm 0.397$ & $\phantom{-}0.679\pm 0.067\pm 0.090$& %%@
$\phantom{-}0.080 \pm 0.183 \pm 0.196$ \\
3& $\phantom{-} 0.869\pm 0.142\pm 0.165$& $-0.890\pm 0.329\pm 0.352$& $\phantom{-}0.278\pm 0.106\pm 0.112$ & %%@
$\phantom{-} 1.090 \pm 0.106 \pm 0.119$\\
4& $-0.786\pm 0.047\pm 0.052$& $-0.137\pm 0.154 \pm 0.159$ & $-0.446\pm 0.116 \pm 0.128$ & $\phantom{-}0.455\pm %%@
0.219\pm 0.235$ \\
5& $-0.139\pm 0.089\pm 0.093$ & $\phantom{-} 0.921 \pm 0.126 \pm 0.132$ & $-0.824\pm 0.051\pm 0.056$ & $-0.194\pm %%@
0.136\pm 0.140$\\
6& $\phantom{-}0.654\pm 0.135 \pm 0.152$ & $\phantom{-}0.832\pm 0.326\pm 0.362$ & $-0.240\pm 0.116\pm 0.123$ & %%@
$-0.557\pm 0.201 \pm 0.211$\\
7& $\phantom{-} 0.901 \pm 0.034 \pm 0.047$ & $-0.076 \pm 0.132 \pm 0.137$ & $\phantom{-}0.480\pm 0.106\pm 0.113$& %%@
$-0.975 \pm 0.104 \pm 0.141$\\
8& $\phantom{-}0.817\pm 0.090\pm 0.095$ & $-0.157\pm 0.281\pm 0.306$ & $\phantom{-}0.708\pm 0.066\pm 0.083$ & %%@
$-0.285\pm 0.150\pm 0.158$ \\ \hline
    & \multicolumn{2}{c|}{Modified optimal}   & \multicolumn{2}{c}{Equal $\Delta\delta_{D}$ Belle} \\
$i$ & $c^{\prime}_i$ & $s^{\prime}_i$   & $c^{\prime}_i$   & $s^{\prime}_i$ \\ \hline
1& $-0.049\pm 0.108\pm 0.125$ & $-0.386\pm 0.218\pm 0.226$ & $\phantom{-}0.817\pm 0.035\pm 0.048$& $-0.070\pm 0.095 %%@
\pm 0.103$\\
2& $\phantom{-}0.935\pm 0.056\pm 0.072$ & $-0.028\pm 0.168 \pm 0.191$ & $\phantom{-}0.668\pm 0.079\pm 0.104$ & %%@
$-0.219\pm 0.176\pm 0.209$\\
3& $\phantom{-}0.614\pm 0.082\pm 0.089$& $-0.398\pm 0.160\pm 0.175$ & $\phantom{-}0.197\pm 0.082\pm 0.090$ & %%@
$\phantom{-}0.935 \pm 0.117 \pm 0.130$ \\
4& $-0.876\pm 0.047\pm 0.051$ & $-0.249\pm 0.139 \pm 0.149$ & $-0.592\pm 0.117\pm 0.125$ & $\phantom{-}0.520\pm %%@
0.254\pm 0.268$\\
5& $-0.357 \pm 0.094 \pm 0.100$ & $\phantom{-}0.980 \pm 0.131 \pm 0.151$ & $-0.720 \pm 0.056\pm 0.062$ & $-0.192\pm %%@
0.135\pm 0.142$\\
6& $\phantom{-}0.584\pm 0.094\pm 0.103$ & $\phantom{-}0.963\pm 0.200\pm 0.214$& $-0.121\pm 0.108\pm 0.117$& %%@
$-0.630\pm 0.194\pm 0.203$\\
7& $\phantom{-}0.789\pm 0.058\pm 0.070$& $-0.091\pm 0.145\pm 0.149$& $\phantom{-}0.426\pm 0.104 \pm 0.113$& %%@
$-0.922\pm 0.179 \pm 0.194$\\
8& $\phantom{-}0.717\pm 0.080\pm 0.089$& $-0.219\pm 0.201\pm 0.209$& $\phantom{-}0.641\pm 0.071 \pm 0.089$& %%@
$-0.095\pm 0.164\pm 0.172$\\ \hline\hline
\end{tabular}
\end{center}
\end{table}

\section{Systematic Uncertainties}
\label{sec:systematic}
The systematic uncertainties on the measured values of $c_i^{(\prime)}$ and $s_i^{(\prime)}$ come from a variety of %%@
sources. Tables~\ref{tab:syst_K0SKK} and \ref{tab:syst_kmat} give examples of the individual sources of uncertainty %%@
for the $\mathcal{N}=3$ equal $\Delta\delta_D$ binning of the $D^{0}\to K^{0}_{S}K^{+}K^{-}$ Dalitz plot and the %%@
equal $\Delta\delta_D$ binning of the $D^{0}\to K^{0}_S\pi^{+}\pi^{-}$ Dalitz plot, respectively. The breakdown of %%@
the systematic uncertainty is similar for the other binnings considered for $D^{0}\to K^{0}_{S}K^{+}K^{-}$ and %%@
$D^{0}\to K^{0}_{S}\pi^{+}\pi^{-}$ data. Many of the systematic uncertainties are from common sources for the two %%@
final-states; these are described in Sec.~\ref{sec:syst_common}. The sources of uncertainty considered exclusively %%@
for the $D^{0}\to K^{0}_{S}K^{+}K^{-}$ analysis, or those that are evaluated in a significantly different fashion %%@
than $D^{0}\to K^{0}_{S}\pi^{+}\pi^{-}$, are described in Sec.~\ref{sec:syst_K0SKK}. A discussion of the %%@
uncertainties related to the acceptance and background to the $D^{0}\to K^{0}_{S}\pi^{+}\pi^{-}$ analysis are given %%@
in Sec.~\ref{sec:syst_K0Spipi}. A brief discussion of the uncertainty related to the constraints $\Delta c_i$ and %%@
$\Delta s_i$ is given in Sec.~\ref{sec:syst_k0l}.  

\begin{table}[htb]
 \begin{center}
  \caption{Statistical and systematic uncertainties on $c_i$ and $s_i$ determined for the $\mathcal{N}=3$ equal %%@
$\Delta\delta_D$ binning of $D^{0}\to K^{0}_{S}K^{+}K^{-}$ data.}\label{tab:syst_K0SKK}
 \begin{tabular}{l|cccccc}
\hline\hline
Uncertainty                                                   & $c_1$       &$c_2$       & $c_3$      &$s_1$       %%@
& $s_2$      &$s_3$ \\ 
\hline
(Pseudo-)flavor statistics                                    & \phantom{-}0.005\phantom{-}  &   %%@
\phantom{-}0.007\phantom{-}  &   \phantom{-}0.055\phantom{-}  &   \phantom{-}0.015\phantom{-}  &   %%@
\phantom{-}0.013\phantom{-}  &   \phantom{-}0.039\phantom{-}  \\
Momentum resolution                                           &   0.002  &   0.004  &   0.012  &   0.018  &   0.025  %%@
&   0.032  \\
Mode-to-mode normalization                                    &   0.004  &   0.008  &   0.017  &   0.001  &   0.010  %%@
&   0.004  \\
Multiple-candidate selection                                  &   0.004  &   0.003  &   0.015  &   0.004  &   0.008  %%@
&   0.002  \\
DCS correction                                                &   0.001  &   0.001  &   0.003  &   0.002  &   0.005  %%@
&   0.002  \\
$K^{0}_{S,L}\pi^{+}\pi^{-}$ $(c_i^{(\prime)},s_i^{(\prime)})$ &  0.006  &   0.011  &   0.036  &   0.132  &   0.063  %%@
&   0.135  \\
Fitter assumptions                                            &   0.008  &   0.001  &   0.013  &   0.013  &   0.003  %%@
&   0.043  \\
MC statistics for determining \bf{U}                          &   0.005  &   0.007  &   0.057  &   0.024  &   0.051  %%@
&   0.048  \\
Parameterization of non-$K^{0}_{L}$ final state background     &   0.001  &   0.001  &   0.006  &   0.000  &   %%@
0.008  &   0.003  \\
Parameterization of $K^{0}_{L}$ final state background        &   0.034  &   0.020  &   0.061  &   0.038  &   0.015  %%@
&   0.071  \\
Background Dalitz space distribution                          &   0.006  &   0.015  &   0.062  &   0.005  &   0.029  %%@
&   0.022  \\
Assumed background $\mathcal{B}$                              &   0.004  &   0.014  &   0.032  &   0.001  &   0.007  %%@
&   0.009  \\
\hline
Total systematic                                              &   0.038  &   0.034  &   0.131  &   0.142  &   0.094  %%@
&   0.175  \\
Statistical plus $K^{0}_{L}K^{+}K^{-}$ model                  &   0.063  &   0.092  &   0.329  &   0.222  &   0.234  %%@
&   0.432  \\
\hline
$K^{0}_{L}K^{+}K^{-}$ model alone                             &   0.000  &   0.000  &   0.136  &   0.007  &   0.000  %%@
&   0.039  \\
\hline\hline
Total                                                         &   0.073  &   0.098  &   0.354  &   0.264  &   0.252  %%@
&   0.466  \\
\hline\hline
\end{tabular}
 \end{center}
\end{table}
 
\begin{table}[htb]
 \begin{center}
  \caption{Statistical and systematic uncertainties on $c_i$ and $s_i$ determined for the equal binning of %%@
$D^{0}\to K^{0}_{S}\pi^{+}\pi^{-}$ data.}\label{tab:syst_kmat}
   \begin{tabular}{l|cccccccc}
\hline\hline
Uncertainty  & $c_1$&$c_2$& $c_3$&$c_4$& $c_5$&$c_6$& $c_7$&$c_8$\\ \hline
(Pseudo-)flavor statistics & \phantom{-}0.005\phantom{-}  & \phantom{-}0.010\phantom{-}  & %%@
\phantom{-}0.009\phantom{-}  & \phantom{-}0.012\phantom{-}  & \phantom{-}0.005\phantom{-}  & %%@
\phantom{-}0.013\phantom{-}  & \phantom{-}0.013\phantom{-}  & \phantom{-}0.010\phantom{-} \\
Momentum resolution & 0.007  & 0.013  & 0.016  & 0.022  & 0.007  & 0.021  & 0.021  & 0.016 \\
Mode-to-mode normalization & 0.007  & 0.010  & 0.015  & 0.018  & 0.008  & 0.014  & 0.024  & 0.013 \\
Multiple-candidate selection & 0.014  & 0.014  & 0.024  & 0.022  & 0.008  & 0.014  & 0.032  & 0.019 \\
DCS correction  & 0.001  & 0.002  & 0.001  & 0.002  & 0.002  & 0.004  & 0.003  & 0.003 \\
Dalitz plot acceptance & 0.004  & 0.005  & 0.009  & 0.008  & 0.006  & 0.009  & 0.011  & 0.006 \\
Tag-side background & 0.024  & 0.032  & 0.049  & 0.059  & 0.027  & 0.046  & 0.079  & 0.046 \\
\kspp signal-side background & 0.014  & 0.020  & 0.028  & 0.034  & 0.016  & 0.025  & 0.049  & 0.026 \\
\klpp signal-side background & 0.017  & 0.035  & 0.032  & 0.047  & 0.017  & 0.022  & 0.046  & 0.032 \\
Continuum background  & 0.020  & 0.026  & 0.031  & 0.038  & 0.017  & 0.029  & 0.049  & 0.031 \\
\hline
Total systematic             & 0.042 & 0.063 & 0.080 & 0.098 & 0.042 & 0.072 & 0.124 & 0.075\\
Statistical plus \klpp model & 0.036 & 0.068 & 0.088 & 0.119 & 0.045 & 0.102 & 0.105 & 0.069\\ \hline
\klpp model alone            & 0.013 & 0.018 & 0.039 & 0.068 & 0.024 & 0.040 & 0.068 & 0.034\\
\hline\hline
Total                        & 0.056 & 0.093 & 0.119 & 0.154 & 0.062 & 0.125 & 0.163 & 0.102\\
\hline\hline
Uncertainty  & $s_1$&$s_2$& $s_3$&$s_4$& $s_5$&$s_6$& $s_7$&$s_8$
\\\hline
(Pseudo-)flavor statistics  & 0.008  & 0.014  & 0.022  & 0.021  & 0.012  & 0.019  & 0.042  & 0.017 \\
Momentum resolution  & 0.021  & 0.037  & 0.030  & 0.041  & 0.019  & 0.041  & 0.039  & 0.030 \\
Mode-to-mode normalization  & 0.001  & 0.000  & 0.002  & 0.002  & 0.001  & 0.002  & 0.001  & 0.001 \\
Multiple-candidate selection & 0.016  & 0.007  & 0.008  & 0.010  & 0.004  & 0.007  & 0.010  & 0.007 \\
DCS correction & 0.004  & 0.003  & 0.011  & 0.004  & 0.001  & 0.007  & 0.016  & 0.008 \\
Dalitz plot acceptance  & 0.006  & 0.004  & 0.005  & 0.005  & 0.005  & 0.004  & 0.007  & 0.004 \\
Tag-side background & 0.004  & 0.001  & 0.007  & 0.005  & 0.002  & 0.003  & 0.004  & 0.005 \\
\kspp  signal-side  background & 0.002  & 0.005  & 0.005  & 0.005  & 0.005  & 0.008  & 0.003  & 0.007 \\
\klpp signal-side  background & 0.031  & 0.051  & 0.016  & 0.054  & 0.010  & 0.035  & 0.074  & 0.024 \\
Continuum background  & 0.003  & 0.004  & 0.006  & 0.006  & 0.001  & 0.002  & 0.006  & 0.004 \\
\hline
Total systematic             & 0.043 & 0.066 & 0.044 & 0.072 & 0.026 & 0.059 & 0.096 & 0.045\\
Statistical plus \klpp model & 0.098 & 0.182 & 0.086 & 0.202 & 0.131 & 0.197 & 0.131 & 0.150\\\hline
\klpp model alone            & 0.037 & 0.038 & 0.000 & 0.000 & 0.030 & 0.006 & 0.000 & 0.025\\
\hline\hline
Total                        & 0.106 & 0.193 & 0.097 & 0.214 & 0.133 & 0.206 & 0.162 & 0.157\\
\hline\hline
\end{tabular}
 \end{center}
\end{table}

\subsection{Common systematic uncertainties}
\label{sec:syst_common}

The statistical uncertainties on the measurements of $K^{(\prime)}_i$ using flavor and pseudo-flavor tagged events %%@
are not included in the fit to determine $c_i$, $s_i$, $c_i^{\prime}$, and $s_i^{\prime}$. Even though there are %%@
significantly more flavor and pseudo-flavor tagged events than the $CP$-tagged or $K^{0}_{S,L}h^{+}h^{-}$ $vs.$ %%@
$K^{0}_{S}h^{+}h^{-}$ events, there is a non-negligible uncertainty related to the limited statistics of these %%@
samples. Each input is varied separately by its uncertainty and the fits are repeated. The differences with respect %%@
to the nominal fit are added quadratically to attain the systematic uncertainty.

The unusual shape of the bins, particularly those with narrow regions, allows the possibility that a mismodeling of %%@
the invariant-mass resolution leads to an incorrect description of the bin-to-bin migration. In the analysis of %%@
$K^{0}_{S}K^{+}K^{-}$ the invariant-mass resolution is estimated from the width of the $\phi$ peak in data and MC. %%@
The difference is compatible with zero but it is conservatively assumed that the MC underestimates the resolution %%@
by up to one standard deviation of the measured difference. The invariant masses in the signal MC simulation sample %%@
are smeared by this difference and then fit to extract $c_i$ and $s_i$. This procedure is repeated many times and %%@
the resulting distributions of $c_i$ and $s_i$ returned from the fits are used to determine the systematic %%@
uncertainty on these parameters due to resolution.   The uncertainty related to the invariant mass resolution in %%@
$D^{0}\to K^{0}_{S}\pi^{+}\pi^{-}$ events is evaluated in similar fashion with the momentum of the charged tracks %%@
smeared by the CLEO-c resolution as in the previous analysis \cite{BRIERE}.

The mode-to-mode normalization in the fitter is performed either using  the measured branching fractions of the %%@
modes combined with $N_{D^{0}\overline{D}^{0}}$, or the measured ST yields. For the $K^{0}_{S}K^{+}K^{-}$ analysis %%@
the former method is used apart from normalizing the $K^{0}_{S,L}K^{+}K^{-}~vs.~K^{0}_{S}\pi^{+}\pi^{-}$ yields, %%@
where the pseudo-flavor ST yields are used. 
In the $K^{0}_{S}\pi^{+}\pi^{-}$ analysis the ST yields are used apart from the two final states that contain %%@
unreconstructed particles in the tag candidate ($K^{-}e^{+}\nu$ and $K^{0}_{L}\pi^{0}$). In addition, previous %%@
investigations \cite{DHAD} have determined that there are small differences in the particle-reconstruction %%@
efficiency between data and the MC. These differences lead to systematic uncertainties, and in some cases %%@
corrections, for final states in which $K^0_{S,L}h^+h^-$ is normalized using the branching fraction and the %%@
$N_{D^{0}\overline{D}^{0}}$; when using ST yields to normalize, these corrections and uncertainties cancel. The %%@
reconstruction efficiency of each final-state containing a $\pi^{0}$ or an $\eta$ meson must be corrected. The DT %%@
efficiency has an uncertainty for each final-state particle reconstructed due to the MC modeling. These %%@
uncertainties are: 0.3\% per $\pi^{+}$, 0.6\% per $K^{+}$, 1.3\% per $\pi^{0}$, 4\% per $\eta$, 0.9\% per %%@
$K^{0}_{S}$, and 3\% per $K^{0}_{L}$ candidates. If a final state contains two particles of the same type the %%@
uncertainty on each identical particle is treated as fully correlated. The uncertainties on the ST yields, %%@
branching fractions, $N_{D^{0}\overline{D}^{0}}$, and particle reconstruction efficiency are used to estimate the %%@
systematic uncertainty from the mode-to-mode normalization method.

The method used to select the best candidate in an event when there are multiple candidates can introduce a bias. %%@
To estimate the systematic uncertainty related to the multiple-candidate selection the simulation is used to %%@
determine how often the wrong candidate is selected. This information is used to derive corrections to the yields. %%@
The difference between the $c_i$ and $s_i$ parameters fit with and without this correction applied is taken as the %%@
systematic uncertainty due to the multiple-candidate selection.

The use of $D^{0}\to K^{-}\pi^{+}$, $D^{0}\to K^{-}\pi^{+}\pi^{0}$ and $D^{0}\to K^{-}\pi^{+}\pi^{+}\pi^{-}$ %%@
pseudo-flavor tags has a systematic uncertainty from the values of the three amplitude ratios $r_D$, the three %%@
strong-phase differences $\delta_D$, and the two coherence parameters required to estimate the corrections from DCS %%@
contamination. The uncertainty is estimated by performing the fit to $c_i$ and $s_i$ for each parameter shifted by %%@
plus or minus the uncertainties given in Table~\ref{tab:pseudoftinputs}. The largest change in each $c_i$ and %%@
$s_i$, from either the positive or negative shift in the parameter, is taken as the symmetric systematic %%@
uncertainty on the parameter. The total uncertainty is the sum in quadrature of the individual parameters. 
The uncertainties on the strong-phase differences and coherence parameters dominate the total uncertainty from this %%@
source.

\subsection{\boldmath \KsKK specific systematic uncertainties}
\label{sec:syst_K0SKK}
 There are specific systematic uncertainties related to the $\KsKK$ analysis: the strong-phase parameters of the %%@
$\Kspipi$ events in the analysis, assumptions in the fitting procedure, the determination of the migration matrix, %%@
and the background assumptions. The use of $K_{S,L}^{0}K^{+}K^{-}~vs.~K_{S}^{0}\pi^{+}\pi^{-}$ and %%@
$K_{S}^{0}K^{+}K^{-}~vs.~K_{L}^{0}\pi^{+}\pi^{-}$ events introduces a dependence on the values of $c_i^{(\prime)}$ %%@
and $s_i^{(\prime)}$ for $D^{0}\to K^0_{L,S} \pi^+\pi^-$ derived from the equal $\Delta\delta_D$ binning reported %%@
in this paper. The nominal fit has the values of $c_i^{(\prime)}$ and $s_i^{(\prime)}$ fixed to the central values. %%@
To determine the systematic uncertainty the input values are smeared by their uncertainties and the fit repeated. %%@
This is done many times and the width of the distribution of $c_{i}$ and $s_{i}$ parameters returned for the %%@
particular $K^{0}_{S}K^{+}K^{-}$ binning is taken as the systematic uncertainty from this source.

The $K^{0}_{S}K^{+}K^{-}$ fit was tested on samples of signal MC data. It was found that there were small but  %%@
statistically significant biases in the central values of $c_i$ and $s_i$ returned by the fitter. These biases are %%@
likely to be consequences of the assumptions used in the fit such as the finite granularity of the Dalitz plot %%@
bit-map and DT branching fractions not being corrected for quantum correlations. The whole bias is conservatively %%@
taken as an additional source of systematic and is found to be significantly smaller than both the statistical %%@
uncertainty and the dominant sources of systematic uncertainty. (No such bias is observed in the validation of the %%@
$K^{0}_{S}\pi^{+}\pi^{-}$ fitter \cite{BRIERE}; therefore, no uncertainty is attributed to fitter bias in that %%@
analysis.)

The elements of the migration matrices, $\mathbf{U}$, are determined from MC samples of signal events. The %%@
resulting statistical uncertainty on the elements due to the finite sample size introduces a systematic error on %%@
the $c^{(\prime)}$ and $s^{(\prime)}$ parameters. The uncertainty is determined by smearing the elements by their %%@
statistical error accounting for correlations. Then the fit is repeated with the new migration matrices. The %%@
procedure is repeated many times and the width of the resulting distribution of $c^{(\prime)}$ and $s^{(\prime)}$ %%@
is taken as the systematic uncertainty from this source.

The background parametrization contains several sources of systematic uncertainties. For final states without a %%@
$K^{0}_{L}$ meson the limited statistics of the generic MC sample, which is used to determine the peaking %%@
backgrounds, leads to an uncertainty in the background parameterization. For $K^{0}_{L}$ there are also %%@
uncertainties arising from the ratios of signal-to-background events in the signal region, as well as the high and %%@
low missing-mass sidebands, which are obtained from the simulation. This fact, combined with the more significant %%@
backgrounds in the modes containing a $K^{0}_{L}$ meson, leads to a larger uncertainty due to MC parameterizations %%@
of the background for final states that include a $K^{0}_{L}$ meson. We vary these parameters by their %%@
uncertainties, repeat the fit then compare the result to the nominal fit to determine the uncertainty from this %%@
source.  

 The assumptions about the distribution of the background events over Dalitz space are also varied to determine a %%@
systematic uncertainty. In the nominal fits the distribution for each background type is either assumed to be %%@
uniform or to follow the $K^{0}_{S}K^{+}K^{-}$ signal distribution. We assign a systematic uncertainty to these %%@
assumptions by randomly choosing the fraction of the background that is uniformly distributed for each source of %%@
background; the remainder of that background component is then assumed to follow the distribution of $D^{0}\to %%@
K^{0}_{S}K^{+}K^{-}$ events. The fraction is assumed to have equal probability of taking any value between 0 and 1. %%@
Then the fit is performed with these assumptions about the Dalitz plot distribution. This procedure is repeated %%@
many times and the resulting distributions of $c_i$ and $s_i$ are used to determine the systematic uncertainty %%@
related to the assumed distribution of background over the Dalitz space.

 The final source of uncertainty is related to the branching fractions, $\mathcal{B}$, assumed when generating the %%@
generic MC samples. For modes that contribute background we vary the branching fraction by the uncertainty reported %%@
in Ref.~\cite{PDG} and repeat the fit to determine the systematic uncertainty on $c_i^{(\prime)}$ and %%@
$s_i^{(\prime)}$.

\subsection{\boldmath Uncertainties related to the acceptance and backgrounds in the $K^{0}_{S}\pi^{+}\pi^{-}$ %%@
analysis}
\label{sec:syst_K0Spipi}

The sources of uncertainty related to the acceptance and background in the $K^{0}_{S}\pi^{+}\pi^{-}$ analysis are %%@
identical to those considered in Ref. \cite{BRIERE}. In addition, the same evaluation procedures are adopted, which %%@
we briefly outline here. 

Any difference in the relative efficiency over the Dalitz space can bias the results. In the nominal fit the Dalitz %%@
plot acceptance is taken from simulation.  To account for any difference between data and simulation, the relative %%@
efficiency is smeared in each bin by 2\%, which is the spread of efficiencies observed in simulation, and a sample %%@
of signal MC is fit with the new efficiencies. This procedure is repeated many times and the resulting %%@
distributions of $c_i$ and $s_i$ returned from the fits are used to determine the systematic uncertainty on these %%@
parameters due to modeling the acceptance. (The relative variation of efficiency is an order of magnitude greater %%@
for $D^{0}\to K^{0}_{S}K^{+}K^{-}$ due to the momentum dependence of the charged kaon detection efficiency. %%@
Therefore, adopting a similar approach would be too conservative. However, the observed relative variation in %%@
efficiency from the momentum and migration matrix smearing for $D^{0}\to K^{0}_{S}K^{+}K^{-}$ is around 5\%, which %%@
is greater than the fluctuations assumed for $D^{0}\to K^{0}_{S}\pi^{+}\pi^{-}$.)

Tag-side background yields are evaluated mode-by-mode from sidebands. The distribution over Dalitz space is assumed %%@
to follow that found in the simulation. The fits are repeated assuming the background events are distributed %%@
uniformly over Dalitz space. The difference between the fits is taken as the systematic uncertainty. 

The nominal fit ignores background in the signal $K^{0}_S\pi^{+}\pi^{-}$ Dalitz space. The level of background %%@
estimated from $K^{0}_{S}$ mass sidebands is $1.9\%$ \cite{BRIERE}. The background distribution is estimated from %%@
MC samples that include the effect of quantum correlations and the fit is repeated with the background subtracted. %%@
The difference between the nominal fit and the fit including background is taken as the systematic uncertainty due %%@
to signal-side $K^{0}_S\pi^{+}\pi^{-}$ background.

The nominal fit assumes the signal-side $K^{0}_L\pi^{+}\pi^{-}$ background is distributed over Dalitz space %%@
following the distribution observed in the $K^{0}_{L}$ mass sidebands. The fit is repeated assuming the background %%@
is distributed uniformly over Dalitz space and the difference in the parameter values taken as the systematic %%@
uncertainty.

The continuum background is only significant in $K^{0}_{L}\pi^{+}\pi^{-}$ events tagged by $K^{0}_{S}\pi^{0}$ and %%@
$K^{0}_{S}\pi^{+}\pi^{-}$ decays. The distribution of the continuum events over Dalitz space is taken from the %%@
simulation. The fit is repeated with the continuum events distributed uniformly over Dalitz space. The difference %%@
between the fitted values of $c_i$ and $s_i$ between the two distribution assumptions is taken as the systematic %%@
uncertainty related to the continuum background.

\subsection{\boldmath The $K^{0}_{L}h^{+}h^{-}$ model uncertainty}
\label{sec:syst_k0l}
The statistical uncertainty returned by the fit includes a contribution that is related to the assigned %%@
uncertainties on $\Delta c_i$ and $\Delta s_i$ present in the $\chi^2$ term of Eqs.~(\ref{eq:kspipiextract}) and %%@
(\ref{eq:kskkextract}). In order to isolate this contribution the fit can be repeated fixing $\Delta c_i$ and %%@
$\Delta s_i$, then the variance related to the constraint is the difference between the variances returned by the %%@
fit with fixed or constrained values of $\Delta c_i$ and $\Delta s_i$. However, due to the variations in the %%@
central values of $s_i$ and $c_i$ and the non-Gaussian behavior of the parameters near the edges of the physical %%@
region $(c_i^{2}+s_i^{2}<1)$, the covariance matrix related to the uncertainty on the constraint is found to give %%@
unphysical values in some cases; these are either negative diagonal elements or off-diagonal terms that correspond %%@
to correlations greater than one or less than minus one. Therefore, we present the statistical covariance matrix %%@
from the constrained fit and the systematic covariance matrix from all other sources of systematic uncertainty. 

 To indicate that the size of the uncertainties related to $\Delta c_i$ and $\Delta s_i$ are not dominant, the %%@
estimated values are given in Tables~\ref{tab:syst_K0SKK} and \ref{tab:syst_kmat}. The procedure outlined above is %%@
used to estimate these $K^{0}_{L}h^{+}h^{-}$ model uncertainties. In cases where the diagonal elements of the %%@
covariance matrix are negative the uncertainty is set to zero. 

\subsection{Summary of systematic uncertainties}
In both the $K^{0}_{S}h^{+}h^{-}$ analyses the largest sources of uncertainty is usually the background %%@
parameterization. In some bins there are significant contributions from the momentum resolution, flavor-tag %%@
statistics, and the knowledge of strong-phase parameters. The determinations of $c_i$ have a systematic uncertainty %%@
comparable in size to the statistical uncertainty whereas for $s_i$ measurements the statistical uncertainty %%@
dominates. Correlation matrices for the systematic uncertainties associated with the different binnings are given %%@
in Ref.~\cite{EPAPS}.

\section{\boldmath Final Results, Impact on $\gamma/\phi_3$ and $CP$ content of $D^{0}\to K^{0}_{S}K^{+}K^{-}$}
\label{sec:gamma}
Section~\ref{subsec:cisifinal} presents the final results for $c_i^{(\prime)}$ and $s_i^{(\prime)}$, along with %%@
comparison of the measured values with the amplitude model predictions. The impact of the $c_i$ and $s_i$ results %%@
on the determination of $\gamma/\phi_3$ is discussed in Sec.~\ref{subsec:gammaimpact}. In addition, results for the %%@
$CP$-odd fraction, $\mathcal{F}_{-}$, of $D^{0}\to K^{0}_{S}K^{+}K^{-}$ decays in the region of the $\phi$ %%@
resonance are presented in Sec.~\ref{subsec:cpcont}.

\subsection{\boldmath Final results for $c_i^{(\prime)}$ and $s_i^{(\prime)}$}
\label{subsec:cisifinal}
 
\begin{figure}[htb]
 \begin{center}
\includegraphics*[width=1.0\columnwidth]{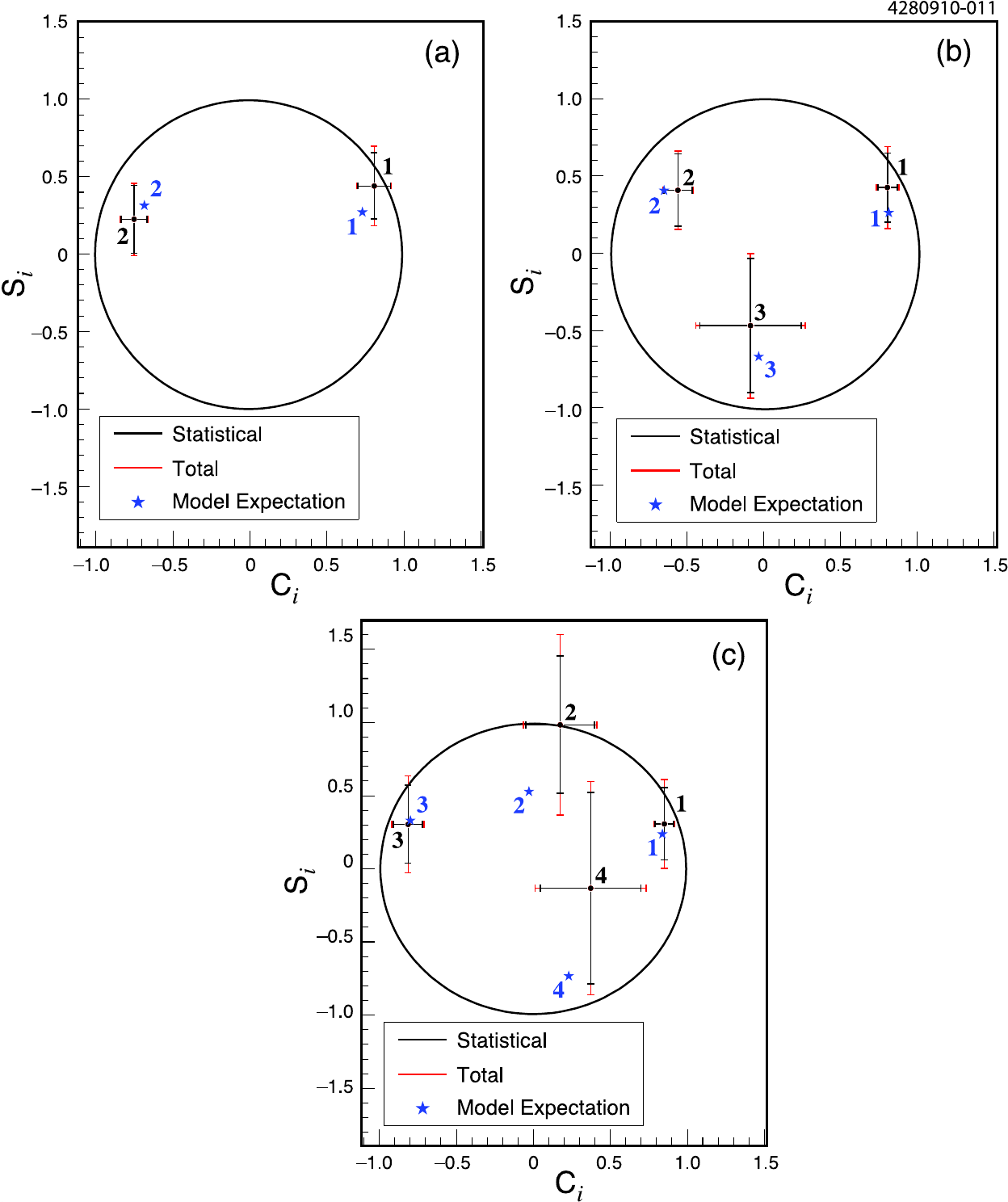}
  \caption{Measured values of $c_i$ and $s_i$ for $D^{0}\to K^{0}_{S}K^{+}K^{-}$ data divided into (a) %%@
$\mathcal{N}=2$, (b) $\mathcal{N}=3$, and (c) $\mathcal{N}=4$ equal $\Delta\delta_D$ bins. The expected values %%@
calculated from the BABAR 2010 model are indicated by the stars. The circle indicates the boundary of the physical %%@
region $c_i^2+s_i^2=1$.}\label{fig:results1} 
 \end{center}
\end{figure}

\begin{figure}[htb]
 \begin{center}
 \includegraphics*[width=1.0\columnwidth]{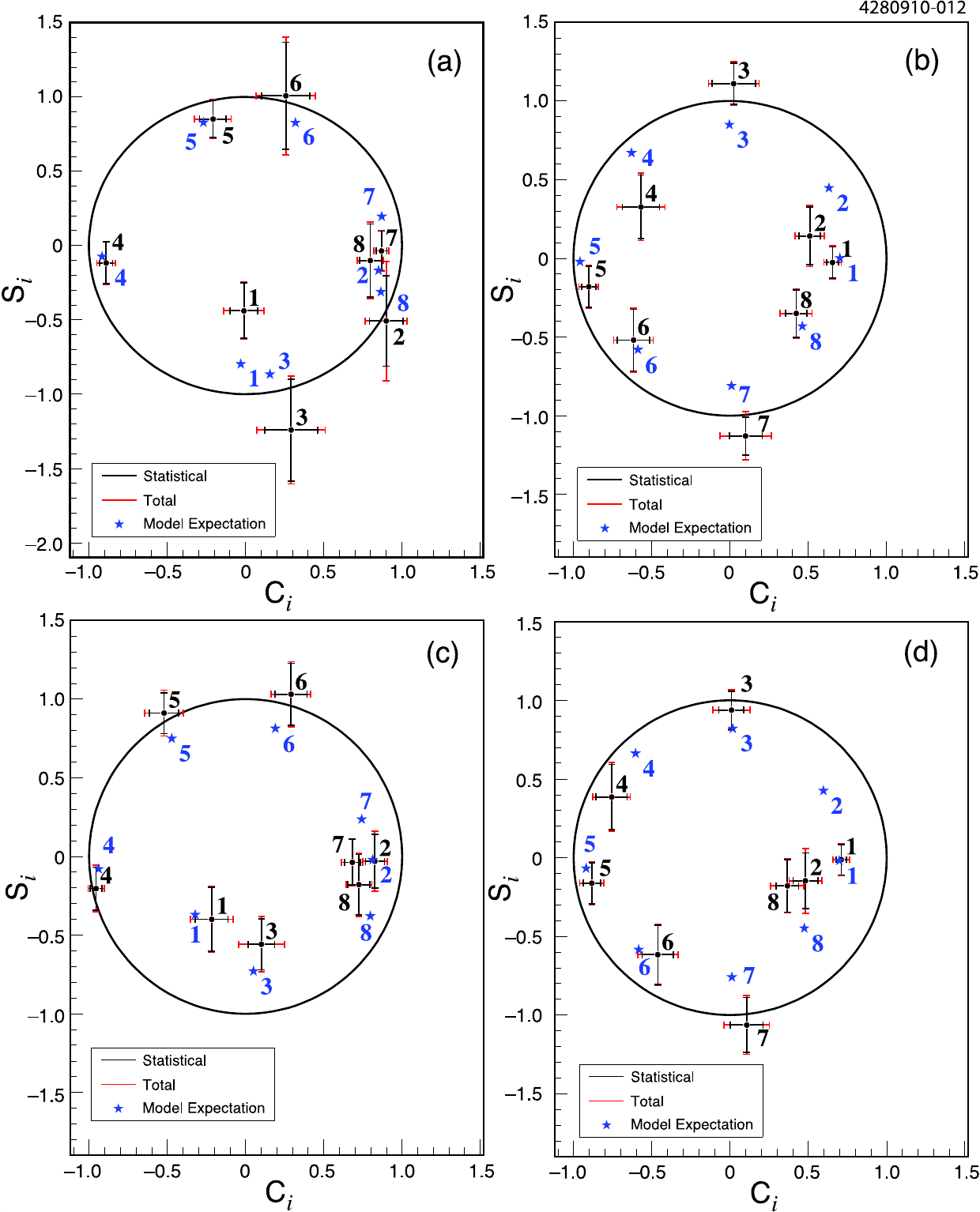}
  \caption{Measured values of $c_i$ and $s_i$ for (a) the optimal binning, (b) the equal $\Delta\delta_D$ binning, %%@
(c) the modified-optimal binning, and (d) the equal $\Delta\delta_D$ binning based on the Belle model. The expected %%@
values calculated from the BABAR 2008 model are indicated by the stars. The circle indicates the boundary of the %%@
physical region $c_i^2+s_i^2=1$.}\label{fig:results2} 
 \end{center}
\end{figure}

The measured values of $c_i$ and $s_i$ are shown in Figs.~\ref{fig:results1} and \ref{fig:results2} for $D^0\to %%@
K^{0}_{S}K^{+}K^{-}$ and $D^{0}\to K^{0}_{S}\pi^{+}\pi^{-}$, respectively. Also, shown are the expectations from %%@
the BABAR 2010 model and BABAR 2008 model for $D^0\to K^{0}_{S}K^{+}K^{-}$ and $D^{0}\to K^{0}_{S}\pi^{+}\pi^{-}$, %%@
respectively. To test the compatibility of our results with the predictions of the models we compute
\begin{equation}
\chi^2=(\mathbf{P}-\mathbf{P}^\mathrm{model})^{T}\mathbf{V}^{-1}(\mathbf{P}-\mathbf{P}^\mathrm{model}) \; ,
\end{equation}
where $\mathbf{P}$ is a vector of the $2\mathcal{N}$ measured values of $c_i$ and $s_i$, %%@
$\mathbf{P}^\mathrm{model}$ is the equivalent vector of $c_i$ and $s_i$ values predicted by the models, and 
$\mathbf{V}$ is the $2\mathcal{N}\times 2\mathcal{N}$ combined statistical and systematic covariance matrix. %%@
Table~\ref{tab:compat} gives the values of the $\chi^{2}$ and the corresponding probabilities. In the case of %%@
$D^{0}\to K^{0}_{S}K^{+}K^{-}$ the agreement between measured and predicted values of $c_i$ and $s_i$ is good for %%@
all binnings. For $D^{0}\to K^{0}_{S}\pi^{+}\pi^{-}$ the compatibility of the model predictions with the optimal %%@
and modified-optimal binning results is very good; the compatibility of the models with the {\it BABAR} 2008 and %%@
Belle model equal $\Delta\delta_D$ binnings is reasonable. The compatibility with the predictions from the Belle %%@
model is also tested and is found to be reasonable. 

\begin{table}[htb]
  \begin{center}
\caption{Values of $\chi^2$ per degree of freedom (DOF) and the corresponding probability for comparison of %%@
measured values of $c_i$ and $s_i$ with those predicted by the BABAR 2010 and BABAR 2008 models for $D^0\to %%@
K^{0}_{S}K^{+}K^{-}$ and $D^{0}\to K^{0}_{S}\pi^{+}\pi^{-}$, respectively.} \label{tab:compat}
 	\begin{tabular}{l c c}\hline \hline
	Binning & $\chi^2/DOF$ & Probability \\ \hline
	\multicolumn{3}{c}{$D^{0}\to K^{0}_{S}K^{+}K^{-}$} \\ \hline
	$\mathcal{N}=2$ equal $\Delta\delta_D$& 1.7/4   & 79\%   \\ 
 	$\mathcal{N}=3$ equal $\Delta\delta_D$& 1.4/6   & 96\%   \\
	$\mathcal{N}=4$ equal $\Delta\delta_D$& 2.2/8   & 98\% \\ \hline
	\multicolumn{3}{c}{$D^{0}\to K^{0}_{S}\pi^{+}\pi^{-}$} \\ \hline
	Optimal & 15.5/16 & 49\% \\
	{\it BABAR} 2008 equal $\Delta\delta_D$ & 25.3/16 & 6.5\% \\
	Modified optimal & 13.8/16 & 61\% \\
	Belle & 26.8/16 & 4.4\% \\ \hline\hline 
 \end{tabular}
 \end{center}
\end{table}

\subsection{\boldmath Impact of $c_i$ and $s_i$ measurements on determining $\gamma/\phi_3$}
\label{subsec:gammaimpact}

Our measurements of $c_i$ and $s_i$ have two consequences on the determination of $\gamma/\phi_3$ from a %%@
model-independent analysis:
\begin{itemize}
\item{the uncertainties on $c_i$ and $s_i$ will result in a systematic uncertainty on the determination of %%@
$\gamma/\phi_3$, and} 
\item{the choice of binning affects the statistical precision of a $\gamma/\phi_3$ measurement.} 
\end{itemize}

Therefore, we investigate the impact on future $\gamma/\phi_3$ measurements using a simplified MC simulation of %%@
$B$-decay data on which we perform the model-independent determination of $\gamma/\phi_3$ \cite{GIRI}. 
The number of $B^-$ decays in a given bin of Dalitz space is dependent on $c_i$, $s_i$, $r_B$, $\delta_B$, %%@
$\gamma/\phi_3$, and normalization parameters. The input values used for $r_B$, $\delta_B$, and $\gamma/\phi_3$ are %%@
0.1, 130$^\circ$, and $60^\circ$, respectively, which are consistent with current %%@
measurements~\cite{CKMFITTER,UTFIT}. 
The values of $K_{(-)i}$ are those predicted by the respective models.
The yield in each bin is generated randomly according to the input values, and a $\chi^2$ fit between the observed %%@
and expected events in each bin is performed to extract the value of $\gamma/\phi_3$. 

To determine the systematic uncertainty in future $\gamma/\phi_3$ measurements due to the uncertainty in the %%@
measured strong phases we generate samples of five million signal events using the measured values of $c_i$ and %%@
$s_i$. The samples are then fit using the measured central values smeared according to their uncertainties, with %%@
the correlation between parameters taken into account. The large number of signal events means that the width of %%@
distribution of fitted $\gamma/\phi_3$ gives the systematic uncertainty due to the uncertainty in $c_i$ and $s_i$, %%@
as the intrinsic width due to the statistical fluctuations for five million events is negligible in comparison. 

For the $\widetilde{D}^0 \to K^0_S K^+ K^-$ mode, the induced uncertainty on $\gamma/\phi_3$ due to the total
uncertainties on $c_i$ and $s_i$ is evaluated from the RMS of $\gamma$ distribution returned by the fits to the %%@
simulated experiments. The induced uncertainty is 3.9$^\circ$, 3.2$^\circ$, and 3.9$^\circ$ for two, three, and %%@
four bins,
respectively. The larger uncertainty for four bins reflects the limited statistics available to determine the %%@
parameters in each bin. It can be noted that there is only a 
limited improvement in the statisitical sensitivity to $\gamma/\phi_3$ by increasing the number of bins, since most %%@
of the
sensitivity is due to the dominant resonances, $K^0_S\phi$ and $K^{0}_S a^0(980)$, which lie in the same region of %%@
the Dalitz plot close to the $m_{K^{+}K^{-}}$ threshold.

For the $\widetilde{D}^0 \to K^0_S \pi^+ \pi^-$ case we find the uncertainty on $\gamma/\phi_3$ related to the %%@
uncertainties on $c_i$ and $s_i$ in a similar manner. The uncertainty varies significantly between the different %%@
binnings. For the equal $\Delta\delta_D$ binning derived from the {\it BABAR} 2008 model, we find an induced %%@
uncertainty  of 2.0$^\circ$, whereas for the optimal-binning it is $3.9^\circ$. These variations reflect the %%@
limited statistics available with which to determine $c_i$ and $s_i$ in some of the optimal bins; these are the %%@
same bins that lead to the reduced statistical performance in determining $\gamma$ from the optimal binning in the %%@
presence of background. This is further emphasized by the expected uncertainty on $\gamma/\phi_3$ from the %%@
modified-optimal binning results being $2.1^\circ$. The values of $c_i$ and $s_i$ measured for the equal %%@
$\Delta\delta_D$ binning derived from the Belle model lead to a 1.7$^\circ$ uncertainty on $\gamma/\phi_3$. In all %%@
cases the error on the predicted uncertainty is less than $0.1^\circ$ and the contribution of the systematic %%@
uncertainties on $c_i$ and $s_i$ is approximately 1.3$^\circ$. For the majority of the $K^{0}_{S}\pi^{+}\pi^{-}$ %%@
binnings, these results demonstrate that, with the exception of the optimal binning choice, the uncertainty on %%@
$\gamma/\phi_3$ arising from the errors on the measured $c_i$ and $s_i$ parameters is less than that of the %%@
assigned model uncertainty in the unbinned analyses \cite{BELLE2,BABAR3}.

For both $\widetilde{D}^0 \to K^0_S K^+ K^-$ and  $\widetilde{D}^0 \to K^0_S \pi^+ \pi^-$, small biases of %%@
$\mathcal{O}(1^\circ)$ are observed in the mean fitted values of $\gamma/\phi_3$. This was also observed in Ref. %%@
\cite{BRIERE}, where the origin of the bias is linked to the assumed values in the fit being unphysical, $c_i^2 + %%@
s_i^2>1$, in some simulated experiments. If the unphysical simulated experiments are removed the bias is not %%@
eliminated because of the non-Gaussian nature of the truncated distributions of $c_i$ and $s_i$. As before it is %%@
found that improvements in the precision of the measurements of $c_i$ and $s_i$ would reduce the bias to a %%@
negligible level. 

We also investigate how the choice of binning for $\widetilde{D}^0 \to K^0_S \pi^+ \pi^-$ affects the statistical %%@
precision on $\gamma/\phi_3$. The optimal and the modified-optimal binnings have been optimized to minimize the %%@
statistical uncertainty on $\gamma/\phi_3$ in the absence of background and with the anticipated background at LHCb %%@
\cite{TABS}, respectively. The impact of these binnings on $B$ data can be assessed by comparing them to the equal %%@
$\Delta\delta_D$ binning derived from the {\it BABAR} 2008 model using simulated $B$ data samples. We generate %%@
simplified MC samples of 5000 $B^{\pm}\to K^{\pm} \widetilde{D}^{0}$ signal events which corresponds to the %%@
expected yield with 2 $\mathrm{fb}^{-1}$ of LHCb data. The statistical uncertainties on $\gamma/\phi_3$ are %%@
8.0$^\circ$ and 7.0$^\circ$ for the equal $\Delta\delta_D$ and optimal binnings, respectively. This demonstrates %%@
the advantage in using the optimized binning, and the improvement is consistent with the increase in the $Q$ %%@
quantity given in Table~\ref{fig:bincomparison}. We also generate samples that include background events with %%@
yields according to the LHCb background model described in Sec.~\ref{subsec:alt_binning}. We find the statistical %%@
uncertainties on $\gamma/\phi_3$ of 15.0$^\circ$, 15.4$^\circ$, and 14.6$^\circ$ for the equal $\Delta\delta_D$, %%@
optimal and modified-optimal binning, respectively. This shows that in the presence of background the %%@
modified-optimal binning has a smaller statistical uncertainty than the other binnings as expected from the $Q'$ %%@
values given in Table~\ref{fig:bincomparison}.  

\begin{table}[htb]
 \begin{center}
  \caption{Values of $Q^{(\prime)}$ derived from the measured values of $c_i$ and $s_i$ for the different binnings. %%@
The uncertainty is determined by varying $c_i$ and $s_i$ within their errors accounting for %%@
correlations.}\label{tab:measureq}
\begin{tabular}{lc} \hline\hline
Binning & $Q^{(\prime)}$ \\ \hline
\multicolumn{2}{c}{$D^{0}\to K^{0}_{S}K^{+}K^{-}$} \\ \hline
$\mathcal{N}=2$ equal $\Delta\delta_D$ &  $0.88^{+0.14}_{-0.08}$              \\
$\mathcal{N}=3$ equal $\Delta\delta_D$ &  $0.82^{+0.15}_{-0.06}$              \\
$\mathcal{N}=4$ equal $\Delta\delta_D$ &  $0.90^{+0.21}_{-0.03}$              \\ \hline
\multicolumn{2}{c}{$D^{0}\to K^{0}_{S}\pi^{+}\pi^{-}$} \\ \hline 
Optimal                           &  $0.93^{+0.12}_{-0.02}$ \\
{\it BABAR} 2008 equal $\Delta\delta_D$ &  $0.81^{+0.05}_{-0.01}$ \\
Modified optimal                  &  $0.95^{+0.10}_{-0.04}$ \\
Belle  equal $\Delta\delta_D$     &  $0.78^{+0.05}_{-0.01}$ \\ \hline\hline
  \end{tabular}
 \end{center}
\end{table}

As an additional cross check of the sensitivity to $\gamma/\phi_3$ we determine $Q^{(\prime)}$ for the different %%@
binnings from the measured values $c_i$ and $s_i$. The measured values of $Q^{\prime}$ are given in %%@
Table~\ref{tab:measureq}. In all cases the values indicate good sensitivity to $\gamma/\phi_3$ relative to the %%@
unbinned method. In addition the values are in reasonable agreement with the predictions presented in %%@
Sec.~\ref{sec:bindef}.

\subsection{\boldmath $CP$ fraction in the region of the $\phi$ resonance in $D^{0}\to K^{0}_{S}K^{+}K^{-}$ decays}
\label{subsec:cpcont}
\begin{table}[htb]
 \begin{center}
 	\caption{Measured values of $\mathcal{F}_{-}$ ($\mathcal{F}_{+}$) for $D^{0}\to K^{0}_{S}K^{+}K^{-}$ ($D^{0}\to %%@
K^{0}_{L}K^{+}K^{-}$) decays with different criteria on $\Delta m_{K^{+}K^{-}}^2$. The predicted value of %%@
$\Delta\mathcal{F}$ and the average value of $\mathcal{F}_{-}$ for $D^{0}\to K^{0}_{S}K^{+}K^{-}$ and $D^{0}\to %%@
K^{0}_{L}K^{+}K^{-}$ are also given. The first uncertainty is statistical and the second is %%@
systematic.}\label{tab:fminus}
\begin{tabular}{ccccc} \hline\hline
Criterion on  & $D^{0}\to K^{0}_{S}K^{+}K^{-}$
& $D^{0}\to K^{0}_{L}K^{+}K^{-}$ & $\Delta\mathcal{F}$ & Combined  \\
$|\Delta m_{K^{+}K^{-}}^2|$ & $\mathcal{F}_{-}$ & $\mathcal{F}_{+}$ & &
$\mathcal{F}_{-}$ \\\hline
$<0.006$ $\mathrm{GeV}^{2}/c^4$& 
\hspace{0.1cm} $ 1.09 \pm 0.09 \pm 0.09 $ \hspace{0.1cm} & 
\hspace{0.1cm} $ 0.98 \pm 0.09 \pm 0.02 $ \hspace{0.1cm} & 
\hspace{0.1cm} $-0.02 \pm 0.04 $ \hspace{0.1cm} & 
\hspace{0.1cm} $ 1.03 \pm 0.07 \pm 0.04 $ \hspace{0.1cm} \\
$<0.010$ $\mathrm{GeV}^{2}/c^4$& $ 1.13 \pm 0.08 \pm 0.08 $ & $ 0.98 \pm 0.07 \pm 0.02 $ & $-0.02 \pm 0.04 $ & $ %%@
1.05 \pm 0.06 \pm 0.04 $ \\
$<0.014$ $\mathrm{GeV}^{2}/c^4$& $ 0.73 \pm 0.27 \pm 0.04 $ & $ 0.90 \pm 0.09 \pm 0.02 $ & $-0.02 \pm 0.05 $ & $ %%@
0.90 \pm 0.10 \pm 0.02 $ \\
$<0.018$ $\mathrm{GeV}^{2}/c^4$& $ 0.75 \pm 0.22 \pm 0.04 $ & $ 0.90 \pm 0.08 \pm 0.02 $ & $-0.03 \pm 0.06 $ & $ %%@
0.90 \pm 0.09 \pm 0.02 $ \\
\hline\hline
  \end{tabular}
 \end{center}
\end{table}

The $CP$-odd fraction of $D^{0}\to K^{0}_{S} K^{+} K^{-}$ decays has been estimated using the expression given in %%@
Eq.~(\ref{eq:fminus}). The values of $M_{(-)i}^{\pm}$ are measured for four different bins defined as having %%@
$|\Delta m_{K^{+}K^{-}}^2|$ less than 0.006, 0.010, 0.014, and 0.018 $\mathrm{GeV^2}/c^4$. Here, $\Delta %%@
m_{K^{+}K^{-}}^2= m_{K^{+}K^{-}}^2-m_{\phi}^2$, where $m_\phi$ is the nominal $\phi$ mass \cite{PDG}. The different %%@
ranges allow the result best suited to the experimental resolution to be used in evaluating any systematic %%@
uncertainty related to the $CP$ content of the $D^{0}\to K^{0}_{S}\phi$ decay. The results are given in %%@
Table~\ref{tab:fminus}. The systematic uncertainty contains contributions from the migration matrix uncertainties, %%@
background parameterizations, and branching fraction uncertainties. The methods for determining the systematic %%@
uncertainties are identical to those used for $c_i^{(\prime)}$ and $s_i^{(\prime)}$.

 To improve the precision of the measurement we also determine the $CP$-even fraction, $\mathcal{F}_{+}$, for %%@
$D^{0}\to K^{0}_{L} K^{+} K^{-}$ decays in the same mass-squared intervals. The measurements are given in %%@
Table~\ref{tab:fminus}; the significant sources of systematic uncertainty are the same as those for the value of %%@
$\mathcal{F}_{-}$ measured from $D^{0}\to K^{0}_{S} K^{+} K^{-}$ decays, but are largely uncorrelated. 
 The value of $\mathcal{F}_{+}$ for $D^{0}\to K^{0}_{L} K^{+} K^{-}$ decays will be slightly different to %%@
$\mathcal{F}_{-}$ for $D^{0}\to K^{0}_{S}K^{+}K^{+}$ decays for the same reason that $c_i$ and $s_i$ differ from %%@
$c_i^{\prime}$ and $s_i^{\prime}$, as discussed in Sec. \ref{sec:cisiextract}. Therefore, we correct the measured %%@
value of $\mathcal{F}_{+}$ before combining it with the measurement of $\mathcal{F_{-}}$. We define %%@
$\Delta\mathcal{F}$ as the value of $\mathcal{F}_{+}$ for $D^{0}\to K^{0}_{L} K^{+} K^{-}$ decays minus the value %%@
of $\mathcal{F}_{-}$ for $D^{0}\to K^{0}_{S} K^{+} K^{-}$ decays predicted by the {\it BABAR} amplitude model %%@
\cite{BABAR3}. The value of $\Delta\mathcal{F}$ is given in Table~\ref{tab:fminus}; the uncertainty on %%@
$\Delta\mathcal{F}$ is determined in identical fashion to those on $\Delta c_i$ and $\Delta s_i$, as described in %%@
Sec.~\ref{sec:cisiextract}. We then subtract the value of $\Delta\mathcal{F}$ from the measured value of %%@
$\mathcal{F}_{+}$ for $D^{0}\to K^{0}_{L} K^{+} K^{-}$ decays and average the result with $\mathcal{F}_{-}$. The %%@
combined values of $\mathcal{F}_{-}$ are given in Table~\ref{tab:fminus}. The systematic uncertainty on the average %%@
value of $\mathcal{F}_{-}$ has a significant contribution from the error on $\Delta\mathcal{F}$ as well as the %%@
other sources already discussed.

The results are consistent with no contamination from $CP$-odd eigenstates in the region of the $\phi$ resonance %%@
for all $m_{K^{+}K^{-}}^2$ intervals. The values of $\mathcal{F}_{-}$ can be greater than one due to the background %%@
subtraction resulting in a negative yield. We calculate the lower limit on $\mathcal F_{-}$ at the 90\% confidence %%@
level (CL), by integrating the Gaussian distribution of the average value of $\mathcal{F}_{-}$ within the physical %%@
region $0\leq\mathcal{F}_{-}\leq 1$. The lower limits on $\mathcal{F}_{-}$ at the 90\% CL are 0.89, 0.91, 0.76, and %%@
0.77 for $|\Delta m_{K^{+}K^{-}}^2|$ less than 0.006, 0.010, 0.014, and 0.018 $\mathrm{GeV}^2/c^4$, respectively.

\section{Summary}
\label{sec:summary}
Using 818~$\rm {pb}^{-1}$ of data collected by the CLEO-c detector at 
the $\psi(3770)$ resonance  we have presented measurements of the amplitude-weighted averages of the 
cosine and sine of the strong phase differences beween $D^0$ and $\overline{D}^0 \to K^0_{S,L} h^+h^-$
($h=\pi,K$) decays in bins of Dalitz space, $c_i^{(\prime)}$ and $s_i^{(\prime)}$.  These results are necessary %%@
input for performing model-independent measurements of the CKM-angle $\gamma/\phi_3$~\cite{GIRI,BONDAR2} and
can also be used in model-independent determinations of the charm-mixing parameters~\cite{BONDARCHARM}.
 
The results for $D^0 \to K^0_S \pi^+\pi^-$ are an update to those of our earlier publication~\cite{BRIERE}.
The measurements presented here benefit from an improved analysis procedure and bin choices informed
by an amplitude model developed by the {\it BABAR} collaboration~\cite{BABAR2} which provides a better description 
of Dalitz space than was previously available.  We have given results for bin choices made with an 
equal division in strong-phase difference (`equal $\Delta \delta_D$') and for a division which optimizes the 
foreseen precision on $\gamma/\phi_3$ in a low background environment (`optimal'), as expected at an 
$e^+e^-$ experiment, and for one (`modified optimal') which  gives the best result under the background conditions %%@
anticipated at a hadron-collider experiment such as LHCb. Results have also been presented for 
an equal $\Delta \delta_D$ binning based on an amplitude model devised by the Belle collaboration~\cite{BELLE2}.
We estimate the uncertainty on $\gamma/\phi_3$ to be between 1.7$^\circ$ and 3.9$^\circ$, depending on the binning, %%@
due to the uncertainties on the measured values of $c_i$ and $s_i$. In most cases, this uncertainty is smaller than %%@
that due to the $D^{0}\to K^{0}_{S}\pi^{+}\pi^{-}$ amplitude model in the most recent analyses %%@
\cite{BELLE2,BABAR3}.
 
The results  for $D^0 \to K^0_S K^+K^-$ are the first measurements of $c_i^{(\prime)}$ and $s_i^{(\prime)}$ for %%@
this
decay.  They have been given for equal $\Delta \delta_D$ divisions of Dalitz space based on the amplitude 
model found in Ref.~\cite{BABAR3} for each half of the Dalitz plot divided into two, three and four bins. The %%@
uncertainty on $\gamma/\phi_3$ from the error on $c_i$ and $s_i$ parameters is between 3.2$^\circ$ and 3.9$^\circ$ %%@
depending on the binnings; these uncertainties are 
comparable to those related to the amplitude model in the unbinned analysis \cite{BABAR3}. 

We test the compatibility of the measured values of $c_i$ and $s_i$ with those predicted by the amplitude models %%@
derived from flavor-tagged samples of $D^{0}\to K^{0}_{S}h^{+}h^{-}$ decays. The agreement is reasonable in all %%@
cases, indicating that there are no large errors in the phase measurements provided by these models. 

In addition, we determine the $CP$-odd fraction, $\mathcal{F}_{-}$, for $D^{0}\to K^{0}_S K^{+}K^{-}$ decays in the %%@
region of the $\phi$ resonance. The results are given in different ranges of invariant mass squared about the %%@
$\phi$ resonance. In all intervals considered $\mathcal{F}_{-}$ is greater than 0.76 at 90\% CL.
This result will better constrain systematic uncertainties related to the the $CP$-even content when $D^{0}\to %%@
K^{0}_{S}\phi$ is used as a $CP$-odd eigenstate in an analysis.

\begin{acknowledgments}
We gratefully acknowledge the effort of the CESR staff
in providing us with excellent luminosity and running conditions.
D.~Cronin-Hennessy thanks the A.P.~Sloan Foundation.
This work was supported by
the National Science Foundation,
the U.S. Department of Energy,
the Natural Sciences and Engineering Research Council of Canada, and
the U.K. Science and Technology Facilities Council.
\end{acknowledgments}

\clearpage

\section{EPAPS addendums}
\subsection{Binning and model look-up tables}
The three binnings of the $D^{0}\to K^{0}_{S}K^{+}K^{-}$ Dalitz plot are provided as look-up tables that 
consist of grids of $(m^2_{-},m^{2}_{K^{+}K^{-}})$, in units of $\mathrm{GeV^2}/c^{4}$, and the bin number at that %%@
point. The granularity of the grid is $0.00179~\mathrm{GeV^2}/c^{4}\times 0.00536~\mathrm{GeV^2}/c^{4}$.
The four binnings of the $D^{0}\to K^{0}_{S}\pi^{+}\pi^{-}$ Dalitz plot are provided as look-up tables that consist %%@
of grids of $(m^2_{+},m^{2}_{-})$, in units of $\mathrm{GeV^2}/c^{4}$, and the bin number at that point. The %%@
granularities of the grids for BABAR model and Belle model derived binnings are and %%@
$0.0054~\mathrm{GeV^2}/c^{4}\times 0.0054~\mathrm{GeV^2}/c^{4}$ and $0.00526~\mathrm{GeV^2}/c^{4}\times %%@
0.00526~\mathrm{GeV^2}/c^{4}$, respectively. Table~\ref{tab:lut} lists the binning and the respective filename. 

In addition, look-up tables are also provided that contain the amplitude and phase information for the different %%@
models over the Dalitz plot. The granularity is the same as the binning look-up tables. For the BABAR models %%@
$\delta_D$ at each point is provided and for the Belle model $|\Delta\delta_D|$. Table~\ref{tab:lut_model} lists %%@
the model and the respective filename.

\begin{table}[htb]
\begin{center}
\caption{Binning look-up table filenames.}\label{tab:lut}
\begin{tabular}{lc} \hline\hline
Binning & Look-up table filename \\ \hline
\multicolumn{2}{c}{$D^{0}\to K^{0}_{S}K^{+}K^{-}$} \\ \hline
Equal $\Delta\delta_D$ $\mathcal{N}=2$ & \hspace{0.5cm}{\tt %%@
BinningLUT\_K0SKK\_BABAR2010\_EqualDeltadeltaD\_2bins.txt} \hspace{0.5cm} \\  
Equal $\Delta\delta_D$ $\mathcal{N}=3$ & {\tt BinningLUT\_K0SKK\_BABAR2010\_EqualDeltadeltaD\_3bins.txt} \\
Equal $\Delta\delta_D$ $\mathcal{N}=4$ & {\tt BinningLUT\_K0SKK\_BABAR2010\_EqualDeltadeltaD\_4bins.txt} \\ \hline
\multicolumn{2}{c}{$D^{0}\to K^{0}_{S}\pi^{+}\pi^{-}$} \\ \hline
{\it BABAR} 2008 equal $\Delta\delta_D$ & {\tt BinningLUT\_K0Spipi\_BABAR2008\_EqualDeltadeltaD.txt} \\ 
Optimal  & {\tt BinningLUT\_K0Spipi\_BABAR2008\_Optimal.txt} \\
Modified optimal & {\tt BinningLUT\_K0Spipi\_BABAR2008\_ModifiedOptimal.txt} \\
Belle equal $\Delta\delta_D$ &  {\tt BinningLUT\_K0Spipi\_Belle\_EqualDeltadeltaD.txt} \\ \hline\hline
\end{tabular} 
\end{center}
\end{table}  

\begin{table}[htb]
\begin{center}
\caption{Amplitude and phase look-up table filenames.}\label{tab:lut_model}
\begin{tabular}{lc} \hline\hline
Model & Look-up table filename \\ \hline
BABAR 2010 $D^{0}\to K^{0}_{S}K^{+}K^{-}$ model & \hspace{0.5cm}{\tt ModelLUT\_K0SKK\_BABAR2010.txt}\hspace{0.5cm} %%@
\\  
BABAR 2008 $D^{0}\to K^{0}_{S}\pi^{+}\pi^{-}$ model & {\tt ModelLUT\_K0Spipi\_BABAR2008.txt} \\
Belle $D^{0}\to K^{0}_{S}\pi^{+}\pi^{-}$ model & {\tt ModelLUT\_K0Spipi\_Belle.txt} \\ \hline\hline
\end{tabular} 
\end{center}
\end{table} 

\subsection{Correlation matrices}
Tables~\ref{tab:corr_K0SKK_2bin_stat} to \ref{tab:corr_K0Spipi_belle_syst} give the correlations among %%@
$c^{(\prime)}_{i}$ and $s^{(\prime)}_{i}$ for the different binnings. The correlations are split into statistical %%@
and systematic uncertainties. The correlations due to the uncertainty on the $\Delta c_i$ and $\Delta s_i$ %%@
constraints are included in the statistical correlations. 

\begin{table}[htb]
\caption{Statistical correlation coefficients (\%) among $c^{(\prime)}_{i}$ and $s^{(\prime)}_{i}$ parameters for %%@
the $\mathcal{N}=2$ equal $\Delta\delta_D$ binning of the $D^{0}\to K^{0}_{S}K^{+}K^{-}$ Dalitz %%@
plot.}\label{tab:corr_K0SKK_2bin_stat}
\begin{center} 
\begin{tabular}{lrrrrrrr} \hline\hline
               & $c_2 $ & $s_1$ & $s_2$ & $c^{\prime}_1$ & $c^{\prime}_2$ & $s^{\prime}_1$ & $s^{\prime}_2$ \\ %%@
\hline 
$c_1$          & $ -1.1 $   & $ -0.3 $   & $ -0.3 $   & $  98.8 $  & $ -1.1 $   & $ -0.3 $   & $ -0.3 $ \\ 
$c_2$          &            & $  0.3 $   & $  3.0 $   & $ -1.1 $   & $  91.4 $  & $  0.3 $   & $  3.0 $ \\ 
$s_1$          &            &            & $  7.5 $   & $ -0.4 $   & $  0.3 $   & $  99.4 $  & $  7.5 $ \\ 
$s_2$          &            &            &            & $ -0.3 $   & $  2.7 $   & $  7.5 $   & $  99.5 $ \\
$c^{\prime}_1$ &            &            &            &            & $ -1.1 $   & $ -0.4 $   & $ -0.3 $ \\ 
$c^{\prime}_2$ &            &            &            &            &            & $  0.3 $   & $  2.7 $ \\ 
$s^{\prime}_1$ &            &            &            &            &            &            & $  7.5 $ \\ 
\hline\hline
\end{tabular}
\end{center}
\end{table}

\begin{table}[htb]
\caption{Systematic correlation coefficients (\%) among $c^{(\prime)}_{i}$ and $s^{(\prime)}_{i}$ parameters for %%@
the $\mathcal{N}=2$ equal $\Delta\delta_D$ binning of the $D^{0}\to K^{0}_{S}K^{+}K^{-}$ Dalitz %%@
plot.}\label{tab:corr_K0SKK_2bin_syst}
\begin{center} 
\begin{tabular}{lrrrrrrr} \hline\hline
               & $c_2 $ & $s_1$ & $s_2$ & $c^{\prime}_1$ & $c^{\prime}_2$ & $s^{\prime}_1$ & $s^{\prime}_2$ \\ %%@
\hline
$c_1$          & $  24.4 $  & $  6.7 $   & $  1.5 $   & $  58.3 $  & $  20.6 $  & $  9.0 $   & $  5.3 $ \\ 
$c_2$          &            & $  1.7 $   & $  0.1 $   & $  22.5 $  & $  60.5 $  & $  5.0 $   & $  5.6 $ \\ 
$s_1$          &            &            & $  60.1 $  & $  4.9 $   & $  2.3 $   & $  95.8 $  & $  52.0 $ \\
$s_2$          &            &            &            & $  5.5 $   & $  4.3 $   & $  58.9 $  & $  87.1 $ \\
$c^{\prime}_1$ &            &            &            &            & $  43.2 $  & $  16.1 $  & $  23.1 $ \\
$c^{\prime}_2$ &            &            &            &            &            & $  13.6 $  & $  21.9 $ \\
$s^{\prime}_1$ &            &            &            &            &            &            & $  56.8 $ \\
\hline\hline
\end{tabular}
\end{center}
\end{table}

\begin{table}[htb]
\caption{Statistical correlation coefficients (\%) among $c^{(\prime)}_{i}$ and $s^{(\prime)}_{i}$ parameters for %%@
the $\mathcal{N}=3$ equal $\Delta\delta_D$ binning of the $D^{0}\to K^{0}_{S}K^{+}K^{-}$ Dalitz %%@
plot.}\label{tab:corr_K0SKK_3bin_stat}
\begin{center} 
\begin{tabular}{lrrrrrrrrrrr} \hline\hline
                & $c_2 $ & $c_3$ & $s_1$ & $s_2$ & $s_3$ & $c^{\prime}_1$ & $c^{\prime}_2$ & $c^{\prime}_3$ & %%@
$s^{\prime}_1$ & $s^{\prime}_2$ & $s^{\prime}_3$ \\ \hline 
$c_1$           & $  0.6 $  & $ -2.9 $  & $ -1.3 $  & $ -0.6 $  & $ -0.2 $  & $  97.6 $ & $  0.6 $  & $ -2.1 $  & $ %%@
-1.3 $  & $ -0.6 $  & $ -0.1 $ \\ 
$c_2$           &           & $  0.8 $  & $  1.1 $  & $  3.2 $  & $  -0.0 $ & $  0.6 $  & $  98.0 $ & $  0.4 $  & $  %%@
1.1 $  & $  3.2 $  & $  -0.0 $ \\
$c_3$           &           &           & $ -0.3 $  & $  0.2 $  & $  2.5 $  & $ -2.8 $  & $  0.8 $  & $  66.4 $ & $ %%@
-0.3 $  & $  0.2 $  & $  2.4 $ \\ 
$s_1$           &           &           &           & $ -2.0 $  & $  6.1 $  & $ -1.3 $  & $  1.1 $  & $ -0.1 $  & $  %%@
99.4 $ & $ -2.0 $  & $  5.8 $ \\ 
$s_2$           &           &           &           &           & $ -3.3 $  & $ -0.5 $  & $  3.2 $  & $  0.1 $  & $ %%@
-1.9 $  & $  99.9 $ & $ -3.0 $ \\ 
$s_3$           &           &           &           &           &           & $ -0.2 $  & $  -0.0 $ & $  2.2 $  & $  %%@
6.1 $  & $ -3.3 $  & $  93.7 $ \\
$c^{\prime}_1$  &           &           &           &           &           &           & $  0.6 $  & $ -2.0 $  & $ %%@
-1.3 $  & $ -0.5 $  & $ -0.1 $ \\ 
$c^{\prime}_2$  &           &           &           &           &           &           &           & $  0.4 $  & $  %%@
1.1 $  & $  3.2 $  & $  -0.0 $ \\
$c^{\prime}_3$  &           &           &           &           &           &           &           &           & $ %%@
-0.1 $  & $  0.1 $  & $  2.2 $ \\ 
$s^{\prime}_1$  &           &           &           &           &           &           &           &           &           %%@
& $ -1.9 $  & $  5.8 $ \\ 
$s^{\prime}_2$  &           &           &           &           &           &           &           &           &           %%@
&           & $ -3.0 $ \\ 
\hline\hline
\end{tabular}
\end{center}
\end{table}

\begin{table}[htb]
\caption{Systematic correlation coefficients (\%) among $c^{(\prime)}_{i}$ and $s^{(\prime)}_{i}$ parameters for %%@
the $\mathcal{N}=3$ equal $\Delta\delta_D$ binning of the $D^{0}\to K^{0}_{S}K^{+}K^{-}$ Dalitz %%@
plot.}\label{tab:corr_K0SKK_3bin_syst}
\begin{center} 
\begin{tabular}{lrrrrrrrrrrr} \hline\hline
                & $c_2 $ & $c_3$ & $s_1$ & $s_2$ & $s_3$ & $c^{\prime}_1$ & $c^{\prime}_2$ & $c^{\prime}_3$ & %%@
$s^{\prime}_1$ & $s^{\prime}_2$ & $s^{\prime}_3$ \\ \hline 
$c_1$           & $  21.9 $  & $  6.4 $   & $  7.2 $   & $ -2.2 $   & $  2.4 $   & $  65.2 $  & $  19.9 $  & $  %%@
14.5 $  & $  7.5 $   & $  0.8 $   & $  2.4 $ \\ 
$c_2$           &            & $  11.9 $  & $  1.2 $   & $ -15.6 $  & $ -0.5 $   & $  14.0 $  & $  49.5 $  & $  %%@
16.6 $  & $  1.2 $   & $ -14.5 $  & $ -0.8 $ \\ 
$c_3$           &            &            & $ -0.5 $   & $  12.6 $  & $ -5.5 $   & $  7.3 $   & $  10.3 $  & $  %%@
73.5 $  & $ -0.3 $   & $  13.4 $  & $ -6.1 $ \\ 
$s_1$           &            &            &            & $  38.5 $  & $  8.0 $   & $  7.4 $   & $  4.5 $   & $ -0.1 %%@
$   & $  99.4 $  & $  38.0 $  & $  7.9 $ \\ 
$s_2$           &            &            &            &            & $  7.7 $   & $ -0.3 $   & $ -6.1 $   & $  6.0 %%@
$   & $  38.4 $  & $  95.5 $  & $  7.8 $ \\ 
$s_3$           &            &            &            &            &            & $  9.9 $   & $  10.4 $  & $  2.0 %%@
$   & $  8.4 $   & $  10.6 $  & $  96.9 $ \\
$c^{\prime}_1$  &            &            &            &            &            &            & $  41.9 $  & $  %%@
30.7 $  & $  8.5 $   & $  10.5 $  & $  9.8 $ \\ 
$c^{\prime}_2$  &            &            &            &            &            &            &            & $  %%@
35.7 $  & $  5.7 $   & $  5.8 $   & $  10.1 $ \\
$c^{\prime}_3$  &            &            &            &            &            &            &            &            %%@
& $  0.8 $   & $  14.2 $  & $  1.3 $ \\ 
$s^{\prime}_1$  &            &            &            &            &            &            &            &            %%@
&            & $  38.3 $  & $  8.3 $ \\ 
$s^{\prime}_2$  &            &            &            &            &            &            &            &            %%@
&            &            & $  10.7 $ \\
\hline\hline
\end{tabular}
\end{center}
\end{table}

\begin{table}[htb]
\caption{Statistical correlation coefficients (\%) among $c^{(\prime)}_{i}$ and $s^{(\prime)}_{i}$ parameters for %%@
the $\mathcal{N}=4$ equal $\Delta\delta_D$ binning of the $D^{0}\to K^{0}_{S}K^{+}K^{-}$ Dalitz %%@
plot.}\label{tab:corr_K0SKK_4bin_stat}
\begin{center} 
\begin{tabular}{lrrrrrrrrrrrrrrr} \hline\hline
      & $c_2 $ & $c_3$ & $c_4$ & $s_1$ & $s_2$ & $s_3$ & $s_4$ & $c'_1$ & $c'_2$ & $c'_3$ & $c'_4$ & $s'_1$ & %%@
$s'_2$ & $s'_3$ & $s'_4$ \\ \hline 
$c_1$ &$  0.6 $ &$ -0.2 $ &$ -2.6 $ &$ -0.9 $ &$  0.1 $ &$  -0.0 $&$  0.1 $ &$  98.2 $&$  0.6 $ &$ -0.2 $ &$ -1.7 $ %%@
&$ -0.9 $ &$  0.1 $ &$  -0.0 $&$  0.0 $ \\ 
$c_2$ &         &$  0.8 $ &$ -0.1 $ &$  0.6 $ &$  1.8 $ &$ -0.7 $ &$  0.1 $ &$  0.6 $ &$  99.2 $&$  0.8 $ &$ -0.2 $ %%@
&$  0.6 $ &$  1.8 $ &$ -0.7 $ &$  0.1 $ \\ 
$c_3$ &         &         &$ -0.1 $ &$  0.4 $ &$ -0.3 $ &$  6.4 $ &$  0.1 $ &$ -0.2 $ &$  0.8 $ &$  98.0 $&$ -0.1 $ %%@
&$  0.4 $ &$ -0.3 $ &$  6.4 $ &$  0.1 $ \\ 
$c_4$ &         &         &         &$  0.6 $ &$ -0.1 $ &$  0.1 $ &$  8.8 $ &$ -2.5 $ &$ -0.1 $ &$ -0.1 $ &$  63.5 %%@
$&$  0.6 $ &$ -0.1 $ &$  0.1 $ &$  8.4 $ \\ 
$s_1$ &         &         &         &         &$ -8.0 $ &$  8.5 $ &$  16.0 $&$ -0.9 $ &$  0.6 $ &$  0.4 $ &$  0.5 $ %%@
&$  99.7 $&$ -8.0 $ &$  8.5 $ &$  14.9 $ \\
$s_2$ &         &         &         &         &         &$ -6.6 $ &$ -1.3 $ &$  0.1 $ &$  1.8 $ &$ -0.3 $ &$ -0.1 $ %%@
&$ -7.9 $ &$  99.9 $&$ -6.6 $ &$ -1.3 $ \\ 
$s_3$ &         &         &         &         &         &         &$  1.5 $ &$  -0.0 $&$ -0.7 $ &$  6.2 $ &$  0.0 $ %%@
&$  8.5 $ &$ -6.6 $ &$  99.8 $&$  1.4 $ \\ 
$s_4$ &         &         &         &         &         &         &         &$  0.0 $ &$  0.1 $ &$  0.1 $ &$  6.6 $ %%@
&$  15.9 $&$ -1.3 $ &$  1.5 $ &$  93.2 $ \\
$c'_1$&         &         &         &         &         &         &         &         &$  0.6 $ &$ -0.2 $ &$ -1.7 $ %%@
&$ -0.9 $ &$  0.1 $ &$  -0.0 $&$  0.0 $ \\ 
$c'_2$&         &         &         &         &         &         &         &         &         &$  0.8 $ &$ -0.2 $ %%@
&$  0.6 $ &$  1.8 $ &$ -0.7 $ &$  0.1 $ \\ 
$c'_3$&         &         &         &         &         &         &         &         &         &         &$ -0.1 $ %%@
&$  0.4 $ &$ -0.3 $ &$  6.2 $ &$  0.1 $ \\ 
$c'_4$&         &         &         &         &         &         &         &         &         &         &         %%@
&$  0.5 $ &$ -0.1 $ &$  0.0 $ &$  6.4 $ \\ 
$s'_1$&         &         &         &         &         &         &         &         &         &         &         %%@
&         &$ -8.0 $ &$  8.5 $ &$  14.8 $ \\
$s'_2$&         &         &         &         &         &         &         &         &         &         &         %%@
&         &         &$ -6.6 $ &$ -1.3 $ \\ 
$s'_3$&         &         &         &         &         &         &         &         &         &         &         %%@
&         &         &         &$  1.4 $ \\ 
\hline\hline
\end{tabular}
\end{center}
\end{table}

\begin{table}[htb]
\caption{Systematic correlation coefficients (\%) among $c^{(\prime)}_{i}$ and $s^{(\prime)}_{i}$ parameters for %%@
the $\mathcal{N}=4$ equal $\Delta\delta_D$ binning of the $D^{0}\to K^{0}_{S}K^{+}K^{-}$ Dalitz %%@
plot.}\label{tab:corr_K0SKK_4bin_syst}
\begin{center} 
\begin{tabular}{lrrrrrrrrrrrrrrr} \hline\hline
      & $c_2 $ & $c_3$ & $c_4$ & $s_1$ & $s_2$ & $s_3$ & $s_4$ & $c'_1$ & $c'_2$ & $c'_3$ & $c'_4$ & $s'_1$ & %%@
$s'_2$ & $s'_3$ & $s'_4$ \\ \hline 
$c_1$ &$  14.5 $ &$  19.7 $ &$  4.3 $  &$  6.3 $  &$ -4.0 $  &$  5.6 $  &$  2.2 $  &$  67.7 $ &$  15.2 $ &$  20.0 $ %%@
&$  10.9 $ &$  6.7 $  &$ -3.1 $  &$  6.8 $  &$  4.3 $ \\ 
$c_2$ &          &$  19.1 $ &$  6.5 $  &$  2.2 $  &$ -7.3 $  &$  3.7 $  &$ -1.1 $  &$  12.4 $ &$  60.1 $ &$  17.2 $ %%@
&$  9.6 $  &$  2.4 $  &$ -6.7 $  &$  4.4 $  &$  0.1 $ \\ 
$c_3$ &          &          &$  6.9 $  &$  10.9 $ &$ -13.6 $ &$  24.8 $ &$ -0.4 $  &$  18.3 $ &$  17.8 $ &$  78.5 $ %%@
&$  12.9 $ &$  11.3 $ &$ -12.6 $ &$  25.7 $ &$  1.8 $ \\ 
$c_4$ &          &          &          &$  1.3 $  &$  4.1 $  &$  1.3 $  &$ -1.6 $  &$  8.8 $  &$  10.6 $ &$  10.3 $ %%@
&$  80.9 $ &$  1.7 $  &$  5.0 $  &$  2.6 $  &$ -0.7 $ \\ 
$s_1$ &          &          &          &          &$ -31.0 $ &$  71.0 $ &$  34.7 $ &$  5.4 $  &$  3.0 $  &$  9.4 $  %%@
&$  0.3 $  &$  99.7 $ &$ -30.5 $ &$  70.3 $ &$  33.2 $ \\
$s_2$ &          &          &          &          &          &$ -27.8 $ &$ -0.1 $  &$ -1.8 $  &$ -3.3 $  &$ -9.6 $  %%@
&$  3.9 $  &$ -30.9 $ &$  99.2 $ &$ -27.2 $ &$  1.9 $ \\ 
$s_3$ &          &          &          &          &          &          &$  13.9 $ &$  4.3 $  &$  3.1 $  &$  19.1 $ %%@
&$  0.7 $  &$  70.8 $ &$ -27.5 $ &$  98.7 $ &$  12.3 $ \\
$s_4$ &          &          &          &          &          &          &          &$  2.1 $  &$  0.2 $  &$  0.4 $  %%@
&$ -6.2 $  &$  34.6 $ &$  0.1 $  &$  13.9 $ &$  96.7 $ \\
$c'_1$&          &          &          &          &          &          &          &          &$  37.3 $ &$  35.1 $ %%@
&$  23.2 $ &$  7.1 $  &$  2.3 $  &$  10.0 $ &$  10.9 $ \\
$c'_2$&          &          &          &          &          &          &          &          &          &$  35.4 $ %%@
&$  24.0 $ &$  4.8 $  &$  1.0 $  &$  9.1 $  &$  9.4 $ \\ 
$c'_3$&          &          &          &          &          &          &          &          &          &          %%@
&$  23.1 $ &$  10.9 $ &$ -6.1 $  &$  23.8 $ &$  8.0 $ \\ 
$c'_4$&          &          &          &          &          &          &          &          &          &          %%@
&          &$  1.4 $  &$  6.4 $  &$  4.2 $  &$ -2.6 $ \\ 
$s'_1$&          &          &          &          &          &          &          &          &          &          %%@
&          &          &$ -30.2 $ &$  70.5 $ &$  33.6 $ \\
$s'_2$&          &          &          &          &          &          &          &          &          &          %%@
&          &          &          &$ -26.1 $ &$  3.1 $ \\ 
$s'_3$&          &          &          &          &          &          &          &          &          &          %%@
&          &          &          &          &$  13.9 $ \\
\hline\hline
\end{tabular} 
\end{center}
\end{table}

\begin{table}
\begin{center}
\caption{Statistical correlation coefficients (\%) among $c^{(\prime)}_{i}$ and $s^{(\prime)}_{i}$ parameters for %%@
the optimal binning of the $D^{0}\to K^{0}_{S}\pi^{+}\pi^{-}$ Dalitz plot.}\label{tab:corr_K0Spipi_opt_stat} 
\rotatebox{90}{
\mbox{
{\begin{tabular}{l@{\hspace{6pt}}r@{\hspace{6pt}}r@{\hspace{6pt}}r@{\hspace{6pt}}
r@{\hspace{6pt}}r@{\hspace{6pt}}r@{\hspace{6pt}}r@{\hspace{6pt}}r@{\hspace{6pt}}r@{\hspace{6pt}}
r@{\hspace{6pt}}r@{\hspace{6pt}}r@{\hspace{6pt}}r@{\hspace{6pt}}r@{\hspace{6pt}}r@{\hspace{6pt}}
r@{\hspace{6pt}}r@{\hspace{6pt}}r@{\hspace{6pt}}r@{\hspace{6pt}}r@{\hspace{6pt}}r@{\hspace{6pt}}
r@{\hspace{6pt}}r@{\hspace{6pt}}r@{\hspace{6pt}}r@{\hspace{6pt}}r@{\hspace{6pt}}r@{\hspace{6pt}}
r@{\hspace{6pt}}r@{\hspace{6pt}}r@{\hspace{6pt}}r}
\hline\hline
  & $c_2$& $c_3$& $c_4$& $c_5$& $c_6$& $c_7$& $c_8$& $s_1$& $s_2$& $s_3$& $s_4$& $s_5$& $s_6$& $s_7$& $s_8$& %%@
$c^{\prime}_1$& $c^{\prime}_2$& $c^{\prime}_3$& $c^{\prime}_4$& $c^{\prime}_5$& $c^{\prime}_6$& $c^{\prime}_7$& %%@
$c^{\prime}_8$& $s^{\prime}_1$& $s^{\prime}_2$& $s^{\prime}_3$& $s^{\prime}_4$& $s^{\prime}_5$& $s^{\prime}_6$& %%@
$s^{\prime}_7$& $s^{\prime}_8$\\ \hline 
$c_1$& 1& -1& -1& 0& -1& 0& -2& 0& 2& 0& 2& 0& 1& -1& -5& 26& 1& -1& -1& -3& 0& 0& 0& 0& 3& 0& 2& 0& 1& -1& -4\\
$c_2$&  & 2& 12& 2& 0& 0& -3& 0& 2& 2& -8& 0& 1& 0& -3& -1& 90& 1& 4& 0& 0& 0& -1& 0& 2& 2& -7& 0& 0& 1& -2\\
$c_3$&  & & 4& 0& 0& 0& -1& 0& 11& 6& 4& -1& 12& 2& -5& 0& 2& 63& 0& -1& 0& 0& 0& 0& 12& 6& 2& -1& 12& 3& -5\\
$c_4$&  &  &  & 0& 5& 8& 0& 0& 3& 0& 3& 0& 2& 0& -2& -1& 13& 7& 24& 1& 12& 8& 4& 0& 3& 0& 2& 0& 2& 0& -2\\
$c_5$& & & & & 3& 0& -1& 0& -6& -6& -8& 1& -3& 4& 4& 1& 2& 0& -3& 24& 5& 0& 0& 0& -6& -6& -7& 0& -2& 3& 2\\
$c_6$& & & & && 0& 0& 0& -8& -12& 9& -4& 7& 0& 1& -1& 0& 0& 0& -1& 43& 0& 0& 0& -9& -13& 8& -3& 3& 0& 1\\
$c_7$& & & & & & & -7& 0& 0& 0& 0& 0& 0& -2& -1& -2& 0& 0& 6& 1& 0& 92& -2& 0& 0& 0& 0& 0& 0& -2& 0\\
$c_8$& & & & & & & & 0& 0& -1& 1& 0& -1& -3& -3& -2& -3& -1& 2& 0& 0& -7& 25& 0& 0& -1& 1& 0& 0& -2& -1\\
$s_1$& & & & & & & & & -8& 8& -2& 6& 10& 7& 18& 0& 0& 0& 0& 0& 0& 0& 0& 94& -7& 8& -4& 6& 8& 6& 14\\
$s_2$& & & & & & & & & & 15& 15& 10& 25& 2& -5& 0& 1& 9& 0& -2& -3& 0& 0& -8& 94& 16& 9& 9& 19& 3& -8\\
$s_3$& & & & & & & & & & & 1& 8& 57& 18& 15& 0& 1& 5& 0& 0& -10& 0& 0& 7& 17& 93& 0& 5& 50& 18& 10\\
$s_4$& & & & & & & & & & & & -5& 17& -9& -7& -1& -9& 3& 0& -1& 19& 0& 1& -1& 18& 0& 86& -4& 25& -9& -11\\
$s_5$& & & & & & & & & & && & -4& 2& 20& 0& 0& -2& 0& -4& -2& 0& 0& 6& 11& 10& -7& 89& -12& 3& 20\\
$s_6$& & & & & & & & & & & & & & 20& 8& 0& 1& 12& 0& 0& 0& 0& 0& 10& 27& 59& 12& -4& 78& 19& 1\\
$s_7$& & & & & & & & & & & & & & & 24& 0& 1& 3& 0& 4& -4& -2& 0& 7& 2& 18& -11& 0& 19& 89& 16\\
$s_8$& & & & & & & & & & & & & & & & -1& -3& -3& -1& 1& -1& -1& 2& 17& -5& 15& -8& 17& 7& 23& 76\\
$c^{\prime}_1$& & & & & & & & & & & & & & & & & 0& 0& -1& 0& -1& -2& 0& 0& 0& 0& -1& 0& 0& 0& -1\\
$c^{\prime}_2$& & & & & & & & & & & & & & & & & & 1& 4& 0& 0& 0& -1& 0& 1& 2& -8& 0& 0& 1& -2\\
$c^{\prime}_3$& & & & & & & & & & & & & & & & & & & 1& 0& 0& 0& 0& 0& 9& 5& 2& -2& 11& 3& -4\\
$c^{\prime}_4$& & & & & & & & & & & & & & & & & & & & 0& 2& 6& 1& 0& 0& 0& 0& 0& 0& 0& -1\\
$c^{\prime}_5$& & & & & & & & & & & & & & & & & & & & & 0& 1& 0& 0& -2& 0& -1& -4& 0& 4& 0\\
$c^{\prime}_6$& & & & & & & & & & & & & & & & & & & & & & 0& 0& 0& -3& -11& 17& -2& -1& -4& -1\\
$c^{\prime}_7$& & & & & & & & & & & & & & & & & & & & & & & -2& 0& 0& 0& 0& 0& 0& -2& 0\\
$c^{\prime}_8$& & & & & & & & & & & & & & & & & & & & & & & & 0& 0& 0& 1& 0& 0& 0& 3\\
$s^{\prime}_1$& & & & & & & & & & & & & & & & & & & & & & & & & -7& 7& -3& 6& 8& 6& 14\\
$s^{\prime}_2$& & & & & & & & & & & & & & & & & & & & && & & & & 17& 12& 9& 22& 3& -9\\
$s^{\prime}_3$& & & & & & & & & & & & & & & & & & & & & & & & & & & -1& 7& 50& 18& 10\\
$s^{\prime}_4$& & & & & & & & & & & & & & & & & & & & & & & & & & & & -6& 20& -12& -11\\
$s^{\prime}_5$& & & & & & & & & & & & & & & & & & & & & & & & & & & & & -11& 0& 18\\
$s^{\prime}_6$& & & & & & & & & & & & & & & & & & & & & & & & & & & & & & 17& 1\\
$s^{\prime}_7$& & & & & & & & & & & & & & & & & & & & & & & & & & & & & & & 15\\
\hline\hline
\end{tabular}
}
}
}
\end{center}
\end{table}

\begin{table}
\begin{center}
\caption{Systematic correlation coefficients (\%) among $c^{(\prime)}_{i}$ and $s^{(\prime)}_{i}$ parameters for %%@
the optimal binning of the $D^{0}\to K^{0}_{S}\pi^{+}\pi^{-}$ Dalitz plot.}\label{tab:corr_K0Spipi_opt_syst} 
\rotatebox{90}{
\mbox{
{
\begin{tabular}{l@{\hspace{6pt}}r@{\hspace{6pt}}r@{\hspace{6pt}}r@{\hspace{6pt}}
r@{\hspace{6pt}}r@{\hspace{6pt}}r@{\hspace{6pt}}r@{\hspace{6pt}}r@{\hspace{6pt}}r@{\hspace{6pt}}
r@{\hspace{6pt}}r@{\hspace{6pt}}r@{\hspace{6pt}}r@{\hspace{6pt}}r@{\hspace{6pt}}r@{\hspace{6pt}}
r@{\hspace{6pt}}r@{\hspace{6pt}}r@{\hspace{6pt}}r@{\hspace{6pt}}r@{\hspace{6pt}}r@{\hspace{6pt}}
r@{\hspace{6pt}}r@{\hspace{6pt}}r@{\hspace{6pt}}r@{\hspace{6pt}}r@{\hspace{6pt}}r@{\hspace{6pt}}
r@{\hspace{6pt}}r@{\hspace{6pt}}r@{\hspace{6pt}}r}
\hline\hline 
&  $c_2$& $c_3$& $c_4$& $c_5$& $c_6$& $c_7$& $c_8$& $s_1$& $s_2$& $s_3$& $s_4$& $s_5$& $s_6$& $s_7$& $s_8$& %%@
$c^{\prime}_1$& $c^{\prime}_2$& $c^{\prime}_3$& $c^{\prime}_4$& $c^{\prime}_5$& $c^{\prime}_6$& $c^{\prime}_7$& %%@
$c^{\prime}_8$& $s^{\prime}_1$& $s^{\prime}_2$& $s^{\prime}_3$& $s^{\prime}_4$& $s^{\prime}_5$& $s^{\prime}_6$& %%@
$s^{\prime}_7$& $s^{\prime}_8$\\ \hline 
$c_1$& 79& 87& 69& 91& 89& 79& 88& -13& 27& -20& 30& -4& 43& 0& -46& 60& 77& 75& 18& 35& 70& 76& 47& -16& 27& -21& %%@
16& -3& 34& 3& -41\\
$c_2$& & 76& 68& 82& 82& 66& 71& 0& 5& -16& 10& -6& 27& 1& -24& 48& 96& 67& 28& 37& 65& 64& 40& -2& 4& -17& 6& -6& %%@
23& 0& -21\\
$c_3$& & & 78& 91& 89& 76& 79& -14& 4& -26& 23& -16& 27& -10& -31& 51& 76& 88& 36& 44& 69& 73& 50& -12& 3& -26& 19& %%@
-14& 25& -11& -25\\
$c_4$& & & & 83& 76& 72& 58& -6& -4& -22& 7& -16& 20& -6& -16& 38& 66& 72& 53& 47& 61& 70& 44& -6& -5& -23& 9& -14& %%@
17& -8& -12\\
$c_5$& & & & & 92& 80& 81& -9& 6& -22& 16& -13& 29& -6& -30& 50& 80& 79& 36& 46& 71& 78& 50& -9& 6& -23& 11& -12& %%@
26& -6& -23\\
$c_6$& & & & & & 75& 82& -6& 8& -22& 14& -10& 31& -3& -32& 49& 81& 78& 28& 36& 76& 73& 48& -6& 8& -23& 10& -10& 26& %%@
-3& -25\\
$c_7$& & & & & & & 74& -15& 20& -20& 24& -7& 34& -8& -41& 40& 64& 66& 30& 41& 58& 99& 59& -16& 20& -20& 12& -6& 28& %%@
-4& -34\\
$c_8$& & & & & & & & -10& 21& -20& 27& -3& 36& -3& -44& 50& 70& 67& 16& 30& 62& 72& 53& -12& 21& -23& 18& -6& 27& %%@
0& -36\\
$s_1$& & & & & & & & & -15& 23& -26& 16& 6& 31& 32& -23& 0& -10& 5& -8& -6& -15& -7& 84& -18& 26& -29& 8& -2& 22& %%@
31\\
$s_2$& & & & & & & & & & 26& 56& 34& 75& 25& -70& 32& 3& 5& -39& -9& 18& 19& -1& -24& 97& 25& 13& 28& 64& 37& -76\\
$s_3$& & & & & & & & & & & 13& 34& 47& 28& 6& -10& -16& -27& -17& -11& -13& -20& -17& 21& 21& 91& 4& 24& 42& 24& %%@
4\\
$s_4$& & & & & & & & & & & & 13& 57& 0& -52& 33& 6& 23& -16& 8& 19& 22& 10& -30& 55& 10& 65& 10& 49& 3& -49\\
$s_5$& & & & & & & & & & & & & 28& 10& -7& 2& -6& -13& -28& -17& -5& -9& -2& 12& 32& 30& 2& 91& 24& 10& -15\\
$s_6$& & & & & & & & & & & & & & 32& -49& 35& 27& 31& -16& 9& 42& 33& 12& -2& 72& 43& 21& 20& 79& 37& -56\\
$s_7$& & & & & & & & & & & & & & & 4& 0& 2& -10& -12& -6& 0& -7& -14& 22& 26& 27& -16& 6& 24& 83& -4\\
$s_8$& & & & & & & & & & & & & & & & -42& -22& -29& 22& -7& -34& -39& -19& 36& -71& 3& -20& -6& -40& -10& 79\\
$c^{\prime}_1$& & & & & & & & & & & & & & & & & 48& 47& 10& 30& 44& 38& 30& -23& 33& -11& 16& 4& 28& 2& -42\\
$c^{\prime}_2$& & & & & & & & & & & & & & & & & & 66& 29& 34& 63& 62& 41& 0& 2& -15& 3& -7& 20& 0& -20\\
$c^{\prime}_3$& & & & & & & & & & & & & & & & & & & 39& 42& 65& 64& 48& -9& 3& -21& 19& -12& 24& -10& -21\\
$c^{\prime}_4$& & & & & & & & & & & & & & & & & & & & 44& 19& 30& 26& 7& -38& -14& -3& -25& -11& -18& 26\\
$c^{\prime}_5$& & & & & & & & & & & & & & & & & & & & & 29& 40& 39& -6& -8& -13& 8& -15& 2& -11& 0\\
$c^{\prime}_6$& & & & & & & & & & & & & & & & & & & & & & 56& 37& -6& 17& -10& 8& -5& 34& 2& -28\\
$c^{\prime}_7$& & & & & & & & & & & & & & & & & & & & & & & 60& -17& 19& -20& 11& -7& 27& -4& -33\\
$c^{\prime}_8$& & & & & & & & & & & & & & & & & & & & & & & & -4& -2& -19& 11& 0& 10& -11& -13\\
$s^{\prime}_1$& & & & & & & & & & & & & & & & & & & & & & & & & -27& 22& -35& 7& -7& 11& 29\\
$s^{\prime}_2$& & & & & & & & & & & & & & & & & & & & & & & & & & 19& 12& 26& 60& 38& -77\\
$s^{\prime}_3$& & & & & & & & & & & & & & & & & & & & & & & & & & & 3& 22& 39& 25& 0\\
$s^{\prime}_4$& & & & & & & & & & & & & & & & & & & & & & & & & & & & 2& 20& -18& -19\\
$s^{\prime}_5$& & & & & & & & & & & & & & & & & & & & & & & & & & & & & 19& 5& -16\\
$s^{\prime}_6$& & & & & & & & & & & & & & & & & & & & & & & & & & & & & & 28& -45\\
$s^{\prime}_7$& & & & & & & & & & & & & & & & & & & & & & & & & & & & & & & -20\\ \hline\hline
\end{tabular}
}
}}
\end{center}
\end{table}

\begin{table}
\begin{center}
\caption{Statistical correlation coefficients (\%) among $c^{(\prime)}_{i}$ and $s^{(\prime)}_{i}$ parameters for %%@
the equal $\Delta\delta_D$ binning, according to the {\it BABAR} 2008 model, of the $D^{0}\to %%@
K^{0}_{S}\pi^{+}\pi^{-}$ Dalitz plot.}\label{tab:corr_K0Spipi_equal_stat} 
\rotatebox{90}{
\mbox{
{
\begin{tabular}{c@{\hspace{6pt}}r@{\hspace{6pt}}r@{\hspace{6pt}}r@{\hspace{6pt}}r@{\hspace{6pt}}
r@{\hspace{6pt}}r@{\hspace{6pt}}r@{\hspace{6pt}}r@{\hspace{6pt}}r@{\hspace{6pt}}r@{\hspace{6pt}}r@{\hspace{6pt}}
r@{\hspace{6pt}}r@{\hspace{6pt}}r@{\hspace{6pt}}r@{\hspace{6pt}}r@{\hspace{6pt}}r@{\hspace{6pt}}r@{\hspace{6pt}}
r@{\hspace{6pt}}r@{\hspace{6pt}}r@{\hspace{6pt}}r@{\hspace{6pt}}r@{\hspace{6pt}}r@{\hspace{6pt}}r@{\hspace{6pt}}
r@{\hspace{6pt}}r@{\hspace{6pt}}r@{\hspace{6pt}}r@{\hspace{6pt}}r@{\hspace{6pt}}r}\hline\hline
& $c_2$& $c_3$& $c_4$& $c_5$& $c_6$& $c_7$& $c_8$& $s_1$& $s_2$& $s_3$& $s_4$& $s_5$& $s_6$& $s_7$& $s_8$& %%@
$c^{\prime}_1$& $c^{\prime}_2$& $c^{\prime}_3$& $c^{\prime}_4$& $c^{\prime}_5$& $c^{\prime}_6$& $c^{\prime}_7$& %%@
$c^{\prime}_8$& $s^{\prime}_1$& $s^{\prime}_2$& $s^{\prime}_3$& $s^{\prime}_4$& $s^{\prime}_5$& $s^{\prime}_6$& %%@
$s^{\prime}_7$& $s^{\prime}_8$\\ \hline 
$c_1$&  -2& -3& 5& 7& 3& 1& -2& 0& 0& -2& 0& 0& 0& -1& 0& 69& -2& -1& 5& 5& 1& -2& -1& 0& 0& -1& 0& 0& 0& -1& 0\\
$c_2$&    & 0& 4& 10& 0& 0& 0& 0& -2& 0& 0& 0& 0& 0& 0& -2& 90& -1& 0& 2& -1& -1& 1& 0& -2& 0& 0& 0& 0& 0& 0\\
$c_3$&  & & 0& 0& 0& 2& -4& 16& -4& 75& 7& -10& 0& 45& 4& -1& 1& 54& 0& -2& -4& 15& -3& 13& -5& 51& 7& -11& -1& 53& %%@
3\\
$c_4$&  & & & 1& 0& 0& 5& 0& -1& 0& 7& -1& 0& 0& 0& 6& 5& 0& 36& 0& -2& 4& 7& 0& -2& 0& 6& 0& 0& 0& 0\\
$c_5$&  & & & & 0& 1& 2& 0& 3& 0& 0& 3& 0& 0& 0& 10& 11& 2& 1& 27& 0& 3& 5& 0& 3& 0& 0& 2& 0& 0& 0\\
$c_6$&  & & & & & -1& -1& 1& 0& 0& 0& 0& 0& 0& 0& 4& 0& 1& 0& -1& 51& -1& 0& 2& 0& 0& 0& 0& 0& 0& 0\\
$c_7$&  & & & & & & 0& 2& 3& 6& 0& 2& 0& 1& 2& 2& 0& 0& -2& 0& -2& 23& 0& 4& 3& 6& 0& 2& 0& 3& 2\\
$c_8$&  & & & & & & & -1& 0& -3& 0& 0& 0& -2& 2& -2& 1& -5& 5& 2& -2& -1& 55& -1& 0& -2& -1& 0& 0& -2& 2\\
$s_1$&  & & & & & & & & -8& 18& 11& -18& -7& 15& 10& 0& 0& 7& -1& -1& 0& 5& 0& 77& -9& 10& 10& -20& -7& 16& 10\\
$s_2$&  & & & & & & & & & -3& 10& 31& -6& -2& 0& 0& -2& 0& 0& 1& 0& 0& 0& -9& 92& -4& 8& 25& -6& -1& -1\\
$s_3$&  & & & & & & & & & & 11& -9& -2& 59& 6& -1& 0& 39& 0& -1& -3& 25& -2& 16& -3& 65& 10& -11& -3& 70& 6\\
$s_4$&  & & & & & & & & & & & 0& -4& 13& 13& 0& 0& 4& 2& 0& 0& -2& 0& 11& 11& 7& 79& 0& -5& 15& 13\\
$s_5$&  & & & & & & & & & & & & 6& -10& -11& 0& 0& 0& 0& 1& -1& -2& 0& -16& 33& -4& 0& 85& 7& -10& -11\\
$s_6$&  & & & & & & & & & & & & & -5& -6& 0& 0& 6& 0& 0& 0& 0& 0& -4& -6& 4& -2& 6& 94& -6& -6\\
$s_7$&  & & & & & & & & & & & & & & 3& 0& 0& 22& 0& -1& -1& 15& -1& 15& -2& 47& 11& -11& -6& 79& 3\\
$s_8$&  & & & & & & & & & & & & & & & 0& 0& 0& -2& 0& 0& 2& 1& 8& -1& 4& 16& -12& -6& 4& 94\\
$c^{\prime}_1$&  & & & & & & & & & & & & & & & & -1& 0& 4& 5& 2& -1& -1& 0& 0& 0& 0& 0& 0& 0& 0\\
$c^{\prime}_2$&  & & & & & & & & & & & & & & & & & 0& 0& 2& -1& -1& 1& 0& -3& 0& 0& 0& 0& 0& 0\\
$c^{\prime}_3$&  & & & & & & & & & & & & & & & & & & 0& 0& -1& 8& -3& 6& 0& 27& 4& -1& 6& 27& 0\\
$c^{\prime}_4$&  & & & & & & & & & & & & & & & & & & & 0& -1& 0& 4& -1& 0& 0& 2& 0& 0& 0& -2\\
$c^{\prime}_5$&  & & & & & & & & & & & & & & & & & & & & 0& 0& 2& -1& 1& -1& 0& 1& 0& -1& 0\\
$c^{\prime}_6$&  & & & & & & & & & & & & & & & & & & & & & -1& 0& 0& 0& -2& 0& -1& 0& -2& 0\\
$c^{\prime}_7$&  & & & & & & & & & & & & & & & & & & & & & & 0& 5& 0& 17& -1& -2& 0& 18& 2\\
$c^{\prime}_8$&  & & & & & & & & & & & & & & & & & & & & & & & 0& 0& -1& 0& 0& 0& -2& 1\\
$s^{\prime}_1$&  & & & & & & & & & & & & & & & & & & & & & & & & -9& 9& 10& -18& -5& 15& 8\\
$s^{\prime}_2$&  & & & & & & & & & & & & & & & & & & & & & & & & & -4& 9& 27& -5& -1& -1\\
$s^{\prime}_3$&  & & & & & & & & & & & & & & & & & & & & & & & & & & 7& -5& 3& 50& 3\\
$s^{\prime}_4$&  & & & & & & & & & & & & & & & & & & & & & & & & & & & 0& -2& 13& 15\\
$s^{\prime}_5$&  & & & & & & & & & & & & & & & & & & & & & & & & & & & & 6& -11& -12\\
$s^{\prime}_6$&  & & & & & & & & & & & & & & & & & & & & & & & & & & & & & -6& -6\\
$s^{\prime}_7$&  & & & & & & & & & & & & & & & & & & & & & & & & & & & & & & 4\\\hline\hline
\end{tabular}
}
}}
\end{center}
\end{table}

\begin{table}
\begin{center}
\caption{Systematic correlation coefficients (\%) among $c^{(\prime)}_{i}$ and $s^{(\prime)}_{i}$ parameters for %%@
the equal $\Delta\delta_D$ binning, according to the {\it BABAR} 2008 model, of the $D^{0}\to %%@
K^{0}_{S}\pi^{+}\pi^{-}$ Dalitz plot.}\label{tab:corr_K0Spipi_equal_syst} 
\rotatebox{90}{
\mbox{
{
\begin{tabular}{l@{\hspace{6pt}}r@{\hspace{6pt}}r@{\hspace{6pt}}r@{\hspace{6pt}}
r@{\hspace{6pt}}r@{\hspace{6pt}}r@{\hspace{6pt}}r@{\hspace{6pt}}r@{\hspace{6pt}}r@{\hspace{6pt}}
r@{\hspace{6pt}}r@{\hspace{6pt}}r@{\hspace{6pt}}r@{\hspace{6pt}}r@{\hspace{6pt}}r@{\hspace{6pt}}
r@{\hspace{6pt}}r@{\hspace{6pt}}r@{\hspace{6pt}}r@{\hspace{6pt}}r@{\hspace{6pt}}r@{\hspace{6pt}}
r@{\hspace{6pt}}r@{\hspace{6pt}}r@{\hspace{6pt}}r@{\hspace{6pt}}r@{\hspace{6pt}}r@{\hspace{6pt}}
r@{\hspace{6pt}}r@{\hspace{6pt}}r@{\hspace{6pt}}r}
\hline\hline
&  $c_2$& $c_3$& $c_4$& $c_5$& $c_6$& $c_7$& $c_8$& $s_1$& $s_2$& $s_3$& $s_4$& $s_5$& $s_6$& $s_7$& $s_8$& %%@
$c^{\prime}_1$& $c^{\prime}_2$& $c^{\prime}_3$& $c^{\prime}_4$& $c^{\prime}_5$& $c^{\prime}_6$& $c^{\prime}_7$& %%@
$c^{\prime}_8$& $s^{\prime}_1$& $s^{\prime}_2$& $s^{\prime}_3$& $s^{\prime}_4$& $s^{\prime}_5$& $s^{\prime}_6$& %%@
$s^{\prime}_7$& $s^{\prime}_8$\\ \hline 
$c_1$& 89& 93& 74& 77& 85& 90& 90& 32& 24& 32& 30& 25& -11& 11& 29& 95& 88& 55& 41& 34& 52& 52& 83& 34& 22& 25& 21& %%@
13& -9& 10& 28\\
$c_2$& & 88& 70& 73& 83& 87& 90& 32& 25& 33& 33& 25& -13& 15& 28& 87& 98& 44& 38& 31& 49& 56& 85& 34& 22& 26& 24& %%@
9& -12& 13& 26\\
$c_3$& & & 73& 77& 86& 91& 91& 34& 22& 37& 31& 23& -9& 13& 29& 90& 86& 60& 39& 32& 51& 49& 80& 34& 20& 29& 22& 10& %%@
-9& 12& 27\\
$c_4$& & & & 90& 80& 84& 79& -11& -22& 13& -12& 0& 24& -31& -1& 70& 68& 45& 72& 62& 48& 38& 75& -12& -21& 17& -8& %%@
5& 21& -31& -2\\
$c_5$& & & & & 82& 83& 81& -5& -14& 16& -6& -1& 16& -23& 2& 74& 71& 48& 67& 61& 52& 43& 76& -5& -14& 19& -4& 3& 14& %%@
-22& 2\\
$c_6$& & & & & & 87& 87& 12& 7& 26& 15& 12& 4& -2& 17& 82& 81& 53& 50& 44& 62& 47& 79& 13& 9& 25& 8& 6& 4& -2& 14\\
$c_7$& & & & & & & 91& 17& 6& 24& 15& 15& 3& -5& 16& 86& 85& 50& 51& 42& 52& 51& 84& 17& 6& 21& 9& 9& 2& -6& 15\\
$c_8$& & & & & & & & 24& 15& 29& 24& 19& -4& 4& 20& 88& 89& 50& 47& 39& 49& 50& 90& 26& 13& 24& 17& 10& -4& 2& 18\\
$s_1$& & & & & & & & & 60& 37& 57& 29& -43& 58& 48& 35& 32& 19& -29& -31& 7& 25& 21& 85& 57& 12& 42& 1& -38& 55& %%@
47\\
$s_2$& & & & & & & & & & 31& 55& 45& -41& 67& 51& 24& 24& 9& -36& -30& 8& 15& 12& 63& 91& 13& 38& 11& -36& 62& 44\\
$s_3$& & & & & & & & & & & 31& 23& -9& 35& 40& 35& 33& 21& -3& 0& 10& 14& 24& 41& 27& 70& 19& 0& -7& 31& 39\\
$s_4$& & & & & & & & & & & & 30& -42& 66& 49& 31& 34& 12& -27& -25& 6& 20& 21& 57& 52& 20& 74& 13& -37& 64& 46\\
$s_5$& & & & & & & & & & & & & -20& 27& 34& 23& 25& 11& -10& -11& 10& 14& 13& 33& 41& 12& 18& 56& -19& 27& 30\\
$s_6$& & & & & & & & & & & & & & -56& -28& -13& -15& -1& 33& 27& 7& -10& -2& -45& -42& 1& -36& -7& 93& -53& -27\\
$s_7$& & & & & & & & & & & & & & & 40& 14& 16& 5& -42& -35& -7& 8& 1& 60& 62& 13& 50& 0& -51& 93& 38\\
$s_8$& & & & & & & & & & & & & & & & 29& 29& 17& -22& -18& 13& 8& 11& 51& 44& 29& 35& 11& -26& 39& 93\\
$c^{\prime}_1$& & & & & & & & & & & & & & & & & 85& 55& 38& 33& 50& 53& 80& 37& 22& 26& 22& 8& -11& 12& 28\\
$c^{\prime}_2$& & & & & & & & & & & & & & & & & & 41& 37& 30& 46& 54& 83& 34& 22& 25& 25& 12& -13& 14& 27\\
$c^{\prime}_3$& & & & & & & & & & & & & & & & & & & 17& 21& 37& 30& 41& 16& 9& 17& 8& 3& -4& 5& 18\\
$c^{\prime}_4$& & & & & & & & & & & & & & & & & & & & 54& 36& 27& 51& -28& -34& 8& -22& 5& 30& -40& -21\\
$c^{\prime}_5$& & & & & & & & & & & & & & & & & & & & & 25& 21& 40& -29& -26& 1& -15& 7& 22& -35& -19\\
$c^{\prime}_6$& & & & & & & & & & & & & & & & & & & & & & 39& 44& 8& 8& 12& 4& 7& 5& -5& 12\\
$c^{\prime}_7$& & & & & & & & & & & & & & & & & & & & & & & 46& 24& 17& 12& 11& 6& -7& 6& 8\\
$c^{\prime}_8$& & & & & & & & & & & & & & & & & & & & & & & & 21& 12& 21& 14& 8& -3& 2& 9\\
$s^{\prime}_1$& & & & & & & & & & & & & & & & & & & & & & & & & 57& 18& 40& 0& -42& 57& 45\\
$s^{\prime}_2$& & & & & & & & & & & & & & & & & & & & & & & & & & 9& 35& 10& -36& 58& 37\\
$s^{\prime}_3$& & & & & & & & & & & & & & & & & & & & & & & & & & & -1& 4& 2& 16& 27\\
$s^{\prime}_4$& & & & & & & & & & & & & & & & & & & & & & & & & & & & 3& -34& 49& 33\\
$s^{\prime}_5$& & & & & & & & & & & & & & & & & & & & & & & & & & & & & -11& -2& 8\\
$s^{\prime}_6$& & & & & & & & & & & & & & & & & & & & & & & & & & & & & & -50& -28\\
$s^{\prime}_7$& & & & & & & & & & & & & & & & & & & & & & & & & & & & & & & 34\\\hline\hline
\end{tabular}
}
}}
\end{center}
\end{table}

\begin{table}
\begin{center}
\caption{Statistical correlation coefficients (\%) among $c^{(\prime)}_{i}$ and $s^{(\prime)}_{i}$ parameters for %%@
the modified-optimal binning of the $D^{0}\to K^{0}_{S}\pi^{+}\pi^{-}$ Dalitz %%@
plot.}\label{tab:corr_K0Spipi_modopt_stat} 
\rotatebox{90}{
\mbox{
{
\begin{tabular}{c@{\hspace{6pt}}r@{\hspace{6pt}}r@{\hspace{6pt}}r@{\hspace{6pt}}r@{\hspace{6pt}}
r@{\hspace{6pt}}r@{\hspace{6pt}}r@{\hspace{6pt}}r@{\hspace{6pt}}r@{\hspace{6pt}}r@{\hspace{6pt}}r@{\hspace{6pt}}
r@{\hspace{6pt}}r@{\hspace{6pt}}r@{\hspace{6pt}}r@{\hspace{6pt}}r@{\hspace{6pt}}r@{\hspace{6pt}}r@{\hspace{6pt}}
r@{\hspace{6pt}}r@{\hspace{6pt}}r@{\hspace{6pt}}r@{\hspace{6pt}}r@{\hspace{6pt}}r@{\hspace{6pt}}r@{\hspace{6pt}}
r@{\hspace{6pt}}r@{\hspace{6pt}}r@{\hspace{6pt}}r@{\hspace{6pt}}r@{\hspace{6pt}}r}\hline\hline
& $c_2$& $c_3$& $c_4$& $c_5$& $c_6$& $c_7$& $c_8$& $s_1$& $s_2$& $s_3$& $s_4$& $s_5$& $s_6$& $s_7$& $s_8$& %%@
$c^{\prime}_1$& $c^{\prime}_2$& $c^{\prime}_3$& $c^{\prime}_4$& $c^{\prime}_5$& $c^{\prime}_6$& $c^{\prime}_7$& %%@
$c^{\prime}_8$& $s^{\prime}_1$& $s^{\prime}_2$& $s^{\prime}_3$& $s^{\prime}_4$& $s^{\prime}_5$& $s^{\prime}_6$& %%@
$s^{\prime}_7$& $s^{\prime}_8$\\ \hline 
$c_1$&  -2& 0& 0& -1& 0& 1& 1& 1& 1& 0& 1& 0& 0& -1& -1& 67& -2& -1& -1& 0& 2& 1& -1& 1& 1& 0& 1& 0& 0& -1& -1\\
$c_2$& & 3& 13& 6& 0& -2& -5& 0& -5& -2& -3& -2& -1& 1& -1& -2& 79& -1& 3& 4& 0& -2& -2& 0& -6& -2& -3& -1& -1& 1& %%@
-1\\
$c_3$& & & 1& 0& -1& 0& -3& 0& 8& -1& 3& 0& 3& -1& -3& 1& 3& 16& 0& -3& -1& 1& -1& 0& 9& -1& 3& 0& 3& 0& -3\\
$c_4$& & & & 0& 1& 5& 0& 0& 4& 1& -2& 0& 1& 0& 0& 0& 17& 4& 21& 0& 3& 7& 1& 0& 4& 1& -2& 0& 0& 0& 0\\
$c_5$& & & & & -1& 1& 0& -3& -5& -9& -13& -15& -5& 5& -5& -1& 7& 6& -1& 26& -3& 2& 3& -3& -6& -9& -13& -15& -7& 5& %%@
-6\\
$c_6$& & & & & & -2& 0& 0& -5& -1& 6& 0& 3& 1& 3& -1& 1& -4& -1& 0& 42& -2& 0& 0& -6& -2& 6& 0& 4& 1& 2\\
$c_7$& & & & & & & -1& 0& 0& 0& 0& 0& 0& -4& 0& 0& -2& 0& 7& 0& -1& 65& -1& 0& 0& 0& 0& 0& 0& -3& 0\\
$c_8$& & & & & & & & 0& -2& 0& 0& 0& 0& 0& -1& 2& -4& -2& -1& 3& 0& -1& 35& 0& -2& 0& 0& 0& 0& 0& 0\\
$s_1$& & & & & & & & & 0& 4& 4& 14& 2& -12& -9& 1& 0& 0& 0& -2& 1& 0& 0& 84& 0& 3& 4& 15& 1& -11& -7\\
$s_2$& & & & & & & & & & 22& 12& 10& 15& -9& 8& 0& -8& 3& 0& -6& 0& 0& -1& 0& 91& 22& 11& 9& 8& -10& 4\\
$s_3$& & & & & & & & & & & 11& 19& 21& -9& 19& 0& -2& 0& 0& -5& 0& 0& 0& 7& 24& 92& 10& 15& 26& -8& 17\\
$s_4$& & & & & & & & & & & & 11& 20& -13& 17& 0& -4& -3& 0& -3& 16& 0& 0& 4& 16& 11& 97& 11& 26& -13& 15\\
$s_5$& & & & & & & & & & & & & 8& -10& 18& 0& -2& -1& 0& -17& 3& 0& 0& 14& 11& 20& 10& 65& 11& -9& 18\\
$s_6$& & & & & & & & & & & & & & 0& 13& 0& -1& 0& 0& -1& 4& 0& 0& 2& 17& 20& 20& 4& 66& 0& 13\\
$s_7$& & & & & & & & & & & & & & & 3& -1& 1& 0& 0& 7& -3& -2& 0& -14& -9& -10& -13& -12& 2& 89& 3\\
$s_8$& & & & & & & & & & & & & & & & -3& -1& -1& 0& -4& 4& 0& 3& -11& 9& 19& 17& 13& 13& 3& 83\\
$c^{\prime}_1$& & & & & & & & & & & & & & & & & -2& 0& 0& 0& 0& 1& 0& 1& 1& 0& 0& 0& 0& -1& -2\\
$c^{\prime}_2$& & & & & & & & & & & & & & & & & & 0& 4& 4& 0& -1& -1& 0& -9& -2& -3& -2& -1& 1& -1\\
$c^{\prime}_3$& & & & & & & & & & & & & & & & & & & 0& 0& -2& 0& 0& 0& 3& 0& -3& -1& 0& 0& -1\\
$c^{\prime}_4$& & & & & & & & & & & & & & & & & & & & 0& 0& 5& 0& 0& 0& 0& 0& 0& 0& 0& 0\\
$c^{\prime}_5$& & & & & & & & & & & & & & & & & & & & & -1& 0& 1& -2& -6& -6& -3& -18& -2& 6& -4\\
$c^{\prime}_6$& & & & & & & & & & & & & & & & & & & & & & -1& 0& 1& 0& 0& 15& 2& 4& -3& 3\\
$c^{\prime}_7$& & & & & & & & & & & & & & & & & & & & & & & 0& 0& 0& 0& 0& 0& 0& -2& 0\\
$c^{\prime}_8$& & & & & & & & & & & & & & & & & & & & & & & & 0& -1& 0& 0& 0& 0& 0& 4\\
$s^{\prime}_1$& & & & & & & & & & & & & & & & & & & & & & & & & 0& 6& 4& 15& 2& -13& -8\\
$s^{\prime}_2$& & & & & & & & & & & & & & & & & & & & & & & & & & 24& 15& 10& 11& -10& 5\\
$s^{\prime}_3$& & & & & & & & & & & & & & & & & & & & & & & & & & & 10& 15& 24& -9& 17\\
$s^{\prime}_4$& & & & & & & & & & & & & & & & & & & & & & & & & & & & 10& 26& -13& 15\\
$s^{\prime}_5$& & & & & & & & & & & & & & & & & & & & & & & & & & & & & 6& -11& 13\\
$s^{\prime}_6$& & & & & & & & & & & & & & & & & & & & & & & & & & & & & & 1& 13\\
$s^{\prime}_7$& & & & & & & & & & & & & & & & & & & & & & & & & & & & & & & 4\\
\hline\hline
\end{tabular}
}
}}
\end{center}
\end{table}

\begin{table}
\begin{center}
\caption{Systematic correlation coefficients (\%) among $c^{(\prime)}_{i}$ and $s^{(\prime)}_{i}$ parameters for %%@
the modified-optimal binning of the $D^{0}\to K^{0}_{S}\pi^{+}\pi^{-}$ Dalitz %%@
plot.}\label{tab:corr_K0Spipi_modopt_syst} 
\rotatebox{90}{
\mbox{
{
 \begin{tabular}{l@{\hspace{6pt}}r@{\hspace{6pt}}r@{\hspace{6pt}}r@{\hspace{6pt}}
r@{\hspace{6pt}}r@{\hspace{6pt}}r@{\hspace{6pt}}r@{\hspace{6pt}}r@{\hspace{6pt}}r@{\hspace{6pt}}
r@{\hspace{6pt}}r@{\hspace{6pt}}r@{\hspace{6pt}}r@{\hspace{6pt}}r@{\hspace{6pt}}r@{\hspace{6pt}}
r@{\hspace{6pt}}r@{\hspace{6pt}}r@{\hspace{6pt}}r@{\hspace{6pt}}r@{\hspace{6pt}}r@{\hspace{6pt}}
r@{\hspace{6pt}}r@{\hspace{6pt}}r@{\hspace{6pt}}r@{\hspace{6pt}}r@{\hspace{6pt}}r@{\hspace{6pt}}
r@{\hspace{6pt}}r@{\hspace{6pt}}r@{\hspace{6pt}}r}
\hline\hline
&  $c_2$& $c_3$& $c_4$& $c_5$& $c_6$& $c_7$& $c_8$& $s_1$& $s_2$& $s_3$& $s_4$& $s_5$& $s_6$& $s_7$& $s_8$& %%@
$c^{\prime}_1$& $c^{\prime}_2$& $c^{\prime}_3$& $c^{\prime}_4$& $c^{\prime}_5$& $c^{\prime}_6$& $c^{\prime}_7$& %%@
$c^{\prime}_8$& $s^{\prime}_1$& $s^{\prime}_2$& $s^{\prime}_3$& $s^{\prime}_4$& $s^{\prime}_5$& $s^{\prime}_6$& %%@
$s^{\prime}_7$& $s^{\prime}_8$\\ \hline 
$c_1$&  85& 88& 74& 82& 80& 90& 85& -10& 25& -15& 8& 5& 19& 0& -56& 94& 77& 40& 24& 26& 33& 86& 49& -8& 24& -14& 7& %%@
1& 17& 4& -47\\
$c_2$& & 86& 75& 84& 86& 84& 80& -7& 16& -17& -2& -4& 13& -3& -52& 82& 94& 42& 27& 30& 43& 82& 41& -8& 15& -15& -2& %%@
-5& 11& -1& -46\\
$c_3$& & & 79& 91& 90& 89& 87& -12& -2& -36& -20& -22& 8& -9& -60& 87& 81& 39& 30& 38& 47& 85& 49& -14& -2& -34& %%@
-20& -21& 7& -6& -50\\
$c_4$& & & & 86& 71& 79& 71& -18& 5& -28& -17& -13& 9& -3& -55& 76& 69& 43& 38& 41& 37& 77& 43& -21& 4& -27& -17& %%@
-14& 6& -3& -45\\
$c_5$& & & & & 85& 86& 81& -13& -5& -37& -28& -26& 6& -5& -57& 82& 79& 39& 37& 47& 47& 84& 50& -16& -6& -35& -28& %%@
-23& 3& -3& -49\\
$c_6$& & & & & & 82& 82& -2& 4& -25& -16& -16& 6& -5& -51& 78& 82& 38& 27& 31& 58& 78& 43& -4& 3& -22& -16& -17& 6& %%@
-4& -44\\
$c_7$& & & & & & & 86& -12& 13& -23& -4& -6& 13& -4& -56& 89& 78& 38& 29& 32& 38& 94& 51& -9& 13& -22& -5& -8& 12& %%@
-2& -49\\
$c_8$& & & & & & & & -7& 7& -25& -9& -11& 11& -7& -56& 85& 74& 31& 26& 31& 33& 82& 64& -6& 7& -25& -9& -12& 8& -7& %%@
-47\\
$s_1$& & & & & & & & & 16& 24& 11& 12& 15& 8& 13& -13& -6& -11& -12& -9& 4& -13& -5& 61& 16& 27& 11& 4& 2& 9& 14\\
$s_2$& & & & & & & & & & 68& 73& 73& 46& 28& 2& 16& 13& 6& -7& -34& -23& 11& 0& 22& 95& 64& 71& 60& 39& 25& -2\\
$s_3$& & & & & & & & & & & 61& 67& 47& 25& 30& -23& -17& -12& -16& -39& -29& -24& -17& 27& 68& 94& 59& 56& 34& 19& %%@
24\\
$s_4$& & & & & & & & & & & & 75& 33& 15& 11& 0& -3& 7& -20& -41& -34& -7& -13& 14& 74& 60& 98& 64& 35& 17& 0\\
$s_5$& & & & & & & & & & & & & 36& 23& 23& -4& -8& -7& -23& -41& -32& -6& -11& 19& 73& 64& 74& 86& 35& 21& 17\\
$s_6$& & & & & & & & & & & & & & 5& 7& 16& 11& 13& 0& -7& -3& 9& 0& 12& 44& 46& 30& 29& 57& 7& 2\\
$s_7$& & & & & & & & & & & & & & & 11& -5& -2& -5& 2& -15& -4& -4& -2& 9& 22& 18& 12& 23& 4& 74& 8\\
$s_8$& & & & & & & & & & & & & & & & -58& -49& -30& -21& -27& -20& -50& -33& 17& 1& 28& 12& 21& 15& 6& 76\\
$c^{\prime}_1$& & & & & & & & & & & & & & & & & 75& 41& 27& 32& 30& 87& 54& -11& 15& -21& 0& -6& 14& -2& -51\\
$c^{\prime}_2$& & & & & & & & & & & & & & & & & & 36& 28& 29& 44& 75& 37& -7& 13& -14& -4& -9& 10& -2& -43\\
$c^{\prime}_3$& & & & & & & & & & & & & & & & & & & 25& 31& 20& 37& 18& -12& 7& -6& 6& -7& 6& -4& -27\\
$c^{\prime}_4$& & & & & & & & & & & & & & & & & & & & 38& 18& 30& 23& -22& -6& -15& -23& -24& 1& -5& -18\\
$c^{\prime}_5$& & & & & & & & & & & & & & & & & & & & & 25& 33& 30& -11& -37& -36& -40& -35& -14& -10& -17\\
$c^{\prime}_6$& & & & & & & & & & & & & & & & & & & & & & 37& 6& -10& -22& -24& -33& -30& 2& -2& -17\\
$c^{\prime}_7$& & & & & & & & & & & & & & & & & & & & & & & 55& -10& 10& -24& -7& -7& 10& -2& -45\\
$c^{\prime}_8$& & & & & & & & & & & & & & & & & & & & & & & & -5& 0& -17& -14& -10& 1& -2& -27\\
$s^{\prime}_1$& & & & & & & & & & & & & & & & & & & & & & & & & 20& 24& 13& 8& 1& 5& 8\\
$s^{\prime}_2$& & & & & & & & & & & & & & & & & & & & & & & & & & 64& 72& 60& 38& 18& -3\\
$s^{\prime}_3$& & & & & & & & & & & & & & & & & & & & & & & & & & & 58& 52& 36& 16& 18\\
$s^{\prime}_4$& & & & & & & & & & & & & & & & & & & & & & & & & & & & 63& 36& 16& 0\\
$s^{\prime}_5$& & & & & & & & & & & & & & & & & & & & & & & & & & & & & 24& 20& 16\\
$s^{\prime}_6$& & & & & & & & & & & & & & & & & & & & & & & & & & & & & & 4& 2\\
$s^{\prime}_7$& & & & & & & & & & & & & & & & & & & & & & & & & & & & & & & 1\\\hline\hline
\end{tabular}
}
}}
\end{center}
\end{table}

\begin{table}
\begin{center}
\caption{Statistical correlation coefficients (\%) among $c^{(\prime)}_{i}$ and $s^{(\prime)}_{i}$ parameters for %%@
the equal $\Delta\delta_D$ binning, according to the Belle model, of the $D^{0}\to K^{0}_{S}\pi^{+}\pi^{-}$ Dalitz %%@
plot.}
\label{tab:corr_K0Spipi_belle_stat} 
\rotatebox{90}{
\mbox{
{
\begin{tabular}{c@{\hspace{6pt}}r@{\hspace{6pt}}r@{\hspace{6pt}}r@{\hspace{6pt}}r@{\hspace{6pt}}
r@{\hspace{6pt}}r@{\hspace{6pt}}r@{\hspace{6pt}}r@{\hspace{6pt}}r@{\hspace{6pt}}r@{\hspace{6pt}}r@{\hspace{6pt}}
r@{\hspace{6pt}}r@{\hspace{6pt}}r@{\hspace{6pt}}r@{\hspace{6pt}}r@{\hspace{6pt}}r@{\hspace{6pt}}r@{\hspace{6pt}}
r@{\hspace{6pt}}r@{\hspace{6pt}}r@{\hspace{6pt}}r@{\hspace{6pt}}r@{\hspace{6pt}}r@{\hspace{6pt}}r@{\hspace{6pt}}
r@{\hspace{6pt}}r@{\hspace{6pt}}r@{\hspace{6pt}}r@{\hspace{6pt}}r@{\hspace{6pt}}r}\hline\hline
&  $c_2$& $c_3$& $c_4$& $c_5$& $c_6$& $c_7$& $c_8$& $s_1$& $s_2$& $s_3$& $s_4$& $s_5$& $s_6$& $s_7$& $s_8$& %%@
$c^{\prime}_1$& $c^{\prime}_2$& $c^{\prime}_3$& $c^{\prime}_4$& $c^{\prime}_5$& $c^{\prime}_6$& $c^{\prime}_7$& %%@
$c^{\prime}_8$& $s^{\prime}_1$& $s^{\prime}_2$& $s^{\prime}_3$& $s^{\prime}_4$& $s^{\prime}_5$& $s^{\prime}_6$& %%@
$s^{\prime}_7$& $s^{\prime}_8$\\ \hline 
$c_1$& -3& -2& 5& 10& 0& 1& -3& 0& 0& 0& 0& 0& 0& 0& 0& 70& -3& 0& 5& 6& 0& 0& -3& 0& 0& 0& 0& 0& 0& 0& 0\\
$c_2$& & 2& -3& 8& 1& 1& 0& 0& -3& 0& 0& -1& 0& 0& 0& -3& 91& 0& -3& 5& 0& -1& 0& 0& -3& 0& 0& -1& 0& 0& 0\\
$c_3$& & & -2& 1& -1& -1& 2& 1& 0& 3& -1& 0& 2& -2& -5& -1& 2& 34& -2& -1& -5& -2& 3& 0& 0& 2& 0& 0& 2& -2& -5\\
$c_4$& & & & 0& 1& 2& 0& 0& 0& 0& 5& 0& 0& 0& 0& 7& -3& 0& 32& 0& 2& 8& 0& 0& 0& 0& 3& 0& 1& 0& 0\\
$c_5$& & & & & 0& 2& 1& 0& 0& 0& 0& 3& 0& 0& 0& 13& 9& 5& 0& 32& -1& 5& 2& 0& 1& 0& 0& 3& 0& 0& 0\\
$c_6$& & & & & & -1& 1& 1& 1& 0& -1& 0& 1& 0& 2& 1& 1& 2& 0& -2& 51& 0& 3& 2& 1& 1& -1& 0& 1& 0& 3\\
$c_7$& & & & & & & 0& 2& 1& 1& -5& 0& 1& -2& 1& 1& 1& 0& -2& 2& -2& 32& 0& 3& 1& 3& -3& -1& 1& -2& 1\\
$c_8$& & & & & & & & 0& 0& 0& 0& 0& 0& 0& 3& -2& 0& -1& 2& 0& -1& 0& 61& 0& 0& 0& 0& 0& 0& 0& 2\\
$s_1$& & & & & & & & & -1& 9& 3& -18& 8& 5& 23& 0& 0& 0& -2& -1& 1& 0& 0& 79& -1& 5& 4& -21& 7& 4& 23\\
$s_2$& & & & & & & & & & -3& -7& 9& -10& -5& -13& 0& -3& 0& 1& 0& 0& 2& 0& -1& 88& -4& -8& 8& -10& -4& -13\\
$s_3$& & & & & & & & & & & 24& 1& 3& 57& 9& 0& 0& 0& -1& 0& 0& 0& 0& 10& -3& 77& 19& 0& 2& 55& 10\\
$s_4$& & & & & & & & & & & & 5& 7& 33& 12& 0& 0& 0& 0& 0& -2& -16& 0& 1& -6& 25& 69& 4& 7& 35& 12\\
$s_5$& & & & & & & & & & & & & -5& 0& -6& 0& -2& 5& 1& 1& 0& -1& 0& -14& 11& 4& 8& 91& -4& 0& -5\\
$s_6$& & & & & & & & & & & & & & -1& 0& 0& 0& 6& 0& 0& 1& -1& 0& 6& -8& 8& 9& -5& 94& -1& 0\\
$s_7$& & & & & & & & & & & & & & & 8& 0& 0& -4& -1& 0& 0& 0& 0& 6& -6& 61& 25& 0& -2& 91& 8\\
$s_8$& & & & & & & & & & & & & & & & 0& 0& -2& -2& 0& 1& 0& 0& 22& -13& 7& 16& -8& 0& 9& 96\\
$c^{\prime}_1$& & & & & & & & & & & & & & & & & -2& 0& 4& 6& 0& 0& -2& 0& 0& 0& 0& 0& 0& 0& 0\\
$c^{\prime}_2$& & & & & & & & & & & & & & & & & & 0& -3& 5& 0& 0& 0& 0& -3& 0& 0& -1& 0& 0& 0\\
$c^{\prime}_3$& & & & & & & & & & & & & & & & & & & 0& 1& 0& 0& 0& 0& 0& 0& 0& 4& 6& -4& -2\\
$c^{\prime}_4$& & & & & & & & & & & & & & & & & & & & 0& 0& 1& 1& -2& 1& 0& -1& 1& 0& -1& -2\\
$c^{\prime}_5$& & & & & & & & & & & & & & & & & & & & & -1& 2& 0& 0& 0& 0& 0& 1& 0& 0& 0\\
$c^{\prime}_6$& & & & & & & & & & & & & & & & & & & & & & 0& 0& 2& 1& 0& -2& 0& 1& 0& 1\\
$c^{\prime}_7$& & & & & & & & & & & & & & & & & & & & & & & 0& 1& 2& 0& -11& -1& -1& 0& 0\\
$c^{\prime}_8$& & & & & & & & & & & & & & & & & & & & & & & & 0& 0& 0& 0& 0& 0& 0& 0\\
$s^{\prime}_1$& & & & & & & & & & & & & & & & & & & & & & & & & -2& 7& 3& -17& 5& 6& 22\\
$s^{\prime}_2$& & & & & & & & & & & & & & & & & & & & & & & & & & -4& -7& 9& -8& -5& -13\\
$s^{\prime}_3$& & & & & & & & & & & & & & & & & & & & & & & & & & & 20& 3& 6& 58& 7\\
$s^{\prime}_4$& & & & & & & & & & & & & & & & & & & & & & & & & & & & 7& 9& 26& 16\\
$s^{\prime}_5$& & & & & & & & & & & & & & & & & & & & & & & & & & & & & -5& 0& -7\\
$s^{\prime}_6$& & & & & & & & & & & & & & & & & & & & & & & & & & & & & & -2& 0\\
$s^{\prime}_7$& & & & & & & & & & & & & & & & & & & & & & & & & & & & & & & 9\\ \hline\hline
\end{tabular}
}
}}
\end{center}
\end{table}

\begin{table}
\begin{center}
\caption{Systematic correlation coefficients (\%) among $c^{(\prime)}_{i}$ and $s^{(\prime)}_{i}$ parameters for %%@
the equal $\Delta\delta_D$ binning, according to the Belle model, of the $D^{0}\to K^{0}_{S}\pi^{+}\pi^{-}$ Dalitz %%@
plot.}\label{tab:corr_K0Spipi_belle_syst} 
\rotatebox{90}{
\mbox{
{
\begin{tabular}{l@{\hspace{6pt}}r@{\hspace{6pt}}r@{\hspace{6pt}}r@{\hspace{6pt}}
r@{\hspace{6pt}}r@{\hspace{6pt}}r@{\hspace{6pt}}r@{\hspace{6pt}}r@{\hspace{6pt}}r@{\hspace{6pt}}
r@{\hspace{6pt}}r@{\hspace{6pt}}r@{\hspace{6pt}}r@{\hspace{6pt}}r@{\hspace{6pt}}r@{\hspace{6pt}}
r@{\hspace{6pt}}r@{\hspace{6pt}}r@{\hspace{6pt}}r@{\hspace{6pt}}r@{\hspace{6pt}}r@{\hspace{6pt}}
r@{\hspace{6pt}}r@{\hspace{6pt}}r@{\hspace{6pt}}r@{\hspace{6pt}}r@{\hspace{6pt}}r@{\hspace{6pt}}
r@{\hspace{6pt}}r@{\hspace{6pt}}r@{\hspace{6pt}}r}
\hline\hline
 & $c_2$& $c_3$& $c_4$& $c_5$& $c_6$& $c_7$& $c_8$& $s_1$& $s_2$& $s_3$& $s_4$& $s_5$& $s_6$& $s_7$& $s_8$& %%@
$c^{\prime}_1$& $c^{\prime}_2$& $c^{\prime}_3$& $c^{\prime}_4$& $c^{\prime}_5$& $c^{\prime}_6$& $c^{\prime}_7$& %%@
$c^{\prime}_8$& $s^{\prime}_1$& $s^{\prime}_2$& $s^{\prime}_3$& $s^{\prime}_4$& $s^{\prime}_5$& $s^{\prime}_6$& %%@
$s^{\prime}_7$& $s^{\prime}_8$\\ \hline 
$c_1$& 87& 93& 72& 85& 87& 85& 92& 31& 21& 29& 0& 20& 0& -4& -6& 95& 86& 72& 12& 51& 52& 68& 90& 33& 21& 29& -6& %%@
16& 0& -1& -7\\
$c_2$&  & 89& 77& 87& 89& 88& 89& 20& 0& 20& -10& 1& 11& -13& 8& 86& 98& 66& 25& 55& 59& 62& 85& 19& 0& 17& -8& 0& %%@
10& -13& 8\\
$c_3$&  & & 73& 88& 90& 89& 94& 29& 17& 29& -3& 13& 6& -6& -1& 90& 87& 77& 15& 51& 56& 66& 89& 28& 16& 28& -9& 10& %%@
4& -5& -1\\
$c_4$& & & & 84& 76& 78& 70& 2& -20& 6& -17& -13& 20& -22& 18& 70& 77& 58& 46& 59& 58& 51& 67& 4& -21& 0& -6& -14& %%@
19& -23& 16\\
$c_5$& & & & & 86& 86& 84& 10& -4& 14& -12& 1& 12& -18& 7& 82& 85& 71& 33& 66& 61& 62& 81& 10& -4& 11& -9& 0& 11& %%@
-17& 6\\
$c_6$& & & & & & 88& 88& 15& 0& 15& -10& 2& 14& -15& 4& 83& 87& 71& 25& 54& 67& 63& 86& 15& 0& 12& -9& 1& 12& -15& %%@
3\\
$c_7$& & & & & & & 84& 15& -8& 11& -15& -7& 20& -18& 13& 81& 87& 68& 29& 58& 64& 62& 82& 12& -7& 7& -9& -8& 17& %%@
-19& 11\\
$c_8$& & & & & & & & 30& 23& 32& 0& 22& 0& -2& -5& 90& 87& 71& 11& 47& 53& 69& 92& 33& 22& 31& -11& 19& -1& -1& %%@
-5\\
$s_1$& & & & & & & & & 43& 42& 20& 19& 0& 32& -6& 32& 18& 17& -19& -2& -3& 24& 31& 78& 44& 39& 3& 17& -2& 33& -4\\
$s_2$& & & & & & & & & & 44& 37& 65& -37& 44& -52& 23& -1& 13& -45& -17& -22& 22& 23& 50& 95& 54& 0& 56& -37& 46& %%@
-47\\
$s_3$& & & & & & & & & & & 21& 43& 0& 51& 0& 34& 19& 23& -17& -4& -4& 28& 29& 47& 41& 81& -4& 35& -3& 51& 1\\
$s_4$&  & & & & & & & & & & & 31& -9& 38& -13& 1& -10& -4& -32& -13& -26& 0& 2& 21& 40& 31& 50& 32& -6& 43& -15\\
$s_5$& & & & & & & & & & & & & -41& 33& -42& 20& 0& 13& -36& -10& -15& 20& 20& 24& 62& 47& -1& 87& -38& 37& -39\\
$s_6$& & & & & & & & & & & & & & -19& 44& 0& 14& 2& 25& 19& 20& -2& 1& -6& -38& -14& 7& -41& 91& -22& 36\\
$s_7$& & & & & & & & & & & & & & & -11& -1& -14& -6& -24& -23& -26& 0& -1& 34& 44& 46& 16& 29& -21& 92& -9\\
$s_8$& & & & & & & & & & & & & & & & -8& 9& -1& 25& 15& 20& -11& -10& -11& -52& -13& 1& -40& 42& -18& 94\\
$c^{\prime}_1$& & & & & & & & & & & & & & & & & 84& 72& 12& 49& 49& 67& 87& 35& 22& 32& -8& 17& -1& 0& -7\\
$c^{\prime}_2$& & & & & & & & & & & & & & & & & & 64& 24& 55& 58& 58& 84& 18& -2& 15& -8& 0& 12& -14& 9\\
$c^{\prime}_3$& & & & & & & & & & & & & & & & & & & 11& 40& 44& 60& 70& 23& 11& 22& -9& 9& 4& -7& -2\\
$c^{\prime}_4$& & & & & & & & & & & & & & & & & & & & 37& 44& 18& 13& -20& -44& -20& -5& -31& 27& -28& 23\\
$c^{\prime}_5$& & & & & & & & & & & & & & & & & & & & & 49& 39& 49& -4& -15& -7& -4& -11& 16& -20& 12\\
$c^{\prime}_6$& & & & & & & & & & & & & & & & & & & & & & 44& 52& -7& -20& -4& -14& -10& 15& -27& 20\\
$c^{\prime}_7$& & & & & & & & & & & & & & & & & & & & & & & 66& 23& 23& 31& -10& 14& -1& 1& -9\\
$c^{\prime}_8$& & & & & & & & & & & & & & & & & & & & & & & & 31& 22& 28& -6& 18& 0& 0& -11\\
$s^{\prime}_1$& & & & & & & & & & & & & & & & & & & & & & & & & 47& 41& 1& 14& -10& 34& -11\\
$s^{\prime}_2$& & & & & & & & & & & & & & & & & & & & & & & & & & 50& 1& 54& -37& 47& -49\\
$s^{\prime}_3$& & & & & & & & & & & & & & & & & & & & & & & & & & & -6& 44& -15& 50& -13\\
$s^{\prime}_4$& & & & & & & & & & & & & & & & & & & & & & & & & & & & -2& 7& 17& 4\\
$s^{\prime}_5$& & & & & & & & & & & & & & & & & & & & & & & & & & & & & -41& 31& -35\\
$s^{\prime}_6$& & & & & & & & & & & & & & & & & & & & & & & & & & & & & & -25& 32\\
$s^{\prime}_7$& & & & & & & & & & & & & & & & & & & & & & & & & & & & & & & -18\\ \hline\hline
\end{tabular}
}
}}
\end{center}
\end{table}

\end{document}